%% file: search-to-decision.tex
\documentclass[11pt,fleqn,dvipsnames]{article}
\usepackage[T1]{fontenc}
\usepackage{amssymb, amsmath, graphicx, subfigure, xparse}
\usepackage{float}
\usepackage{physics}
\usepackage{fullpage}
\usepackage{hyperref}
\usepackage{qcircuit}
\usepackage{mdframed}
\usepackage{mathpazo}
\usepackage{todonotes}
\usepackage{qcircuit}
\usepackage{thmtools}
\usepackage{thm-restate}
\usepackage{cleveref}
\usepackage{cancel}

\usepackage{setspace}

\usepackage{booktabs}
\usepackage{tabularx}

\usepackage{pifont}
\newcommand{\cmark}{\ding{51}\,}%
\newcommand{\xmark}{\ding{55}\,}%

\usepackage{complexity}

\usepackage{tikz}
\usetikzlibrary{shapes,backgrounds}

\newtheorem{theorem}{Theorem}[section]
\newtheorem{definition}[theorem]{Definition}

\newtheorem{lemma}[theorem]{Lemma}
\newtheorem{conjecture}[theorem]{Conjecture}
\newtheorem{corollary}[theorem]{Corollary}
\newtheorem{proposition}[theorem]{Proposition}

\newtheorem{fact}[theorem]{Fact}

\numberwithin{equation}{section}

\input{berkeleymacros.tex}

\newenvironment{proof}{\noindent{\bf Proof:} \hspace*{1mm}}{
    \hspace*{\fill} $\Box$ }
\newenvironment{proof_of}[1]{\noindent {\bf Proof of #1:}
    \hspace*{1mm}}{\hspace*{\fill} $\Box$ }

\newcommand{\oracle}{\mathcal{O}}

\def\END{{\bf end}\ }
\def\IF{{\bf if}\ }

\def\FOR{{\bf for}\ }
\newcommand{\calb}{{\cal{B}}}

\newcommand{\bphi}{{\bar{\phi}}}

\newcommand{\proj}[1]{\ket{#1}\bra{#1}}

\title{Quantum search-to-decision reductions \\ and the state synthesis problem}

\author{Sandy Irani$^1$\thanks{\texttt{irani@ics.uci.edu}} \quad Anand Natarajan$^2$\thanks{\texttt{anandn@mit.edu}} \quad Chinmay Nirkhe$^{34}$\thanks{\texttt{nirkhe@cs.berkeley.edu}} \quad Sujit Rao$^2$\thanks{\texttt{sujit@mit.edu}} \quad Henry Yuen$^5$\thanks{\texttt{hyuen@cs.columbia.edu}} \\
\ \\
\small \emph{The authorship order is alphabetical by surname and reflects equals contribution.}
\ \\
\small $^1$Department of Computer Science, University of California, Irvine\\
\small $^2$CSAIL, Massachusetts Institute of Technology \\
\small $^3$Challenge Institute for Quantum Computation, University of California, Berkeley\\
\small $^4$Department of Computer Science, University of California, Berkeley \\
\small $^5$Department of Computer Science, Columbia University
}

\date{}

\begin{document}

\maketitle{}

\begin{abstract}
It is a useful fact in classical computer science that many search problems are reducible to decision problems; this has led to decision problems being regarded as the \emph{de facto} computational task to study in complexity theory. 
In this work, we explore search-to-decision reductions for quantum search problems, wherein a quantum algorithm makes queries to a classical decision oracle to output a desired quantum state. In particular, we focus on search-to-decision reductions for $\QMA$, and show that there exists a quantum polynomial-time algorithm that can generate a witness for a $\QMA$ problem up to inverse polynomial precision by making one query to a $\PP$ decision oracle. We complement this result by showing that $\QMA$-search does \emph{not} reduce to $\QMA$-decision in polynomial-time, relative to a quantum oracle.

We also explore the more general \emph{state synthesis problem}, in which the goal is to efficiently synthesize a target state by making queries to a classical oracle encoding the state. We prove that there exists a classical oracle with which any quantum state can be synthesized to inverse polynomial precision using only one oracle query and to inverse exponential precision using two oracle queries. This answers an open question of Aaronson~\cite{aaronson2016complexity}, who presented a state synthesis algorithm that makes $O(n)$ queries to a classical oracle to prepare an $n$-qubit state, and asked if the query complexity could be made sublinear.


\end{abstract}

\newpage

\tableofcontents

\newpage

\pagenumbering{arabic}

\section{Introduction}

\input{introduction}

\paragraph{Preliminaries.} Preliminaries and definitions necessary for the proofs are listed in \Cref{sec:prelim}.

\section{Search-to-decision for \textsf{QMA} problems}
\label{sec:QMA}

In the traditional search-to-decision paradigm, the goal is to create
a witness $\ket{\psi}$ which could convince a verifier that indeed
some string is in a particular $\QMA$ language. The creation of this
witness should be carried out by a quantum machine running in polynomial time with access to a $\QMA$ oracle. There are multiple ways to relax this paradigm; here we consider
using a $\PP$ oracle instead of a $\QMA$ oracle and show
that there is a polynomial time quantum algorithm which makes
only one $\PP$ oracle call\footnote{The improvement over the algorithm of Aaronson \cite{aaronson2016complexity} is in the number of oracle queries.} and generates a solution to a
$\QMA$-complete problem. 

Our algorithm proceeds from two observations:
\begin{enumerate}
  \item Any phase state
$\ket{p_f} = 2^{-n/2} \sum_{x \in \{0,1\}^n} (-1)^{f(x)} \ket{x}$ for which the function $f$ is computable in $\PP$ may be
prepared by a single quantum query to a $\PP$ oracle.
\item \emph{Any} state $\ket{\tau}$, after applying a random unitary
  $U$, looks like a phase state: in particular, with high probability
  over the choice of $U$, the state $U\ket{\tau}$ has constant overlap with some
  phase state.
\end{enumerate}
Our results heavily rely on the preliminaries defined in Section \ref{subsec:prelim-phase-state-prelims} and \ref{subsec:prelim-phase-estimation}.

\subsection{One-query search-to-decision for \textsf{QMA}}

We now consider the $\QMA$ search problem with respect to phase oracles. In general, the statement of the $\QMA$ search problem is to construct, given a verification circuit for some $\QMA$ language, and an input $\chi$ in the language, a state $\ket{\psi_\chi}$ that is accepted by the verification circuit with high probability. Rather than working with general verifiers, we will restrict to verifiers that measure the energy of a local Hamiltonian on the witness state up to inverse-polynomial precision. This restriction is almost without loss of generality, for two reasons. First, the local Hamiltonian problem with this precision is $\QMA$-complete, so any $\QMA$ language has a verifier of this form. And secondly, the reduction to local Hamiltonian can be performed so that every low-energy state is very close to an accepting witness $\ket{\psi}$ for the original verifier. More precisely, given a general $\QMA$ verification circuit $V$, we can apply the padding trick of Nirkhe, Vazirani and Yuen \cite{nirkhe_et_al:LIPIcs:2018:9095} to generate a local Hamiltonian instance $H$ such that any ground-state $\rho$ of $H$, $\norm{\rho - \sigma \otimes \Phi} \leq \delta$ where $\sigma$ is an accepting witness of $V$ and $\Phi$ is a fixed state independent of the instance. The size of the Hamiltonian instance $H$ scales as $\poly(1/\delta)$ and therefore the approximation can be chosen as any inverse polynomial function of the system size.

Assume the input to the problem is an instance $\chi = (H, a, b)$ of the Local Hamiltonian problem with
Hamiltonian $H$ on $n$ qubits and two thresholds $a < b$ such that $b
- a = 1/\poly(n)$. Moreover, we assume that $\chi$ is a YES instance,
so the minimum eigenvalue of $H$ is at most $a$: $\lambda_{\min}(H)
\leq a$. The goal is to construct a state $\ket{\phi}$ such that
\begin{align}
    \ev{H}{\phi} \leq \frac{a+b}{2}. \label{eq:witness-satisfied}
\end{align}
While it would be ideal to construct a state for which the energy is at most $a$ (since one exists), this may drastically increase the computational complexity of the function $\fn{f}{\bits^n}{\bits}$ defining the oracle. 
Instead, due to the promise gap in the problem, it suffices to construct a \emph{witness} state which proves that the Hamiltonian has a state with energy at most $< b$. 
A state $\ket{\phi}$ satisfying \eqref{eq:witness-satisfied} is a proof that $\chi$ is a yes instance. We now prove the formal version of Theorem \ref{thm:1-query-qma-informal}.

\begin{theorem}[$\QMA$-search to $\PP$-decision reduction]
There exists a probabilistic polynomial time quantum algorithm with access to a single $\PP$ phase oracle query that, given as input an instance $(H, a, b)$ of the local Hamiltonian problem on $n$ qubits, either aborts or outputs a witness $\ket{\phi}$ with $\ev{H}{\phi} \leq (a + b)/2$ for $b - a = 1/\poly(n)$. The algorithm will succeed in outputting a witness (i.e. not abort) with probability\footnote{We will later argue that this probability can be amplified through a variation of parallel repetition to improve to any function $1-\exp(-n)$.} $\ge 1/1024$.
\label{thm:1-query-qma}
\end{theorem}


\paragraph{Remark} 

In fact, we shall see that the algorithm achieves something stronger: if the algorithm does not abort, then the output state
is almost entirely supported on states of energy less than $a + (b-a)/4 + \eps$, where $\eps = 1/\poly(n)$ is a precision parameter much smaller than $b -a$. This is performed by using phase estimation to ``check" the outcome of the query algorithm by measuring the energy.

At this point, it is useful to remember that in general, the notion of a $\QMA$ \emph{witness} is defined only with reference to a specific verifier. The guarantee we achieve ensures that the ``standard" verifier, which measures the energy of the local Hamiltonian $H$, has a high chance of accepting the given state. If one is willing to use a more sophisticated verifier, e.g. a verifier that performs the Marriott-Watrous amplification procedure~\cite{Marriott-Watrous-amplification}, a witness of considerably worse quality could still be acceptable. Our theorem also sidesteps the issue of unique witnesses: we only guarantee that the energy of our state is low, not that it is the \emph{unique} such state. 

One can easily boost probability that our algorithm does not abort to $1 - \exp(-n)$ by repeating the construction in parallel with independent randomness and selecting any witness which did not cause the algorithm to abort. Furthermore, from the design of the algorithm, one can merge the oracle queries into a single larger $\PP$ query\footnote{One way to see that the merged oracle is also definable in $\PP$ is through the connection $\PP = {\PostBQP}$. The merged oracle can be seen as the logical exclusive-or (XOR) of multiple $\PP$ functions, and it is easy to create a new ${\PostBQP}$ function equal to the logical XOR of multiple ${\PostBQP}$ functions.}, so the query complexity does not increase.  \\

\begin{proof}
Assume, without loss of generality, that problem is stated for a normalized Hamiltonian: $0 \leq H \leq 1$. We will use $[C]$ to denote the classical description of the quantum circuit $C$. Likewise $\chi = (H,a,b)$ describes an encoding of the problem.

\paragraph{Algorithm. }The following quantum circuit describes the algorithm:
$$
\Qcircuit @C=1em @R=1em {
    \lstick{[\chi]}             & \cw                  & \cghost{\Oo_f}         & \cw  & \cw & \cw & \cw \\
    \lstick{\$\$\$}              & \gate{\textsf{Samp}} & \cghost{\Oo_f}         & \cctrl{1} & \cw & \cw & \cw \\
    \lstick{\ket{0}^{\otimes n}} & \gate{H^{\otimes n}} & \multigate{-2}{\Oo_f}  & \gate{C^\dagger} & \multigate{1}{\textsf{Phase Estimation}} & \qw & \qw \\
    \lstick{\ket{0}^{\otimes O(\log \frac{1}{b-a})}} & \qw & \qw & \qw & \ghost{\textsf{Phase Estimation}} & \meter & \cw
}
$$
Here, the collection of random coins ($\$\$\$$) are $O(\poly(n))$ bits of uniform randomness and the gate $\textsf{Samp}$ use the input randomness to sample\footnote{This can be done, for example, with the algorithm of van den Berg \cite{sample-cliffords}.} the descriptions $\ket{C}$ and $\ket{D}$ of two independent random Clifford unitaries described as circuits. $H^{\otimes n}$ is the tensor-product of the Hadamard gate. The controlled gate $C^\dagger$ denotes the application of the inverse circuit $C^\dagger$ according to the classical description $\ket{C}$. The oracle gate $\Oo_f$ applies a phase oracle
\begin{align} 
\ket{[\chi],C,D,x} \mapsto (-1)^{f([\chi],C,D,x)} \ket{[\chi],C,D,x} 
\end{align}
in superposition for a function $f$ specified in \eqref{eq:pp-fn}. The phase estimation unitary applies the energy estimation algorithm algorithm of \cite{rall-phase-estimation} which writes the energy of the third register state with respect to $H$ in the contents of the fourth register. The fourth register contains $O(\log \frac{1}{b-a})$ qubits, enough to write the energy down to a granularity finer than $b - a$. Then the last gate is a measurement deciding if the measured energy is $\leq a + (b-a)/4 + \eps$, where $\eps$ is the error incurred in rounding the energy ($\eps$ can be chosen to be much smaller than $(b-a)$). We \emph{abort} the algorithm if the measurement fails and otherwise output the state on the third register.

\paragraph{Correctness} 
Let $\ket{\tau} = (1-H)^p D \ket{0^n}$ be an unnormalized state, where $p$ is a parameter that will be chosen later, satisfying $p = \Omega(n/(b-a))$. Let $\ket{\tilde{\tau}} = \ket{\tau}/ \|\ket{\tau}\|$ be the normalization of this state. We will choose our oracle function $f$ such that the state after the oracle is applied yields a phase-state approximation to $C \ket{\tilde{\tau}}$ for  the sampled random Clifford unitary $C$.
Specifically, we define the oracle function $f$ as
\begin{align}
    f(H,C,D,x) &\defeq \sgn \left( \Re \left( \bra{x} C(1-H)^{p} D \ket{0^n} \right) \right). \label{eq:pp-fn}
\end{align}
For a fixed $H, C, D$, we write $f_{H, C, D}(x) = f(H, C, D, x)$, and denote the corresponding phase state by $\ket{p_{f_{H, C, D}}}$. Observe that the state of the algorithm immediately after the call to $\mathcal{O}_f$ is proportional to
\begin{align}
    \ket{\chi} \otimes \left(\sum_{\$\$\$} \ket{C(\$\$\$), D(\$\$\$)} \otimes \ket{p_{f_{H, C, D}}} \otimes \ket{0}^{\otimes O(\log(1/(b-a)))} \right).
\end{align}

At this point, we may imagine that the register containing the random coins $\$\$\$$ has been measured. To show that the algorithm succeeds with constant probability over the random coins (equivalently, over the random choice of $C$ and $D$), we must show the following
\begin{enumerate}
    \item The state $\ket{\tilde{\tau}}$ has almost all its mass in the low-energy subspace, with respect to $H$.
    \item For a random choice of $C$ and $D$, $C^\dagger \ket{p_{f_{H, C, D}}}$ is close to $\ket{\tilde{\tau}}$.
    \item That the phase estimation algorithm accepts $C^\dagger \ket{p_{f_{H, C, D}}}$ with sufficiently high probability, and the post-measurement state upon a successful measurement has low energy with respect to $H$. The post-measurement state is the state $\ket{\phi}$ in the theorem statement.
\end{enumerate}

We formulate Item~1 precisely as a separate lemma:

\begin{lemma}
Let $\Pi_{<}$ be the projector onto the eigenspaces of $H$ with eigenvalue at most $a + (b-a)/4$  (see \eqref{eq:def-of-low-energy-projector} for formal definition). Then, with probability $\geq \frac{1}{8}$ over the choice of $D$, it holds that  $\ev{\Pi_{<}}{\tilde{\tau}} \geq 1- (b-a)/2$.
\label{lem:concentration-in-low-energy-subspace}
\end{lemma}
This lemma is proven directly after the end of this proof.

To show Item~2, we will use \Cref{lem:design-approximation}, which we state here for convenience:
\begin{restatable*}[Random states have good overlap]{lemma}{overlap}
\label{lem:design-approximation}
Let $\ket{\tau} \in \qubits{n}$ be a unit vector. Let $\Cc$ be any 2-design (ex. $\Cc = \Cliff$). Sample a random $C \in \Cc$ and set $\ket{\psi} = C \ket{\tau}$. Then with constant probability, it has good overlap with the phase state defined by the function $f(x) = \sgn( \Re( \braket{x}{\psi}))$. Formally for $0 \leq \gamma \leq \frac{1}{4}$,
\begin{align}
    \Pr_{\substack{C \in \Cc \\ \ket{\psi} = C \ket{\tau}}} \left( \abs{\braket*{\psi}{p_f}}^2 \geq \gamma \right) \geq \frac{1}{2} - 2\gamma.
\end{align}
\end{restatable*}

\medskip

\noindent We prove \Cref{lem:design-approximation} in the Appendix. Let $\gamma \in (0,1)$ be a constant to be chosen later.  Then, by \Cref{lem:design-approximation}, we have that
\begin{align}
    \Pr_{C \in \Cliff} \left( \abs{\bra*{p_{f_{H,C,D}}} C \ket{\tilde \tau}}^2 \geq \gamma \right) \geq \frac{1}{2} - 2 \gamma. \label{eq:p-close-to-tau}
\end{align}


Let us fix a ``good" choice of $C$, such that the event in \eqref{eq:p-close-to-tau} occurs, and write
\newcommand{\pf}{p_{f_{H,C,D}}}
\begin{align}
\ket{\psi} &= C^\dagger \ket{\pf} = \alpha \ket{\tilde{\tau}} + \beta \ket*{\tilde{\tau}^\perp},
\end{align}
where $|\alpha|^2 \geq \gamma$.
Combining \eqref{eq:p-close-to-tau} with \Cref{lem:concentration-in-low-energy-subspace}, we obtain that $\ket{\psi}$ has high overlap with $\Pi_{<}$.
\begin{align}
    \eta &\defeq \ev{\Pi_{<}}{\psi} \\
    &= |\alpha|^2 \ev{\Pi_{<}}{\tilde{\tau}} + |\beta|^2 \ev{\Pi_{<}}{\tilde{\tau}^\perp} + \alpha^*\beta \bra{\tilde{\tau}}\Pi_{<} \ket*{\tilde{\tau}^\perp} + \alpha \beta^* \bra*{\tilde{\tau}^\perp} \Pi_{<} \ket{\tilde{\tau}} \\
    &\geq |\alpha|^2\left(1 - \frac{b-a}{2}\right) - 2 |\alpha \beta| \sqrt{b-a} \\
    &\geq \gamma\left(1 - \frac{b-a}{2}\right) - 2 \sqrt{\gamma(1-\gamma)(b-a)} .
\end{align}
Now in order to prove Item 3, let $\eps = \log\left(\frac{1}{100(b-a)}\right)$ and $\delta = \frac{1}{(b-a)^{100}}$. By \Cref{lem:rall-energy-post}, the energy estimation succeeds with probability at least $\eta - \delta$, and the post-measurement state has overlap $1 - 2\delta/\eta$ with the subspace of energy at most $a + (b-a)/4 + \eta$. A straightforward calculation shows that this implies that the post-measurement state has expected energy at most $a + (b-a)/2$, as desired.

It remains to set the parameter $\gamma$ so that the overall probability of not aborting is at least a constant, and so that $\eta$ is at least a constant for good $C, D$. We choose $\gamma = 1/8$, yielding $\eta \geq 1/16$ and $\eta - \delta \geq 1/32$ for all sufficiently large $n$ such that $(b-a)$ is sufficiently small. This yields a probability of not aborting
\begin{align}
    \Pr[\text{not abort}] &= \sum_{C,D}\Pr[C] \cdot \Pr[D] \cdot \Pr[\text{not abort} \mid C, D] \\
    &\geq \sum_{C, D \text{ good}} \Pr[C] \cdot \Pr[D] \cdot \Pr[\text{not abort} \mid C, D] \\
    &\geq\left(\frac{1}{2} - 2\gamma\right)  \cdot \frac{1}{8}   \cdot \left(\eta - \delta\right) \\
    &\geq \frac{1}{1024}.
\end{align} 
(We note that this bound is extremely loose; our only goal here is to show a \emph{some} constant lower bound on the non-abort probability as we can amplify with parallel repetition.)
\paragraph{Complexity of $\mathcal{O}_f$}
To show that the function $f$ can be calculated in $\PP$, it suffices to give a counting algorithm for calculating $f(H,C,x)$. Let us assume that $H = \sum_{i = 1}^m H_i$ where each $H_i$ is a local Hamiltonian term and $C = g_t \cdot \ldots \cdot g_1$ where each $g_j$ is a 2-qubit Clifford gate. Expanding the Feynman path integral gives
\begin{align}
    &\bra{x} C (1-H)^p C^\dagger \ket{0^n} \\
    &= \Exp_{i_1, \ldots, i_p \in [m]} \bra{x} g_t \ldots g_1 (1-H_{i_1}) \ldots (1-H_{i_p}) g_1^\dagger \ldots g_t^\dagger \ket{0^n} \\
    &= \Exp_{i_1, \ldots, i_p \in [m]} \sum_{y_1, \ldots, y_{2t+p-1} \in [2^n]} \bra{x} g_t \ketbra{y_1} \ldots g_1 \ldots \ketbra{y_{2t+p-1}} g_t^\dagger \ket{0^n} \label{eq:expanded-sum}
\end{align}
Note that we only need to calculate the sign of the real component of \eqref{eq:expanded-sum}. Each quantity of the form $\bra{y}g_j\ket{y'}$ or $\bra{y}(1-H_{i_q})\ket{y'}$ is known to exact precision (in the first case since $g_{j}$ is a Clifford gate, and in the second since it is an input to the problem). This also implies that each element of the sum can be efficiently calculated as $\bra{y}g_j\ket{y'}$ and $\bra{y}(1-H_{i_q})\ket{y'}$ are efficiently calculable. Since its length is at most $\poly(n)$ bits, the aforementioned computation can be calculated by a $\SHARPP$ or $\PP$ oracle since it is a counting problem involving terms of size $\exp(n)$.

\end{proof}

\begin{proof_of}{Lemma \ref{lem:concentration-in-low-energy-subspace}}
To prove that $\ev{\Pi_{<}}{\tilde \tau} \leq 1- (a+b)/2$ occurs with constant
probability, let us first introduce some notation. Let $\Delta = b - a$, and $m = a+ \Delta/4$. Let $\lambda_1 \leq \lambda_2 \leq
\ldots \leq \lambda_{2^n}$ be a sorted list of the eigenvalues of $H$
with corresponding eigenvectors $\ket{\lambda_1}, \ket{\lambda_2},
\ldots, \ket{\lambda_{2^n}}$ and let $j$ be the first $j$ such that
$\lambda_j \geq m \defeq a + \Delta/4$. Then, the projector $\Pi_{<}$ from the statement of the lemma is given by 
\begin{equation} \Pi_{<} \defeq \sum_{j = 1}^{j-1} \ketbra*{\lambda_j}.
\label{eq:def-of-low-energy-projector}\end{equation}
We define a complementary projector $\Pi_{\geq } = \II - \Pi_{<}$ as the projector onto the ``high''-energy subspace. 

Using these projectors, we now write the (unnormalized) state $\ket{\tau}$ as a sum of a ``low-energy" part and a ``high-energy" part:
\begin{align}
    \ket{\tau} = \ket{\tau_{<}} + \ket{\tau_{\geq}} \text{ where } \ket{\tau_{<}} \defeq \Pi_{<} \ket{\tau} \text{ and } \ket{\tau_{>}} \defeq \Pi_{\geq} \ket{\tau}.
\end{align}
Our goal is to show that 
\begin{align}
    q \defeq \frac{\braket{\tau_{<}}}{\braket{\tau_{<}} + \braket{\tau_{\geq}}} \geq 1 - \Delta/2
\end{align}
This event will always occur when
\begin{align}
    \norm{\ket{\tau_{\geq}}}^2 \leq \frac{\Delta}{2} \cdot \norm{\ket{\tau_{<}}}^2.
    \label{eq:ratio-of-high-low-energy}
\end{align}
Now we show that this event occurs with probability $\geq \frac{1}{8}$. It follows from Fact \ref{fact:ell_4_norm_of_haar} that
\begin{align}
    \Pr_D \left( \norm{\Pi_< D \ket{0^n}}^2 \geq \frac{1}{2 \cdot 2^n} \right) \geq \frac{1}{8}.
    \label{eq:good-enough-seed-D}
\end{align}
To see this, apply Fact \ref{fact:ell_4_norm_of_haar} with the unitary design $\mathcal{C}$ in Fact \ref{fact:ell_4_norm_of_haar} taken to be $U \cdot D$, where $U$ is a unitary such that $H = U \cdot \mathrm{diag}(\lambda_1, \lambda_2, \dots) \cdot U^\dagger$, and $x = 0^n$, and $\theta = 1/2$. This yields a lower bound of $1/{2\cdot 2^n}$ on $\| \proj{\lambda_1} D \ket{0^n}\|^2$ with probability at least $1/4$. Therefore,
since $\Pi_< \succeq \ketbra{\lambda_1}$, \eqref{eq:good-enough-seed-D} follows. Henceforth, we will condition on this event. Since $\Pi_<$ commutes with $(1-H)^p$,
\begin{align}
    \norm{\ket{\tau_<}}^2 = \norm{\Pi_< (1-H)^p D \ket{0^n}}^2 \geq (1 - \lambda_1)^{2p} \cdot \frac{1}{2 \cdot 2^n}.
\end{align}
On the other hand, 
\begin{align}
    \norm{\ket{\tau_\geq}}^2 = \norm{\Pi_\geq (1-H)^p D \ket{0^n}}^2 \leq (1-m)^{2p}
\end{align}
since $\Pi_\geq$ projects onto the low-energy subspace of $(1-H)^p$.
Therefore, in order to satisfy \eqref{eq:ratio-of-high-low-energy}, it suffices to pick $p$ such that
\begin{align}
    (1-m)^{2p} \leq \frac{\Delta}{2} \cdot (1-\lambda_1)^{2p} \cdot \frac{1}{2 \cdot 2^n}. \label{eq:choice-of-p}
\end{align}
Since $\lambda_1 \leq a$,
\begin{align}
    \frac{1-m}{1-\lambda_1} \leq \frac{1-m}{1-a} \leq 1 - (m-a) = 1 - \Delta/4.
\end{align}
Therefore our threshold for $p$ simplifies to 
\begin{align}
    \left( 1 - \frac{\Delta}{4} \right)^{2p} \leq \frac{\Delta}{2 \cdot 2^n} \label{eq:energy-exponential-decay}
\end{align}
which is satisfied for a choice of 
\begin{align}
    p = \Omega \left( \frac{n}{\Delta} \log\frac{1}{\Delta} \right).
\end{align}
(To see this, observe that the left-hand side of \eqref{eq:energy-exponential-decay} can be made less than $1/2$ with $p = \Omega(\Delta^{-1})$. The bound of $\Delta/(2\cdot 2^n)$ can be achived by multiplying $p$ by $\Omega(n \log(1/\Delta))$.) Therefore, with probability $\geq \frac{1}{8}$, we have that $\ev{\Pi_{<}}{\tilde \tau} \leq 1 - \Delta/2 = 1 - (b-a)/2$. 
\end{proof_of}

\paragraph{Difficulty in improving the oracle complexity. }
In the previous section, we detail a proof that a witness to any $\QMA$ problem can be constructed with access to a single $\PP$ oracle query. This does not satisfy our desire to understand search-to-decision for $\QMA$ since it is widely suspected that $\QMA \neq \PP$. We also note that without any additional assumptions, we cannot simplify that search-to-decision algorithm (Theorem \ref{thm:1-query-qma}) to utilize a $\QMA$ oracle.
Consider the class of local Hamiltonians problems for which the gap between the completeness ($c$) and soundness ($s$) thresholds is inverse exponentially small, i.e. $c - s = 2^{-\poly(n)}$ but we are promised that the spectral gap of $H$ is at least $1/\poly(n)$. The complexity of deciding this class of problems (called $\mathsf{PGQMA}$) is $\PP$-complete by a result of Deshpande, Gorshkov, and Fefferman \cite{spectral-gap-for-precise-qma}. We refer to their work for more details and precise definitions. We note that there is a simple modification\footnote{More specifically, the previous algorithm ignores the differences between spectral and promise gap and conflates the two notions. We notice that whenever the spectral gap is $1/\poly(n)$, $(1-H)^p$ is still a ``good'' approximate projector onto the ground-space and for a choice of $p = \poly(n)$, it projects onto the ground-space up to exponentially small error. One can derive that for $\mathsf{PGQMA}$ local Hamiltonians, the equivalent lower bound on $p$ as that in \eqref{eq:choice-of-p}  is
\begin{align}
    \left(1 - O(\lambda_{\textsf{spectral}})\right)^p \leq \lambda_{\textsf{promise}} \cdot (1-\lambda_1)^p \cdot \frac{1}{2 \cdot 2^n}.
\end{align}
In which case, it is easy to check that $p =\poly(n)$ suffices. Lastly, we note that the phase estimation step also works due to the spectral gap promise.} of Theorem \ref{thm:1-query-qma} that can  be applied to this class of $\PP$-complete Hamiltonians which provides a search-to-decision algorithm. Therefore, we cannot reduce the complexity of the oracle from $\PP$ without critically assuming that we are only interested in finding witnesses for a promise gap precision of $1/\poly(n)$.


\subsection{Search-to-decision for $\QMA_\exp$}
\label{sec:qma_exp}
Instead, we note that our proof can be adjusted slightly to construct a search-to-decision reduction for the class $\QMA_\exp$ which is the class of quantum Merlin-Arthur promise problems for which the gap between the completeness ($c$) and soundness ($s$) thresholds is inverse exponentially small, i.e. $c - s = 2^{-\poly(n)}$ but there is no promise on the spectral gap of the Hamiltonian. The canonical complete problem for this class is, unsurprisingly\footnote{This follows directly from Kitaev' s proof \cite{kitaev-qma-complete} of $\QMA$-completeness for local Hamiltonians.}, the local Hamiltonian problem for a promise gap of $2^{-\poly(n)}$. An earlier work of Fefferman and Lin \cite{qma_exp} showed that $\QMA_\exp = \PSPACE$.

\begin{theorem}[\cite{qma_exp}]
The problem of deciding, for a local Hamiltonian $H$ on $n$ qubits, if $\lambda_{\min}(H) \leq a$ or $> b$ with a promise gap of $b - a = 2^{-\poly(n)}$ is $\PSPACE$-complete.
\end{theorem}

Using this interesting connection between complexity classes, we can actually create a 1 query search-to-decision algorithm for $\QMA_\exp$.

\begin{theorem}
\label{th:qma_exp}
Pick a constant $0 < \gamma < 1/4$. There exists a probabilistic polynomial time quantum algorithm with access to a single $\QMA_\exp$ phase oracle query that given an instance $(H, a, b)$ of the local Hamiltonian problem on $n$ qubits for $b - a = 1/\exp(n)$, outputs a state $\ket{\psi}$ with the following property. With probability $\gamma/8$, there exists a state $\ket{\tilde \tau}$ such that $\ev{H}{\tilde \tau} \leq (a+b)/2$ and $\abs{\ip{\tilde \tau}{\psi}}^2 \geq 1/2 - 2 \gamma$.
\end{theorem}

\begin{proof}
The proof follows almost identically to that of Theorem \ref{thm:1-query-qma} except for two critical variations. First, we remove the phase estimation step from the procedure and use the quantum register as the output without any possibility of aborting. This is because running phase estimation to the $2^{-\poly(n)}$ precision necessary for determining if the quantum register is a witness would require an exponentially long runtime \cite{Atia2017FastforwardingOH}. By eschewing the phase estimation step, the protocol will run in polynomial time but at the cost that it cannot detect if it has produced a witness. This is not surprising because unless $\QMA = \PSPACE$, the quantum witness should not be verifiable in polynomial time and, in effect, the phase estimation routine is verifying if the witness is valid. 



And second, we need to prove that the complexity of the oracle function $f$ is now solvable in $\PSPACE$ (instead of $\PP$). The function $f$ is the same as in the previous proof (see \eqref{eq:pp-fn}). But now the exponent of $(1-H)$ is $p = \Omega(n/(b-a))$ = $2^{\poly(n)}$. Therefore, the Feynman path integral used in \eqref{eq:expanded-sum} is over an exponential number of variables. To solve this issue, we simply rewrite the Feynman path integral to not include $(1-H)^p$.
\begin{align}
    \bra{x} C (1-H)^p C^\dagger \ket{0^n} = \sum_{y_1,\ldots, y_{2t} \in [2^n]} \bra{x}g_t \ket{y_1} \ldots \bra{y_t} (1-H)^p \ket{y_{t+1}} \ldots \ket{y_{2t}} g_t^\dagger \ket{0^n}.
\end{align}
We now only need to show that $\bra{y_t} (1-H)^p \ket{y_{t+1}}$ is computable in $\PSPACE$. We proceed by induction. Notice
\begin{align}
    \bra{y_t} (1-H) \ket{y_{t+1}} = \Exp_{i = 1}^m \bra{y_t} (1-H_i) \ket{y_{t+1}}
\end{align}
which is easily computable in polynomial space. For the induction step, notice
\begin{align}
    \bra{y_t} (1-H)^{2k} \ket{y_{t+1}} = \sum_{z \in [2^n]} \bra{y_t} (1-H)^{k} \ket{z} \cdot \bra{z} (1-H)^{k} \ket{y_{t+1}}.
\end{align}
By reusing the space for both parts of the sub-computation, the computation for exponent $2k$ only uses $\poly(n)$ additive additional step over the computation for exponent $k$. Since, we only need to calculate the result for $p = 2^{\poly(n)}$, it follows that $\bra{y_t} (1-H)^p \ket{y_{t+1}}$ is computable in $\PSPACE$ and the overall function $f$ is computable in $\PSPACE$. Since $\QMA_\exp = \PSPACE$, the proof is complete.
\end{proof}

\subsection{Search-to-decision for $\QCMA$ with one oracle query}
\label{sec:qcma-search-to-decision}

With a completely different procedure, we can show a one oracle query search-to-decision procedure for $\QCMA$. More specifically, we show that a polynomial time quantum algorithm can produce a \emph{classical} witness for any $\QCMA$ problem with a single query. Note that since the witness is classical, as intuition suggests, one can also show that a polynomial time randomized \emph{classical} algorithm can produce, with high probability, a classical witness for any $\QCMA$ problem with $\poly(n)$ classical queries to a $\QCMA$ oracle based on the binary-search algorithm\footnote{The reduction involves a randomized mapping the $\QCMA$ instance to a $\UQCMA$ instance and then applying binary-search for the witness to the $\UQCMA$ problem.}. 
We show that a variation of the Bernstein-Vazirani algorithm \cite{bernstein-vazirani} allows us to reduce the query complexity to a single query which succeeds with probability $\Omega(\inv{m})$, where $m$ is the size of the witness. This is particularly pertinent because if $\QCMA = \QMA$, then there exists a single query search-to-decision reduction for $\QMA$, albeit with sub-constant probability. 


Recall the Bernstein-Vazirani algorithm \cite{bernstein-vazirani} for learning a hidden string $d \in \bits^m$ given quantum access to a specific $m$-bit oracle. If the oracle $\fn{f}{\bits^m}{\bits}$ is defined by $f(x) = x^\top d$ (over $\FF_2$), then
\begin{align}
    H^{\otimes m} \Oo_f H^{\otimes m} \ket{0^m} &= H^{\otimes m} \Oo_f \frac{1}{\sqrt{2^m}} \sum_x \ket{x} \label{eq:bernstein-vazirani} \\ 
    &= H^{\otimes m} \frac{1}{\sqrt{2^m}}  \sum_x (-1)^{x^\top d} \ket{x} \\
    &= \frac{1}{2^m} \sum_{x,y} (-1)^{x^\top (y + d)} \ket{y} \\
    &= \ket{d}.
\end{align}
Therefore, a polynomial time quantum algorithm  with a single oracle query can produce a $m = \poly(n)$ output string. To show that $\QCMA$ search has a single query algorithm, it suffices to show that the function $f(x) = x^\top d$ is $\QCMA$-computable when $d$ is the classical $\QCMA$ witness to the instance $\chi$.

Let us notice that the witness $d$ needs to be made explicit if we are to define such a function $f$. In particular, $f(x)$ should only depend on $x$ and $\chi$. For example, consider defining $f(x) = 1$ iff there exists a witness $d$ such that $x^\top d = 1$. Since there can be many witnesses $d$, it is even possible that $f = 1$ everywhere. One idea is to define the description $d$ as that of the \emph{lexicographically} first witness. However, calculating the lexicographically first witness is not known to be contained in $\QCMA$ and the best upper bound on the complexity of calculating $d$ is $\P^\QCMA$. This gives us one search-to-decision result.

\begin{theorem}
Consider any $\QCMA$ problem optimizing over witnesses of length $m$. There exists a probabilistic polynomial time quantum algorithm with access to a single $\P^\QCMA$ phase oracle that outputs a satisfying witness $d \in \bits^m$ with probability 1. \label{thm:p-qcma-oracle-version}
\end{theorem}

In order to consider an oracle of only $\QCMA$ complexity, however, we will need to isolate a specific witness $d$ through a different manner.
To achieve this we recall the randomized reduction from $\QCMA$ to $\UQCMA$ formalized in \cite{quantum-valiant-vazirani}. Here $\UQCMA$ is the class of $\QCMA$ problems for which there holds an additional promise that in the case of yes instances, there exists a unique accepting classical witness $d \in \bits^{\poly(n)}$. 

\begin{theorem}[{\cite[Corollary 40]{quantum-valiant-vazirani}}]
$\UQCMA \subseteq \RP^{\QCMA}$.
\end{theorem}

When presented with a $\UQCMA$ instance, the question of $f(x) = 1$ iff there exists a witness with description $d$ such that $x^\top d = 1$ is equivalent to $f(x) = x^\top d^0$ where $d_0$ is the unique satisfying witness.

The reduction from $\QCMA$ to $\UQCMA$ follows a standard technique originating from the Valiant-Vazirani theorem and can be used to reduce $\NP$ and $\MA$ to their unique counterparts. By adding additional constraints to the $\QCMA$ problem, we can reduce the number of accepting witnesses until a unique witness remains. More formally, we can pose a constraint of the form $A d = 0$ where $A \in \FF_2^{k \times m}$ is a uniformly random matrix and arithmetic is over the field $\FF_2$. Each row of $A$ is an additional constraint which roughly halves the number of accepting witnesses. Therefore, is known that when $2^k$ is approximately the number of accepting witnesses, with high probability there will only be one witness which satisfies the original $\QCMA$ problem as well as the constraint $A d = 0$. A critical difficulty in this approach is that the verifier does know how many accepting witnesses the original $\QCMA$ instance has; they need this to select an appropriate $k$. 

What we instead do is uniformly randomly guess a choice of $k \in \{1, \ldots, n\}$. Notice that if the verifier were to make a guess for $k$ that is too large, then with high probability, the new instance will be unsatisfied. On the other hand, if the guess is too small, then the new instance will have many solutions and the algorithm will not necessarily output a valid witness. However, note that whatever purported witness results from applying the Bernstein-Vazirani algorithm \cite{bernstein-vazirani} to the quantum oracle (\eqref{eq:bernstein-vazirani}) and then measuring the register, one can easily verify the validity of said witness. 

Therefore, we have all the ingredients for a one-query search-to-decision algorithm for $\QCMA$. Let $V$ be the original instance  and $m$ the witness length. Namely, one chooses a uniformly random $k \in \{1,\ldots, n\}$ and then a uniformly random matrix $A \in \FF_2^{k \times m}$. Let $V'$ be the instance $V$ appended with the additional test that $Ad = 0$ for witness $d$. The oracle function $f(x)$ is defined as
\begin{align}
    f(x) = 1 \iff \exists \ \text{witness } d \text{ for  } V' \text{ and } x^\top d = 1.  \label{eq:qcma-oracle-def}
\end{align}
Then, the algorithm computes the left-hand side of \eqref{eq:bernstein-vazirani} and measures in the standard basis to obtain a witness $d$. Then, the algorithm checks that $d$ is a solution to $V$ and accordingly answers with $d$ or aborts.

\begin{theorem}
Consider any $\QCMA$ problem optimizing over witnesses of length $m$. There exists a probabilistic polynomial time quantum algorithm with access to a single $\QCMA$ phase oracle query that either aborts or outputs a satisfying witness $d \in \bits^m$. The algorithm will output a witness with probability $\geq \Omega(1/m)$. \label{thm:qcma-query}

Likewise there is a probabilistic polynomial time quantum algorithm with access to a single $\NP$ ($\MA$, respectively) phase oracle query that either aborts or outputs a satisfying witness $d \in \bits^m$. The algorithm will output a witness with probability $\geq \Omega(1/m)$.
\end{theorem}

\paragraph{Remarks. } One could also perform a similar \emph{classical} query algorithm to generate a witness for $\QCMA$ using the search-to-decision paradigm for $\textsf{UNP}, \textsf{UMA},\UQCMA$ of learning the witness bit-by-bit with $O(m)$ queries. However, to extend this algorithm to $\NP, \MA, \QCMA$, this protocol requires an expected $O(m^2)$ classical queries due to the guessing required to apply the reduction from $\QCMA$ to $\UQCMA$. 

Secondly, one can improve the success probability of this algorithm to any $1 - 2^{-t}$, by running the algorithm independently in parallel $k = O(n+t)$ times. Ideally, one would like to be able to merge the oracle queries into a singular, albeit larger, oracle query. This would involve defining a function
\begin{align}
    f(x_1, \ldots, x_{k}) = \prod_{i = 1}^k \left[ \exists \text{ witness } d_i \text{ for } V_i \text{ and } x_i^\top d_i = 1  \right] 
\end{align}
where $V_1, \ldots, V_k$ are the $k$ independent instances that the reduction outputs. However, it is not necessarily the case that $f$ is a $\QCMA$ decidable function since it is a product of $\QCMA$ problems. Instead, the complexity upper-bound we can place on $f$ is at most $\P^\QCMA$. However, this results in an algorithm strictly worse than Theorem \ref{thm:p-qcma-oracle-version}. Thus, without increasing the complexity of the oracle, we do not know of a method of probability amplification without multiple queries.

\begin{corollary}
Theorem \ref{thm:qcma-query} can be amplified to a success probability of $1 - 2^{-t}$ with $O(n+t)$ parallel queries to a $\QCMA$ phase oracle.
\end{corollary}

\input{oracle-no-go}

\input{one-query}

\section{Swap test distillation procedure}
\label{sec:swaptest}

 If a synthesis protocol is able to produce a state with at least constant overlap with the target state and the target state 
is a witness for a $\QMA$ verifier, then phase estimation can be used to boost the overlap  and the probability of success.
If  the target state is an arbitrary state,  we may not have the means to directly measure whether the output state is close to the target. In this section we describe a procedure that can take
the output of $m$ parallel applications of a state synthesis protocol, each of which has a constant overlap with the target state and apply a procedure to increase  the overlap. 
The algorithm begins with $m$ states, $\ket{\psi_1},\ldots\ket{\psi_{m}}$, each of which is stored in an $n$-qubit register.
We show that if the number of states $m$ is a sufficiently large polynomial in $n$, then the overlap of the final output state will be at least $1-1/\poly(n)$, with high probability. 
The distillation process is based on the Swap Test and works subject to two conditions on the collection
of input states:
\begin{enumerate}
    \item For all $j$, $|\braket{\psi_j}{\tau}|^2 \ge a$, for some constant $a$.
    \item For all $i \neq j$ $|\braket{\phi_j}{\phi_i}|^2 \le \delta$, where $\delta$ is exponentially small in $n$ and for all $j$
    \begin{align}
        \ket{\phi_j} \defeq \frac{\ket{\psi_j} - \braket{\psi_j}{\tau} \ket{\tau}}
        {\norm{\ket{\psi_j} - \braket{\psi_j}{\tau} \ket{\tau}}}.
    \end{align}
\end{enumerate}
The second condition is satisfied if the portion of each state $\ket{\psi_j}$ that is perpendicular to the target state $\ket{\tau}$ is essentially random. If the $\ket{\psi_j}$'s are generated {according to some independent randomness}, one might expect that the overlap between these perpendicular components to be {(exponentially)} small. 
In this section, we analyze the behavior of the Swap Test distillation procedure subject to these two properties. At the end of the section, we will show that the first condition can be relaxed to a lower bound on the expectation of the overlap as long as the $m$ states are generated {according to some independent randomness}. In Section \ref{sec:one-query}, we {showed} how the algorithm can be used in conjunction with a $1$-query protocol to produce a state that has $1-1/\poly(n)$ overlap with the target state.

\subsection{The Algorithm}

\begin{figure}[ht]
  \centering
  \includegraphics[width=6.0in]{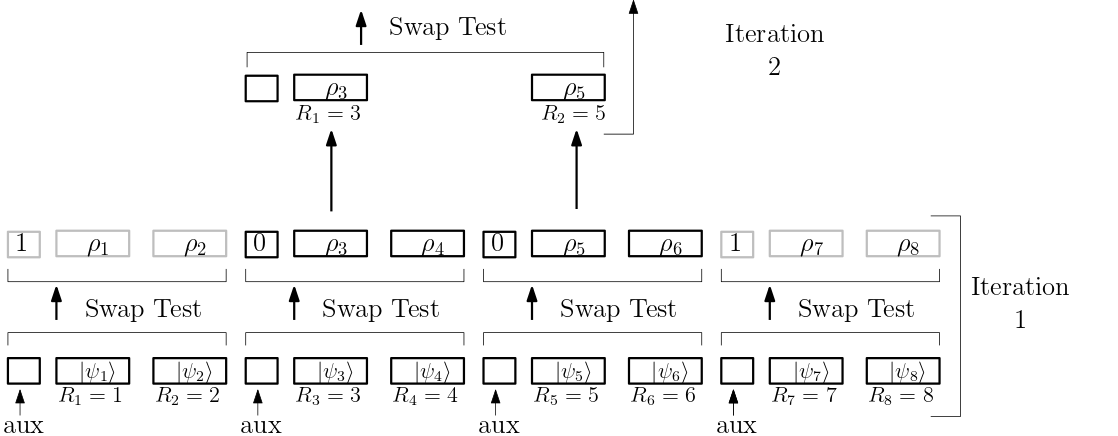}
\caption{The Swap Test Distillation algorithm applied to eight input registers. The value measured in the auxiliary qubit indicates whether an application of the Swap Test is successful.
In the first iteration, the swap tests applied to pairs
$\ket{\psi_3}, \ket{\psi_4}$ and
$\ket{\psi_5}, \ket{\psi_6}$ are successful.
The resulting states in registers $3$ and $5$ advance to the next round. In each iteration, the sequence
$(R_1, R_2, \ldots)$ indicates the indices of the surviving registers from left to right.}
\label{fig:SwapTest}
\end{figure}

Each round of the algorithm begins with some set of surviving registers. The surviving registers are paired up and the swap test is applied to each pair. An auxiliary qubit is used in each application of the swap test which is measured at the end of the swap test. If the outcome is $0$ (a {\em successful} outcome), then one of the two registers is selected to survive to the next iteration. If the outcome is $1$ (an {\em unsuccessful} outcome), neither register survives.
Figure \ref{fig:IterativeSwap} shows the pseudocode for the procedure. 
Figure \ref{fig:SwapTest} shows an example of the procedure
for one iteration applied to eight input states.
Note that the state in a surviving register
may be entangled with the other registers. If $\rho$ is the reduced density matrix of the state in one
of the surviving registers obtained by
tracing out the other registers, we will refer to $\tr{\rho \ketbra{\tau}{\tau}}$ as
the {\em overlap} of $\rho$ with $\ket{\tau}$.
We will show that for $m$ sufficiently large, with high probability the overlap of a surviving register
with $\ket{\tau}$ is
at least $1 - 1/\poly(n)$.

Consider one round of the algorithm applied to a particular pair of registers.
We will prove that if the swap test succeeds, then
the surviving register has an overlap with $\ket{\tau}$
that is at least the average of the overlap of the states in the two registers before the round. 
Moreover, if each of the two  registers at the beginning of a round have overlap at least $\gamma$, then
the overlap of a surviving register  is strictly larger than $\gamma$ and with enough successful rounds will tend towards $1$. 

\begin{figure}[ht]
\noindent
\begin{center}
\fbox{\begin{minipage}{\textwidth}
\begin{tabbing}
{\sc SwapTestDistillation}\\
Input: $m$ states $\ket{\psi_1},\ldots\ket{\psi_{m}}$ stored in $n$-qubit registers numbered $1$ through $m$\\
(1)~~~~ \= Initialize $(R_1, \ldots, R_{m}) \leftarrow (1, \ldots, m)$ \\
(2) \> $\ell = \lfloor \log_6 (m/n) \rfloor$\\
(3) \> \FOR $k = 1, \ldots, \ell$:\\
(4)   \> ~~~~ \= count $ = 0$\\
(5)  \> \> \FOR $j = 1, \ldots, \lfloor m/2 \rfloor$\\
(6)  \>\> ~~~~~ \=  \IF {\sc SwapTest}$(R_{2j-1}, R_{2j})$ returns $0$\\
(7)  \>\>\> ~~~~\=  count = count +1 \\
(8)  \> \> \> \> $R_{count} = R_{2j-1}$\\
(9) \>\> \END \\
(10) \>\> $m = $ count\\
(11) \END
\end{tabbing}
\end{minipage}}

\vspace{.1in}

\fbox{\begin{minipage}{\textwidth}
\begin{tabbing}
{\sc SwapTest}$(R, R')$\\
Start with auxiliary qubit $b$ initialized to $\ket{0}$\\
(1)~~~~ \= Apply:\\
(2)\> ~~~~ \=  $H_b \otimes I_{R, R'}$\\
(3)\> \> Controlled {\sc SWAP} operation on Registers $R$ and $R'$, controlled by qubit $b$\\
(4) \> \> $H_b \otimes I_{R, R'}$\\
(5)   \> Measure qubit $b$ and return the result\\
(6) \END
\end{tabbing}
\end{minipage}}
\end{center}
\caption{Pseudo-code for {\sc SwapTestDistillation} algorithm.}
\label{fig:IterativeSwap}
\end{figure}

\subsection{Analysis: no overlap case}

We will first analyze the algorithm for the idealized case in which there is no overlap between the portion of each input state that is perpendicular to the target $\ket{\tau}$. That is, for the remainder of this subsection, we assume that $\braket{\phi_i}{\phi_j} = 0$
for $i \neq j$. We also assume that Condition $1$ above is met, namely, for all $j$,
$\abs{\braket{\psi_j}{\tau}}^2 \ge a$, for some constant $a$.

We can track which original states  are used to create the state in each surviving register after every iteration of the while loop. Originally, $S_i = \{ i\}$.
Whenever a swap test succeeds in Line (6), we can update the set associated with $R_{count}$ to be $S_{count} \leftarrow S_{2j-1} \cup S_{2j}$.
Thus, at the end of each iteration of the while loop, the collection of sets $\{S_j\}$ is pairwise disjoint. Moreover, the reduced density matrix of the state $\rho_j$ in register $R_j$
has support contained in the space spanned by $\{\ket{\tau}\} \cup \{\ket{\phi_i} \mid j \in S_j\}$.
Define $\Pi \defeq \II - \ketbra{\tau}{\tau}$ o be the projector onto the perpendicular space to $\ket{\tau}$. Then
$\tr (\rho_i \Pi \rho_j \Pi) = 0$ for any two surviving registers $R_i$ and $R_j$.
Also, the combined state is a tensor product state: $\rho_i \otimes \rho_j$.
The  lemma below gives an expression for the resulting state in two registers after a successful swap test is applied. 

\begin{lemma}[Result of a Successful Swap Test]
\label{lem:swapresult}
Suppose that the {\sc SwapTest} is performed on $\ketbra{0}{0} \otimes \rho_1 \otimes \rho_2$. Let $S$ be the Swap operation for the second and third registers 
that contain $\rho_1$ and $\rho_2$, respectively.
If the measurement outcome is $0$, then the resulting state is
$\ketbra{0}{0} \otimes \tilde{\rho}/\tr(\tilde{\rho})$, where
\begin{align}
\label{eq:rho-tilde-in-swap-analysis}
\tilde{\rho} = (\rho_1 \otimes \rho_2) + S(\rho_1 \otimes \rho_2)S + S(\rho_1 \otimes \rho_2) + (\rho_1 \otimes \rho_2)S.\end{align}
\end{lemma}

\begin{proof}
After step $(2)$ the state is $\ketbra{+}{+} \otimes \rho_1 \otimes \rho_2$.
The controlled swap operation is $\ketbra{0}{0} \otimes I + \ketbra{1}{1} \otimes S$. After the controlled swap is applied in Step $(3)$, the state equals
\begin{align}
    \frac{\ketbra{0}{0} \otimes (\rho_1 \otimes \rho_2)
    + \ketbra{1}{1} \otimes S(\rho_1 \otimes \rho_2)S
    + \ketbra{0}{1} \otimes (\rho_1 \otimes \rho_2)S
    + \ketbra{1}{0} \otimes S(\rho_1 \otimes \rho_2)}{2}.
\end{align}
After the Hadamard gate, $H$, is applied again to the first qubit, the component of the state corresponding to the $\ketbra{0}{0}$ term (which is the result after a $0$ is measured) will equal
\begin{align}
    \frac{\ketbra{0}{0} \otimes (\rho_1 \otimes \rho_2)
    + \ketbra{0}{0} \otimes S(\rho_1 \otimes \rho_2)S
    + \ketbra{0}{0} \otimes (\rho_1 \otimes \rho_2)S
    + \ketbra{0}{0} \otimes S(\rho_1 \otimes \rho_2)}{4}.
\end{align}
\end{proof}

The following lemma will be useful in understanding the contribution of the $S(\rho_1 \otimes \rho_2)$ and $(\rho_1 \otimes \rho_2)S$
terms from Lemma \ref{lem:swapresult}.
\begin{lemma}
\label{lem:oneside}
Consider a tensor product state $\rho_i \otimes \rho_j$. Let $S$ be the swap operator on the registers holding $\rho_i$ and $\rho_j$.
Suppose also that $\tr{\rho_i (\II - \ketbra{\tau }{\tau }) \rho_j (\II - \ketbra{\tau }{\tau})} = 0$. Then 
\begin{align}\tr{ S (\rho_i \otimes \rho_j ) (\II - \ketbra{\tau ,\tau})} = \tr{ (\rho_i \otimes \rho_j )S (\II - \ketbra{\tau ,\tau})} = 0\end{align}
and
\begin{align}\tr{S (\rho_i \otimes \rho_j )\ketbra{\tau ,\tau}} = \tr{ (\rho_i \otimes \rho_j )S \ketbra{\tau, \tau}} = \tr{ \rho_i \ketbra{\tau}}) \cdot
\tr{ \rho_j \ketbra{\tau}}.\end{align}
\end{lemma}

\begin{proof}
Since $\tr{\rho_i (I - \ketbra{\tau }{\tau }) \rho_j (I - \ketbra{\tau }{\tau})} = 0$, there is an orthonormal basis of the Hilbert space of a single register $\calb = \{ \ket{\tau} \} \cup \{\ket{1}, \ldots, \ket{D-1}\}$
where the support of $\rho_i$ is contained in the space spanned by $\{\ket{\tau}\} \cup \{ \ket{1} \ldots, \ket{r} \}$ for some $r$ and
the support of $\rho_j$ is contained in the space spanned by $\{\ket{\tau}\} \cup \{ \ket{r+1}, \ldots, \ket{D-1} \}$. 
When $\rho_i$ and $\rho_j$ are expressed in the basis $\calb$, row and column $0$ correspond to $\ket{\tau}$, so
$\rho_i [0, 0] = \tr{\rho_i \ketbra{\tau}{\tau}}$ and $\rho_j [0, 0] = \tr{\rho_j \ketbra{\tau}{\tau}}$. The figure below shows the matrices $\rho_i$ and $\rho_j$. The non-zero entries are contained within the shaded regions in each matrix.

\begin{center}
  \includegraphics[width=4.0in]{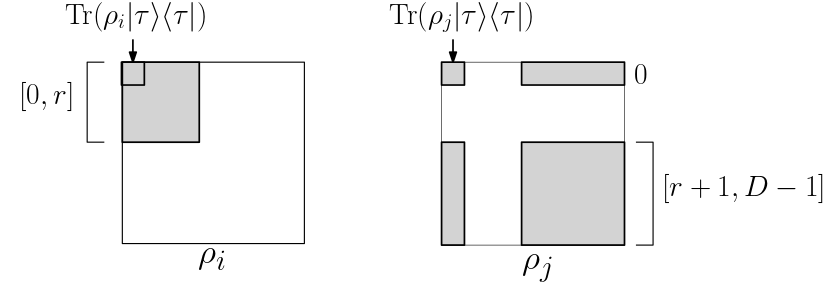}
  \end{center}

Now express $\rho_i \otimes \rho_j$ in the $\calb \otimes \calb$ basis.
This matrix consists of $D^2$ submatrices of size $D \times D$. A row is indexed by a pair $(a, b) \in [D] \times [D]$, where $a$ denotes the 
row of submatrices and $b$ denotes the row within the row of submatrices. Similarly a column  is indexed by a pair $(c, d) \in [D] \times [D]$, where $c$ denotes the 
column of submatrices and $d$ denotes the collumn within a the column of submatrices. The entry corresponding to row $(a,b)$ and column $(c,d)$ is $\rho_i [a,c] \cdot \rho_j [b,  d]$.
When the Swap operator, $S$, is applied to the left ($S (\rho_i \otimes \rho_j)$), the rows, but not the columns, of the matrix are permuted. Thus, the entry corresponding
to row $(a,b)$ and column $(c,d)$ becomes $\rho_i [b,c] \cdot \rho_j [a,  d]$. A diagonal entry has $(a,b) = (c,d)$ which is $\rho_i [b,a] \cdot \rho_j [a,  b]$.
If $a = b = 0$, then the entry corresponding to row $(0,0)$ and column $(0,0)$ is 
\begin{align}
\rho_i [0,0] \cdot \rho_j [0,0] = \tr ( \rho_i \ketbra{\tau}{\tau}) \cdot \tr ( \rho_j \ketbra{\tau}{\tau}).
\end{align}
Now consider the case where $a \neq 0$ or $b \neq 0$. Assume, without loss of generality, that $a \neq 0$. If $a \in \{1, \ldots, r\}$ then $\rho_j [a,  b] = 0$
because the support of $\rho_j$ is contained in the space spanned by $\{ \ket{\tau} \} \cup \{ \ket{r+1}, \ldots ,\ket{D-1} \}$. Similarly, if $a \in \{r+1, \ldots, D-1 \}$ then 
$\rho_i [b, a] = 0$
because the support of $\rho_i$ is contained in the space spanned by $\{ \ket{\tau} \} \cup \{ \ket{1}, \ldots ,\ket{r} \}$.
In either case $\rho_i [b,a] \cdot \rho_j [a,b] = 0$.
A similar argument holds for $ (\rho_i \otimes \rho_j)S$ where the columns are permuted instead of the rows.
\end{proof}

\begin{lemma}[Change in Overlap After One Iteration]
\label{lem:swaptestrecurrence}
Suppose that the Swap Test is performed on $\ketbra{0}{0} \otimes \rho_1 \otimes \rho_2$. The first register holds the control qubit,
and the second and third registers are $n$-qubit registers that hold $\rho_1$ and $\rho_2$, respectively. 
Suppose also that $\tr(\rho_1 \ketbra{\tau}{\tau} ) = a_1$ and $\tr(\rho_2 \ketbra{\tau}{\tau} ) = a_2$.
If the measurement outcome is $0$, and $\rho$ is the result of tracing out the first and third registers, then
\begin{enumerate}
    \item $\tr(\rho \ketbra{\tau}{\tau} ) \ge (a_1 + a_2)/2$
    \item If $a_1 \ge a$ and $a_2 \ge a$ for some constant $a$, then $\tr(\rho \ketbra{\tau}{\tau} ) \ge \frac{a (1 + a)}{1 + a^2}$.
\end{enumerate}
\end{lemma}

\begin{proof}
According to Lemma \ref{lem:swapresult}, the contents of the second and third registers after the $0$ is measured in the first register is
$\ketbra{0}{0} \otimes \tilde{\rho}/\tr(\tilde{\rho})$, where $\tilde{\rho}$ is given by \eqref{eq:rho-tilde-in-swap-analysis}.
We will first calculate $\tr(\tilde{\rho})$. 
Since $\tr{\rho_1} = \tr{\rho_2} = 1$, we have that
$\tr{\rho_1 \otimes \rho_2} = 1$ and $\tr{S(\rho_1 \otimes \rho_2)S} = \tr{\rho_2 \otimes \rho_1} = 1$.
Lemma \ref{lem:oneside} implies that $\tr{S(\rho_1 \otimes \rho_2)} = \tr{(\rho_1 \otimes \rho_2)S} = a_1 a_2$. Therefore
$\tr{\tilde{\rho}} = 2(1 + a_1 a_2)$.

Now consider the state $\tilde{\rho}/\tr{\tilde{\rho}}$ in registers $2$ and $3$. Let $\{\ket{
\tau}\} \cup \{ \ket{1}, \ldots, \ket{D-1}\}$ be a basis for the Hilbert
space of each register. We can express the state $\tilde{\rho}/\tr{\tilde{\rho}}$ as a $D^2 \times D^2$ matrix, which can be divded into $D \times D$
submatrices of dimension $D \times D$. When we trace out the third register, the result is $\rho$, which is a $D \times D$ matrix, where each entry is the
trace of the corresponding submatrix in $\tilde{\rho}/\tr{\tilde{\rho}}$. The value of $\tr{\rho \ketbra{\tau}{\tau}}$ is the trace of the submatrix in the upper left
corner. We will consider the four terms in $\tilde{\rho}$ separately. The contribution from $\rho_1 \otimes \rho_2$ is $\rho_1[0,0] \cdot \tr{\rho_2} = a_1$.
The contribution from $S(\rho_1 \otimes \rho_2)S$ is $\rho_2[0,0] \cdot \tr{\rho_1} = a_2$.
From Lemma \ref{lem:oneside}, $S(\rho_1 \otimes \rho_2)$ and $(\rho_1 \otimes \rho_2)S$ have the same diagonal, which consists of $a_1 a_2$ in the
upper left corner and $0$ elsewhere. Therefore the contribution from each of these matrices is $a_1 a_2$. The result is that
\begin{align}\tr{\rho \ketbra{0}{0}} = \frac{ a_1 + a_2 + 2 a_1 a_2}{2(1 + a_1 a_2)}
= \frac{ \frac{a_1 + a_2}{2} +  a_1 a_2}{1 + a_1 a_2}.\end{align}
To establish the first fact in the lemma, since $(a_1 + a_2)/2 \in [0,1]$,
\begin{align}\tr{\rho \ketbra{0}{0}} 
= \frac{ \frac{a_1 + a_2}{2} +  a_1 a_2}{1 + a_1 a_2} \ge
\frac{a_1 + a_2}{2}.\end{align}
Now for the second fact, note that
\begin{align}\tr{\rho \ketbra{0}{0}} = \frac{ \frac{a_1 + a_2}{2} +  a_1 a_2}{1 + a_1 a_2} \ge \frac{ \sqrt{a_1 a_2} +  a_1 a_2}{1 + a_1 a_2},
\end{align}
where the inequality follows from the fact that the arithmetic mean/geometric mean inequality.
Define $\gamma \defeq \sqrt{a_1 a_2}$, so $\tr{\rho \ketbra{0}{0}} \ge (\gamma + \gamma^2)/(1 + \gamma^2)$.
Since $a_1 \ge a$ and $a_2 \ge a$, we have that $1 \ge \gamma \ge a \ge 0$ and we would like to show that
\begin{align}\frac{\gamma + \gamma^2}{1 + \gamma^2} 
\ge \frac{a + a^2}{1 + a^2}.
\end{align}
After some rearranging, this inequality is equivalent to
\begin{equation}
\label{eq:ineq}
\gamma^2 + (\gamma - a) \ge a^2 + \gamma a(\gamma - a).
\end{equation}
Since $\gamma \ge a$ and $\gamma a \le 1$, it follows that $(\gamma - a) \ge \gamma a(\gamma - a)$.
The fact that $\gamma \ge a$ also implies that $\gamma^2 \ge a^2$.
The sum of the two inequalities implies the inequality in \eqref{eq:ineq}.
\end{proof}

We are now ready to analyze the result of the entire {\sc SwapTestDistillation} algorithm.
First we need some definitions to describe the state of the algorithm after each iteration.

\begin{definition}
Let $m_k$ be the number of surviving registers after the $k^{th}$
iteration of the algorithm. Initially $m_0 = m$, the number of states at the beginning of the algorithm. 
We will denote the state in the $j^{th}$ surviving register after $k$ iterations as $\rho_{j,k}$, for $1 \le j \le m_k$.
\end{definition}

\begin{lemma}
\label{lem:overlapbound}
{\bf (Overlap After $\ell$ Rounds)}
Suppose  that the  {\sc SwapTestDistillation} algorithm  lasts for $\ell$ rounds, then for every $1 \le j \le m_\ell$, 
\begin{align}1 - \tr{\rho_{j, \ell} \ketbra{\tau}{\tau}} \le \frac 1 2 \cdot \left( \frac 4 5 \right)^{\ell - 2/a^2}.\end{align}
where $a$ is a lower bound on the overlap of every input state with the target state.
\end{lemma}

\begin{proof}
Let $\gamma_i$ be the smallest overlap of any of the surviving registers after $i$ rounds.
We start with $\gamma_0 \ge a$. Using Lemma \ref{lem:swaptestrecurrence}, we have by induction that
\begin{align}\gamma_{i+1} \ge \gamma_i \cdot \frac{1 + \gamma_i}{1 + (\gamma_i)^2}.\end{align}
Since $\gamma_i \le 1$,   this recurrence implies that the $\gamma_i$'s are strictly increasing. 
We start by analyzing the number of iterations to reach $\gamma \ge 1/2$. 
If $\gamma_i \le 1/2$, then
\begin{align}
\frac{\gamma_i + (\gamma_i)^2}{1 + (\gamma_i)^2} &= (\gamma_i + (\gamma_i)^2) \left( 1 - \frac{(\gamma_i)^2}{1 + (\gamma_i)^2} \right)\\
& \ge (\gamma_i + (\gamma_i)^2) (1 - (\gamma_i)^2)\\
& = \gamma_i + (\gamma_i)^2  - (\gamma_i)^3  - (\gamma_i)^4\\
& \ge \gamma_i + (\gamma_i)^2 - (\gamma_i)^2 (1/2 + 1/4)\\
& \ge \gamma_i + (\gamma_i)^2/4.
\end{align}
Since $\gamma_{i+1} \ge \gamma_i + (\gamma_i)^2/4$, 
we know that $\gamma_{i+1} - \gamma_i \ge (\gamma_0)^2/4$.
Therefore, the number of iterations to reach
$1/2$ from $\gamma_0$ is at most $4(1/2 - \gamma_0)/(\gamma_0)^2 \le 2/a^2$. 
Next we show that once $\gamma_i$ reaches $1/2$, then $1 - \gamma_i$ decreases by a constant factor in each iteration. 
\begin{align}1 - \gamma_{i+1} \le 1 - \gamma_i \cdot \frac{1 + \gamma_i}{1 + (\gamma_i)^2} \le \frac{1 - \gamma_i}{1 + (\gamma_i)^2}
\le \frac{4}{5} (1 - \gamma_i),\end{align}
where the last inequality uses the assumption that $\gamma_i \ge 1/2$.

After the next $2/a^2$ iterations,
$1 - \gamma_k \le 1/2$. After every successive iteration, 
$1 - \gamma_k$ decreases by a factor of $4/5$. The lemma follows.
\end{proof}

Finally, $\ell$ the number of rounds must be chosen  so that given the number of input states,
with high probability, the algorithm lasts for $\ell$ rounds.

\begin{lemma}[Probability of a Surviving Register After $\ell$ Rounds]
\label{lem:probbound}
For $m \ge n \cdot 6^\ell$ and $n \ge 12$, the probability that there are no surviving registers after $\ell$ rounds is at most
$2 \exp(-n/12)$.
\end{lemma}

\begin{proof}
First we analyze the probability that $m_k \ge n \cdot 6^{\ell-k}$ conditioned on the event that $m_{k-1} \ge n 6^{\ell-k+1}$.
There will be $m_{k-1}/2$ pairs in round $k$ to which the Swap Test is applied. Each pair succeds with probability at least $1/2$.
The number of registers surviving to round $k$ is the number of pairs that succed the swap test. Thus, the expected number of surviving
registers is $m_{k-1}/4$. Using Chernoff's Inequality, the probability that there are fewer than $m_{k-1}/6$ is
at most 
\begin{align}\exp[ -(m_{k-1}/4) (1/3)^2 (1/2))] = \exp(-m_{k-1}/72)
\end{align}
The probability that any of the iterations fails to keep $1/6$ of their registers is at most
\begin{align}\sum_{k=0}^{\ell-1} \exp(-n \cdot 6^{\ell-k}/72) \le 2 \exp(-n/12).
\end{align}
\end{proof}

\begin{theorem}
\label{th:one-query-no-overlap}
{\bf (Swap Test Distillation Performance: No Overlap)}
For every pair of constants $c$ and $a \in [0,1]$, there is a polynomial $p$ such that
if the {\sc DistillationSwapTest} algorithm starts with $m$ states $\ket{\psi_1},\ldots\ket{\psi_{m}}$ 
 such
that  $m \ge p(n)$ and
\begin{enumerate}
\item for all $j$, $ \abs{\braket{\psi_j}{\tau}}^2  \ge a$
\item for all $i \neq j$, $\bra{\psi_j} (\II - \ketbra{\tau}{\tau}) \ket{\psi_i} = 0$,
\end{enumerate}
then the algorithm succeeds with probability at least $1 - \exp(-n/12)$ in producing a state $\rho$ such that
$\tr{\rho \ketbra{\tau}{\tau}} \ge 1 - 1/n^c$.
\end{theorem}

\begin{proof}
The proof follows directly from Lemmas \ref{lem:overlapbound} and \ref{lem:probbound}
and selecting the polynomial $p(n) \ge n \cdot 6^\ell$, where $\ell = c \log_{5/4}(2n) + 2/a^2$.
The probability of success is at least $1 - 2\exp(-n/12)$.
\end{proof}

\paragraph{Relaxing the Conditions for Swap Test Distillation.}

We can relax the requirements for Theorem \ref{th:one-query-no-overlap} so that
the expected overlap (instead of the actual overlap) of each input state is at least some constant $a$ as long as the input states are orthogonal outside the $\ket{\tau}$ component. The algorithm will require some additional rounds in this case, which in turn requires a larger number of input states. The analysis for small overlap in the next section applies to both versions.

\begin{theorem}
\label{th:one-query-no-overlap-relax}
{\bf (Swap Test Distillation Performance: Relaxed Conditions, No Overlap)}
For every pair of constants $c$ and $a \in [0,1]$, there is a polynomial $p$ such that
if the {\sc DistillationSwapTest} algorithm starts with $m$ states $\ket{\psi_0},\ldots\ket{\psi_{m-1}}$ 
generated according to some independent randomness such
that  $m \ge p(n)$ and
\begin{enumerate}
\item for all $j$, $\Exp \left[ \abs{\braket{\psi_j}{\tau}}^2 \right] \ge a$
\item for all $i \neq j$, $\bra{\psi_j} (\II - \ketbra{\tau}{\tau}) \ket{\psi_i} = 0$,
\end{enumerate}
then the algorithm succeeds with probability at least $1 - 2\exp(-n/12) - m 2^{-n}$ in producing a state $\rho$ such that
$\tr{\rho \ketbra{\tau}{\tau}} \ge 1 - 1/n^c$.
\end{theorem}

\begin{proof}
A state $\rho$ in a register that survives after $\ell$ rounds depends on exactly $2^\ell$ input states.
After $\ell'$ rounds, there are $2^{\ell-\ell'}$ surviving registers that will be used to create the final surviving state $\rho$. 
Consider an arbitrary $\rho'$ in one of these registers after $\ell'$ rounds. 
$L'$ input states are used to create $\rho'$, where $L' = 2^{\ell'}$.
Let $a_1, \ldots, a_{L'}$ be the overlap of each of those input states with the target state.
Then by Lemma \ref{lem:swaptestrecurrence},
$\tr{\rho \ketbra{\tau}}$ is at least the average of $a_1, \ldots, a_{L'}$.
Note that the event that the register containing $\rho$ survives is not independent from the
$a_1, \ldots, a_{L'}$, but the probability of a successful swap test increases with the average of the overlaps of the pair of registers to which it is applied. Therefore, the probability that an arbitrary
$L'$ registers have an average overlap of $a/2$ is a lower bound on the probability when we condition on the register surviving to a particular round. 

Let $S_j = \left( \sum_{i=1}^j a_i \right) - ja$,
where $a$ is the expected overlap of each initial state with the target states. 
The sequence $S_0, \ldots, S_{L'}$ forms a Martingale with $|S_{j+1} - S_j| \le 1 - a$.
The event that the average of the $a_i$'s is less than $a/2$ is the same as $S_{L'} \le - a L'/2$.
We can apply Azuma's Inequality (Lemma \ref{lem:azuma}) to
get that the probability $S_{L'} \le - a L'/2$ is at most 
$\exp( -L' a^2 / 8(1-a)^2)$. We will select $L' = 8n (1-a)^2/a^2$ so that the probability of
a surviving register after $L'$ rounds
not having overlap at least $a/2$ is at most $2^{-n}$. Since $\ell' = \log_2 (L')$,  
$\ell' = \log_2 n + \log_2 (8 (1-a)^2/a^2)$. 
The probability that all of the $2^{\ell-\ell'}$ registers that are used to create $\rho$ have an overlap of $a/2$ with the target state is at least  $1 - 2^{\ell -\ell' - n}$. 

Assuming  that this condition is true, we can now apply Lemmas \ref{lem:overlapbound} and \ref{lem:probbound} using a lower bound of $a/2$ on the overlap  instead of $a$.
By selecting the polynomial $p(n) \ge n \cdot 6^\ell$, where $\ell = c \log_{5/4}(2n) + 8/a^2 + \log_2 n + \log_2(8(1-a)^2/a^2)$.
The probability of success is at least $1 - 2\exp(-n/12) -  2^{\ell-n} \ge 1 - 2\exp(-n/12) - m 2^{-n}$.
\end{proof}

\subsection{Analysis: small overlap case}

Now we extend the analysis for the {\sc DistillationSwapTest} algorithm to the situation where there is a small overlap between the portion of each state that is orthogonal to $\ket{\tau}$. 
We have expressed each state $\ket{\psi_j}$ as $\alpha_j \ket{\tau} + \beta_j \ket{\phi_j}$, where $\ket{\phi_j}$ lies in the space perpendicular to
$\ket{\tau}$. Previously, we assumed that for $i \neq j$, $\braket{\phi_i}{\phi_j} = 0$. Now we will assume that  $\abs{\braket{\phi_i}{\phi_j}}^2 \le \delta$. Note that this is a weaker
assumption that Condition $2$: $|\bra{\psi_i}(\II - \ketbra{\tau}{\tau}\ket{\psi_j}|^2 \le \delta$.

We will use the Gram-Schmidt procedure to produce
a new orthonormal set $\{ \ket{1}, \ldots, \ket{m}\}$,
all of which are orthogonal to $\ket{\tau}$, and 
such that for each $j$, the space spanned by  $\{ \ket{\phi_1}, \ldots, \ket{\phi_j}\}$ lies in the space spanned by
$\{ \ket{1}, \ldots, \ket{j}\}$. Specifically:
\begin{align}\ket{j} = \frac{\ket{\phi_j} - \sum_{i=1}^{j-1} \ket{i} \braket{i}{\phi_j} }{
\norm{ \ket{\phi_j} - \sum_{i=1}^{j-1} \ket{i} \braket{i}{\phi_j}}}.\end{align}
We will use $\ket{0}$ to represent the target state $\ket{\tau}$.
We can now break up $\ket{\phi_j}$ into to components: $\ket{j}$ and $\ket{\bphi_j}$, where $\ket{\bphi_j}$ is orthogonal to $\ket{j}$:
\begin{align}\ket{\psi_j} = \alpha_j \ket{0} + \beta_j \ket{\phi_j} = \alpha_j \ket{0} + \beta_j  \left[ \braket{j}{\phi_j} \ket{j} + \bar{\beta}_j \ket{\bphi_j} \right],\end{align}
where
\begin{align}\ket{\bphi_j} =  \frac{\sum_{i=1}^{j-1} \ket{i} \braket{i}{\phi_j} }{
\norm{ \sum_{i=1}^{j-1} \ket{i} \braket{i}{\phi_j}}}.\end{align}
First, we would like to show that for all of the states $\ket{\psi_j}$,
most of the weight is on $\ket{\tau}$ and $\ket{j}$ so that the initial  collection of states closely approximates a tensor product of states with no overlap as assumed in the previous section. Thus we would like to bound
$|\bar{\beta}_j|^2$ for every $j$.

\begin{lemma}
If $\abs{\braket{\phi_j}{\phi_i}}^2 \le \delta$ for all $i \neq j$
and $\sqrt{\delta} \le 1/8m$,
then for all $j$, $|\bar{\beta}_j|^2 \le (2j-1) \delta$.
\end{lemma}

\begin{proof}
We will define a parameter $a_i$ that is an upper bound for $|\braket{\phi_j}{i}|^2$ for any $j > i$
and define $b_j \defeq a_1 + \cdots + a_j$. Since  
\begin{align}\ket{\phi_i} = \braket{i}{\phi_i} \ket{i} + \braket{i-1}{\phi_i} \ket{i-1} + \cdots + \braket{1}{\phi_i} \ket{1},\end{align}
$b_{i-1}$ is an upper bound for $|\bar{\beta}_i|^2$. 

We will derive a recurrence to upper bound the $b_j$'s and show by induction that $b_j \le (2j-1)\delta$. 
For the base case, $\ket{\phi_1} = \ket{1}$, for all $j > 1$, $|\braket{\phi_j}{1}|^2 \le \delta$, so $a_1 = b_1 = \delta$.
Now we will derive a recurrence relation for the $b_j$'s.
For $j > i$, we can express the inner product $\braket{\phi_j}{\phi_i}$ as
\begin{align}\braket{\phi_j}{\phi_i} = \braket{i}{\phi_i} \braket{\phi_j}{i} + \braket{i-1}{\phi_i} \braket{\phi_j}{i-1} + \cdots + \braket{1}{\phi_i} \braket{\phi_j}{1},\end{align}
and solve for $|\braket{\phi_j}{i}|^2$:
\begin{align}
|\braket{\phi_j}{i}|^2 & = \frac{\left|  \braket{\phi_j}{\phi_i}  - \displaystyle \sum_{k=1}^{i-1} \braket{k}{\phi_i} \braket{\phi_j}{k} \right|^2 }{\left|    \braket{i}{\phi_i} \right|^2 }\\
& \le \frac{|\braket{\phi_j}{\phi_i}|^2  + 2 |\braket{\phi_j}{\phi_i}| \displaystyle \sum_{k=1}^{i-1} | \braket{k}{\phi_i}| | \braket{\phi_j}{k}| +
\left[ \displaystyle \displaystyle \sum_{k=1}^{i-1} | \braket{k}{\phi_i}| | \braket{\phi_j}{k}| \right]^2  }{1 -   \displaystyle \sum_{k=1}^{i-1} \left| \braket{k}{\phi_i} \right|^2 }.
\end{align}
$\sqrt{\delta}$ is an upper bound for $|\braket{\phi_j}{\phi_i}|$ and $\sqrt{a_k}$ is an upper bound for both $|\braket{k}{\phi_i}|$ and $|\braket{\phi_j}{k}|$. Therefore,
\begin{align}
|\braket{\phi_j}{i}|^2  \le \frac{\delta + 2 \sqrt{\delta} \displaystyle \sum_{k=1}^{i-1} a_k + \left[ \sum_{k=1}^{i-1} a_k\right]^2 }
{1 - \displaystyle \sum_{k=1}^{i-1} a_k}
= \frac{\delta + 2 \sqrt{\delta} b_{i-1} + \left[ b_{i-1} \right]^2 }{1 - b_{i-1} }.
\end{align}
We will use the expression above on the right as $a_i$ to get:
\begin{equation}b_i = b_{i-1} + a_i = b_{i-1} + \frac{\delta + 2 \sqrt{\delta} b_{i-1} + \left[ b_{i-1} \right]^2 }{1 - b_{i-1} }.\end{equation}
We will show that if $b_{i-1} \le (2i-3)\delta$, then $b_i \le (2i-1)\delta$. The assumption $\sqrt{\delta} \le 1/8m$ in the statement of the lemma implies that $\sqrt{\delta} \le 1/8$. Also since by the inductive hypothesis, $b_{i-1} \le (2i-1)\delta \le 2m\delta$,
then $b_{i-1} \le \sqrt{\delta}/4 \le 1/32$.
These inequalities along with the inductive hypothesis can be used to bound $b_i$: 
\begin{align}
b_i & = b_{i-1} + \frac{\delta + 2 \sqrt{\delta} b_{i-1} + \left[ b_{i-1} \right]^2 }{1 - b_{i-1} }\\
& \le (2i-3)\delta + ( \delta + 2 \sqrt{\delta} b_{i-1} + \left[ b_{i-1} \right]^2) \cdot \frac{32}{31}\\
& \le (2i-3)\delta + \left( \delta + \frac{\delta}{2} + \frac{\delta}{16} \right) \cdot \frac{32}{31}\\
& \le (2i-3)\delta + 2 \delta = (2i-1)\delta.
\end{align}
\end{proof}

We are now ready to bound the effect of the small overlap between the components of each state
that is orthogonal to the target state.

\begin{theorem}[Swap Test Distillation Performance: Small Overlap]
\label{th:one-query-small-overlap}
For every pair of constants $c$ and $a \in [0,1]$, there is a polynomial $p$ such that
if the {\sc DistillationSwapTest} algorithm starts with $m$ states $\ket{\psi_1},\ldots\ket{\psi_{m}}$ 
 such that  $m \ge p(n)$ and
\begin{enumerate}
\item for all $j$, $ |\braket{\psi_j}{\tau}|^2  \ge a$
\item for all $i \neq j$, $|\bra{\psi_j} (I - \ketbra{\tau}{\tau}) \ket{\psi_i}|^2 \le \delta$,
\end{enumerate}
then the algorithm succeeds with probability at least $1 - 2\exp(-n/12) - 4m \sqrt{\delta}$ in producing a state $\rho$ such that
$\tr{\rho \ketbra{\tau}{\tau}} \ge  1 - 1/n^c - 8m^2 \sqrt{\delta}$.
\end{theorem}

\begin{proof}
Let $\epsilon  = 1 - \prod_{j=1}^m \left(1 -  2j \delta\right)$.
Note that $\epsilon \le 2m^2 \delta$.
We can define the state 
\begin{align} \ket{\Omega} = \frac{1}{\sqrt{1 - \epsilon}} ( \alpha_1 \ket{\tau} + \tilde{\beta}_1 \ket{1})
\otimes ( \alpha_2 \ket{\tau} + \tilde{\beta}_2 \ket{2})\otimes 
\cdots \otimes ( \alpha_m \ket{\tau} + \tilde{\beta}_m \ket{m}).\end{align}
The original state
$\ket{\Psi} = \ket{\psi_1} \otimes \cdots \otimes \ket{\psi_m}$
is equal to
\begin{align}(1 - \epsilon)^{1/2} \ket{\Omega} + \sqrt{\epsilon} \ket{\Phi}.
\end{align}
for some state $\ket{\Phi}$ that is orthogonal to $\ket{\Omega}$.
Furthermore, $\ket{\Omega}$ has the properties assumed in the
no-overlap case. 

Now imagine a coherent version of the {\sc DistillationSwapTest} algorithm
in which the measurements are deferred to the end. For coherence, we will need to fix the number of iterations
to $\ell$ so that in the no-overlap case,
there is at least one surviving register after $\ell$ iterations
with some probability $p$ that is exponentially close to $1$.
We will have auxiliary qubits $b_1, \ldots, b_m$ such that $b_j = 1$ if Register $R_j$ survives the $\ell$ rounds, and a final qubit $s$ such that $s = 1$ if any of the registers
succeed after $\ell$ rounds. Thus, if qubit $s = 1$ when measured then the algorithm succeeds. The result of applying the algorithm (before measurement) is
$A ( \ket{\Psi} \otimes \ket{\bar{0}} ),$
where the second register contains the auxiliary qubits initialized to $0$. 
$A$ is the unitary performed by the {\sc DistillationSwapTest} algorithm before measurement.
\begin{eqnarray}
A ( \ket{\Psi} \otimes \ket{\bar{0}} ) & = 
(1 - \epsilon)^{1/2} A ( \ket{\Omega} \otimes \ket{\bar{0}} ) + \sqrt{\epsilon} A ( \ket{\Phi} \otimes \ket{\bar{0}} )\\
& = (1 - \epsilon)^{1/2} [ \alpha_0 \ket{0}_s \ket{\Omega_0} + \beta_0 \ket{1}_s \ket{\Omega_1} ]\\
& + \sqrt{\epsilon} [ \alpha_1 \ket{0}_s \ket{\Phi_0} + \beta_1 \ket{1}_s \ket{\Phi_1} ]
\end{eqnarray}
The probability of success with $\ket{\Omega}$ is at least $p$, so $\abs{\beta_0}^2 \ge p$.
The probability of success with $\ket{\Psi}$ is then at least
\begin{align}[(1 - \epsilon)^{1/2} \sqrt{p} - \sqrt{\epsilon}]^2
\ge p - 2 \sqrt{\epsilon}.\end{align}
In the no-overlap case (Theorem \ref{th:one-query-no-overlap}) if
we select $\ell = c \log_{5/4}(2n) + 2/a^2$
and $m \ge n \cdot 6^\ell$, then
The probability of success $p$ is at least $1 - 2\exp(-n/12)$.
The probability of success with overlap $\delta$ is then
\begin{align}1 - 2\exp(-n/12)- 2 \sqrt{\epsilon} \ge
1 - 2\exp(-n/12)  - 4 m \sqrt{\delta}.\end{align}

Now suppose that we change the algorithm slightly so that when a pair of registers passes the swap test, the register that survives to the next iteration is selected randomly from the pair. Since the $\ket{\psi_j}$ are all generated according to the same process, the probability of any register surviving the the last stage is the same for every register. There are $m$ registers and the probability of success is at least $p$, so the probability that any particular register survives is at least $p/m$.
We are going to analyze the contents of register $R_j$ conditioned on the event that the register survives  and show that the probability that this register contains the target state $\ket{\tau}$ conditioned on survival is roughly the
same, regardless if we start with $\ket{\Psi}$ or $\ket{\Omega}$. 
Qubit $b_j$ contains the information regarding the survival of register $R_j$, so are interested in the probability that $R_j$ contains $\ket{0}$ conditioned on $b_j = 1$. To ease the notation, we will
drop the subscript $j$ and just refer to qubit $b$ and register $R$. We can divide 
$A ( \ket{\Omega} \otimes \ket{\bar{0}} )$ into three components:
\begin{align}\alpha \ket{0}_s \ket{\Psi_0} + \beta_0 \ket{1}_s \ket{\tau}_R \ket{\Psi_{10}} + \beta_1 \ket{1}_s \ket{\phi}_R \ket{\Psi_{11}},\end{align}
where $\ket{\phi}$ is an $n$-qubit state in register $R$ that is orthogonal to $\ket{\tau}$.
The probability that register $R_j$ survives is $\abs{\beta_0}^2 + \abs{\beta_1}^2 \ge p/m$.
The overlap of the state in $R$ with $\ket{\tau}$ conditioned on $R$ surviving is $|\beta_0|^2/(|\beta_0|^2 + |\beta_1|^2)$.
Since $\ket{\Omega}$ satisfied the non-overlapping condition from Theorem \ref{th:one-query-no-overlap}, 
this quantity is some $q$, which is at least $1 - 1/n^c$.
If we start with $A ( \ket{\Psi} \otimes \ket{\bar{0}} )$ instead, this quantity will be at least
\begin{eqnarray}
\label{eq:success}
\frac{ |(1 - \epsilon)^{1/2} \beta_0 - \sqrt{\epsilon_1}|^2}{|(1 - \epsilon)^{1/2} \beta_0 - \sqrt{\epsilon_1}|^2 + |\beta_1 + \sqrt{\epsilon_2}|^2}
\end{eqnarray}
where $\epsilon_1 + \epsilon_2 \le \epsilon$.
$\epsilon_1/\epsilon$ is the probability of measuring $\ket{1}_s\ket{\tau}_R$ with state $A ( \ket{\Phi} \otimes \ket{\bar{0}} )$ and $\epsilon_2/\epsilon$ is the probability of measuring $\ket{1}_s\ket{\phi}_R$, where $\ket{\phi}$ is orthogonal to $\ket{\tau}$. 
This quantity is minimized with
$\epsilon_1 = \epsilon$ and $\epsilon_2 = 0$, which gives:
\begin{align}
\frac{ |(1 - \epsilon)^{1/2} \beta_0 - \sqrt{\epsilon}|^2}{|(1 - \epsilon)^{1/2} \beta_0 - \sqrt{\epsilon}|^2 + |\beta_1 |^2}
 \ge \frac{|\beta_0|^2 - 2 \sqrt{\epsilon} }{ |\beta_0|^2 + |\beta_1 |^2- 2 \sqrt{\epsilon}}\\
 \ge \frac{|\beta_0|^2 - 2 \sqrt{\epsilon} }{ |\beta_0|^2 + |\beta_1 |^2}\\
 \ge q - \frac{2 m\sqrt{\epsilon}}{p}.
 \end{align}
 Thus if $\rho$ is the state in Register $R$ conditioned on $R$ surviving $\ell$ rounds ($b=1$), then
 \begin{align}
\tr{\rho \ketbra{\tau}{\tau}} \ge q - \frac{2 \sqrt{\epsilon}}{p/m}
 \ge 1 - \frac{1}{n^c} - 4m \sqrt{\epsilon} \ge 1 - \frac{1}{n^c} - 8m^2 \sqrt{\delta}.     
 \end{align}
 The first inequality above holds if $n$ is large enough so that the probability of success
 $p = 1 - 2\exp(-n/12)$ is at least $1/2$.
 \end{proof}

\input{wasserstein}

\section{Open Questions}
\label{sec:openq}


We exhibited state synthesis algorithms for $\QMA$ witnesses and arbitrary states that only require a \emph{single} query to a classical oracle, that generate the target state up to inverse polynomial error. We also presented a \emph{two}-query state synthesis algorithm that generates the target state up to inverse \emph{exponential} error. As mentioned, this resolves Open Question 3.3.6 of Aaronson~\cite{aaronson2016complexity}. However, there are several remaining open questions regarding these algorithms.

\begin{enumerate}
    \item The one- and two-query algorithms for arbitrary states use polynomial space, but they aren't time efficient (because their existence is argued by sampling a Haar-random unitary and applying the probabilistic method). Can this probabilistic construction be derandomized (and thus be made time efficient) by using (approximate) unitary designs? 
    \item Is there a one-query algorithm for state synthesis that also achieves inverse exponential error?
\end{enumerate}

\subsection{The power of quantum queries to $\QMA$ oracles}

Our impossibility result in Section \ref{sec:no-go} combined with our reduction of $\QMA$-search to $\PP$-decision problems leaves an interesting gap as to what exactly is the power of $\QMA$ oracle. More specifically, are there interesting computational tasks solvable only with quantum access to a $\QMA$ oracle? One question is to understand the collection of problems which have search-to-decision reductions where the oracle is a $\QMA$ oracle. Is this class strictly larger than $\QCMA$, a class with known search-to-decision reductions (Theorem \ref{thm:qcma-query})?

\subsection{The Unitary Synthesis Problem}

In Aaronson's lecture notes \cite{aaronson2016complexity} and his published list of open questions in quantum query complexity \cite{aaronson-open-qs}, he identifies the unitary synthesis problem as one of the major unresolved questions.

\begin{conjecture}[{\cite[Problem 6]{aaronson-open-qs}}]
For every $n$-qubit unitary transformation $U$, does there exists an oracle $\fn{A}{\bits^\ast}{\bits}$ such that a $\BQP^A$ machine can implement $U$?
\end{conjecture}
While, we do not know how to synthesize the unitary $U$, we do know how to synthesize the Choi - Jamiolkowski state, \cite{CHOI1975285,JAMIOLKOWSKI1972275}
\begin{align}
    \ket{g_U}_{LR} \defeq \sqrt{\frac{1}{2^n}} \sum_{x \in \bits^n} \ket{x}_L U \ket{x}_R,
\end{align}
which can also be seen as applying the unitary $\II_L \otimes U_R$ to the maximally entangled state. This comes from the previously constructed state synthesis algorithms or the state synthesis algorithm of Aaronson \cite{aaronson2016complexity}. While the Choi - Jamiolkowski state contains all the information about $U$, it is unclear how to use $\ket{g_U}$ to apply the unitary $U$. One idea is to use the gate-by-teleportation technique from measurement-based quantum computation to apply $U$. In this procedure, one measures an $n$-qubit input state $\ket{\psi}$ in register $A$ and the half of $\ket{g_U}$ in register $L$ in the generalized Bell basis, i.e. the POVM with elements $\ketbra{g_{X^a Z^b}}_{AL}$ for $a, b \in \bits^n$. It is known that each outcome $(a,b)$ is equally likely to occur and the resulting state on the $R$ register is $U X^a Z^b \ket{\psi}$. Unfortunately, the Pauli twirl, $X^a Z^b$, has been applied inside of the unitary $U$. 

However, note that whenever the measurement outcome $a = b = 0^n$ occurs, the resulting state is $U \ket{\psi}$ as desired. Therefore, there is a \emph{post-selection} algorithm which generates the output $U \ket{\psi}$ by post-selecting on the outcome $a = b = 0^n$. The issue, of course, is that post-selection is a non-unitary operation. However, we note that while post-selection is non-unitary, the classes $\mathsf{PostBQP}$ and $\mathsf{PostQMA}$ have definitions as classical complexity theory classes $\PP$ and $\PSPACE$, respectively. We previously outlined search-to-decision reductions for both of these classes through their equivalences with $\mathsf{PGQMA}$ and $\QMA_\exp$, respectively. While not obvious to us at the moment, we suspect that there may be an insight connecting these ideas together to generate a solution to the unitary state synthesis problem.

\subsection{Improving the construction of witnesses for $\QMA_\exp$}

We leave it as an open question as to whether the Swap Test Distillation algorithm can be used to improve the overlap with a $\QMA_\exp$ witness produced by the 
protocol described in Section \ref{sec:qma_exp}. The challenge is establishing that the conditions for the distillation algorithm are met when $t$-designs (such as Clifford unitaries) are used to randomize the target state instead of Haar-random unitaries.
We know that Clifford unitaries will produce a state whose expected overlap with the target state is at least a constant. Theorem \ref{th:one-query-no-overlap-relax} shows that the Swap Test Distillation algorithm still works under this relaxed condition (instead of requiring that every input state have constant overlap with probability $1$). The problem lies in the second condition: showing  that with high probability, for two independently generated output states, their components orthogonal to the target state are close to orthogonal to each other. The proof in Lemma \ref{lem:swaptestconditions} showing that this holds for the $1$-query protocol that uses Haar-random unitaries relies on the following fact: for any two orthogonal states $\ket{\psi_1}$ and $\ket{\psi_2}$, even when conditioning on the event that $U \ket{\psi_1} = \ket{\phi}$ for some specific $\ket{\phi}$, the state $U \ket{\psi_2}$ is still distributed in a manner that looks close to random. We leave it as an open question whether a similar fact can be shown when $U$ is a $t$-design or whether there is a different way to establish the second requirement for the Swap Test Distillation algorithm. A proof that $t$-designs  satisfy the second requirement for the distillation algorithm would also result in an improvement over the $1$-query protocol for synthesizing arbitrary states shown in Section \ref{sec:one-query}, by reducing the time complexity of the protocol from exponential to polynomial time.

\section*{Acknowledgments}
CN was supported by NSF Quantum Leap Challenges Institute Grant number OMA2016245 and an IBM Quantum PhD internship. HY was supported by AFOSR award FA9550-21-1-0040. SI, CN, and HY were participants in the Simons Institute for the Theory of Computing \emph{Summer Cluster on Quantum Compuatation}. 
Additionally, we thank Aram Harrow and Zeph Landau for insightful discussions.  

\appendix

\input{prelims}
\newpage

\bibliography{references}
\bibliographystyle{alpha}






\end{document}

%% file: berkeleymacros.tex
\usepackage{amsmath, amssymb, graphicx, subfigure}
\usepackage{enumitem}
\usepackage{xparse}
\usepackage{physics}

\newcommand{\CC}{\mathbb{C}}
\newcommand{\FF}{\mathbb{F}}

\newcommand{\II}{\mathbb{I}}

\newcommand{\RR}{\mathbb{R}}

\newcommand{\Cc}{\mathcal{C}}

\newcommand{\Hh}{\mathcal{H}}

\newcommand{\Oo}{\mathcal{O}}

\newcommand{\Exp}{\mathop{\mathbb{E}\hspace{0.13em}}}

\newcommand{\inv}[1]{#1^{-1}}

\newcommand{\sgn}{\mathrm{sgn}}
\renewcommand{\Pr}{\mathop{\mathbf{Pr}\hspace{0.05em}}}

\newcommand{\eps}{\epsilon}

\newcommand{\half}{\frac{1}{2}}

\newcommand{\defeq}{\mathrel{\overset{\makebox[0pt]{\mbox{\normalfont\tiny\sffamily def}}}{=}}}

\usepackage{complexity}
\newcommand{\SHARPP}{\mathsf{\#}\P}

\renewcommand{\exp}{\mathsf{exp}}

\newcommand{\bits}{\{0,1\}}

\newcommand{\fn}[3]{#1: #2 \rightarrow #3}

\newcommand{\qubit}{\CC^{2}}
\newcommand{\qubits}[1]{(\qubit)^{\otimes #1}}

\newcommand{\Cliff}{\mathsf{Cliff}}

\renewcommand{\hat}{\widehat}
\renewcommand{\tilde}{\widetilde}

\newcommand{\UQCMA}{\mathsf{UQCMA}}

\renewcommand{\eqref}[1]{\textrm{eq.~}(\ref{#1})}

\renewcommand{\C}{\CC}

\newcommand{\hilb}{\mathcal{H}}

\newcommand{\qverifier}{\textsc{QCircuitSAT}}

\newcommand{\acc}{\mathrm{acc}}

\renewcommand{\tr}{\Tr}

%% file: introduction.tex
It is a useful fact in classical computer science that \emph{search} problems are often efficiently reducible to \emph{decision} problems.
For example, the canonical way of constructing a satisfying assignment of a given 3SAT formula $\varphi$ (if there exists one) using an oracle for the decision version of 3SAT is to adaptively query the oracle for the satisfiability of $\varphi$ conditioned on some partial assignment to the variables of the formula. Based on the oracle answers, the partial assignment can be extended bit-by-bit to a full assignment. Each oracle query reveals an additional bit of the assignment. This strategy generally works for any problem in $\NP$. Likewise, the optimal value of an optimization problem can be calculated to exponential accuracy using binary search. The main consequence of this is that complexity theory often focuses on decision problems (without losing generality) and less on the complexity of search problems. 

Quantum information and computation has shifted our perspective on these traditional notions of classical complexity theory.
In this paper we consider \emph{quantum} search problems, where the goal is to output a quantum state (as opposed to a classical bit string) satisfying some condition. In the quantum setting, it is no longer apparent that search-to-decision reductions still hold, and thus it is unclear whether the complexity of quantum search problems can be directly related to the complexity of corresponding quantum decision problems. 

To illustrate this, we consider the analogues of $\P$ and $\NP$ in quantum computing, which are the complexity classes $\BQP$ and $\QMA$, respectively\footnote{Technically speaking, $\BQP$ and $\QMA$ are better thought of as the quantum analogues of $\BPP$ and $\sf MA$, respectively. {However,} even in this randomized setting, there are efficient search-to-decision {randomized} reductions.}. The analogue of the $\NP$-complete problem 3SAT for $\QMA$ is the \emph{Local Hamiltonian} problem, in which one has to decide whether the lowest energy state of a local Hamiltonian $H = H_1 + \cdots + H_m$ { acting on $n$ qubits} has energy greater than $a$ or less than $b$ for $a -b = 1/\poly(n)$, where each term $H_i$ acts non-trivially on only a constant number of qubits. This problem was proven to be $\QMA$-complete by Kitaev~\cite{kitaev2002classical}. Is there an efficient search-to-decision reduction for the Local Hamiltonians problem, or more generally for the class $\QMA$? In other words, given {quantum query access to an oracle deciding} the Local Hamiltonians problem, can {a polynomial-time quantum algorithm (i.e. $\BQP$ machine)} efficiently \emph{prepare} a low-energy state $\ket{\psi}$ of a given local Hamiltonian? 

The classical strategy of {incrementally} building a partial assignment does not appear to work in the $\QMA$ setting. 
First, there does not appear to be a natural way of ``conditioning'' a quantum state on a partial assignment.  Second, quantum states are exponentially complex: the description size (complexity) of a general quantum state on $n$ qubits is exponential in $n$, and this is {suspected} to remain true even when considering ground states of local Hamiltonians\footnote{Due to the $\QMA \neq \QCMA$ conjecture \cite{aaronson2007quantum}. Formally, there is no known poly-sized description of a witness (proof) for every local Hamiltonian problem.}. This complexity of quantum states poses a significant challenge to finding a search-to-decision reduction for $\QMA$; it is not clear how yes/no answers to $\QMA$ decision problems (even when obtained in superposition) can be used to construct exponentially-complex $\QMA$ witnesses.



On the other hand, there \emph{is} a natural quantum analogue of the bit-by-bit search-to-decision algorithm for $\NP$ that works for constructing \emph{general} quantum states. This is due to a general algorithm for \emph{state synthesis} described by Aaronson in~\cite{aaronson2016complexity} (for which we give an overview of in \Cref{sec:start}): there exists a polynomial-time quantum algorithm $A$ such that every $n$-qubit state $\ket{\psi}$ can be encoded into a classical oracle $f$ where, by making $O(n)$ superposition queries to the oracle $f$, the algorithm $A$ will output a state that is exponentially close to $\ket{\psi}$. One can observe that for states $\ket{\psi}$ that $\QMA$ witnesses (such as ground states of local Hamiltonians), the oracle $f$ corresponds to a $\PP$ function (which is at least as powerful as a $\QMA$ oracle). This yields a search-to-decision reduction for $\QMA$, albeit with a decision oracle of higher complexity.




{
In this work, we explore the complexity of search-to-decision procedures in the quantum setting, where the goal is a quantum \emph{state synthesis} algorithm that outputs a \emph{target} quantum state (e.g.\ a ground state of a local Hamiltonian) by making quantum queries to a classical decision oracle. We investigate how the complexity of the state synthesis algorithm and the complexity of the decision oracle depend on the type of states we want to generate. We consider both the generalized state synthesis problem for abritrary states in the Hilbert space $\qubits{n}$ as well as the specific task of generating solutions to $\QMA$ problems. 
}

{
We construct state synthesis and search-to-decision procedures for the quantum setting using only one or two superposition queries as opposed to $O(n)$ superposition queries; for $\QMA$ witnesses, the synthesis procedure requires only one query to a $\PP$ oracle. Simultaneously, we prove results suggesting the impossibility of any search-to-decision reduction for $\QMA$. More precisely, we show that there exists a \emph{quantum} oracle $\oracle$ relative to which \emph{all} efficient query algorithms fail to be a good search-to-decision reduction for $\QMA^\oracle$, the relativization of $\QMA$. This stands in contrast to classes such as $\NP$, $\MA$, and $\QCMA$, which all have efficient search to decision reductions, relative to any oracle. As a consequence, proving impossibility of $\QMA$ search-to-decision without an oracle is at least as hard as separating $\QCMA$ and $\QMA$ which is at least as hard as separating $\P$ and $\PP$. We believe that the juxtaposition of our results lend further weight to the view that the complexity of tasks where the outputs (and inputs) are quantum states cannot be directly explained by the traditional study of decision problems (which has been the main focus of quantum complexity theory to date). In particular, we believe our results suggest that the relationship between search and decision problems is much more mysterious in the quantum setting. As suggested by Aaronson in~\cite{aaronson2016complexity} and others in some recent works~\cite{kretschmer2021quantum,rosenthal2021interactive}, the complexity of quantum states (and more generally, quantum state transformations) deserves to be studied more deeply as a subject in its own right. 
}

\subsection{Starting point}
\label{sec:start}

Before describing our results in more detail, we first explain the starting point for our investigations, which is a simple state synthesis algorithm described by Aaronson~\cite{aaronson2016complexity} in his lecture notes.
He shows that there exists a $\poly(n)$-time quantum algorithm $A$ which makes $O(n)$ quantum queries to a classical oracle such that for every $n$-qubit state $\ket{\psi} = \sum_x \alpha_x \ket{x}$, there exists a classical oracle $f$ for which the algorithm $A^{\oracle_f}$ will output a state that is $\exp(-n)$-close to $\ket{\psi}$. 
In ~\cite{aaronson2016complexity}, Aaronson raises the question as to whether his protocol can be improved to a sublinear number of queries. We show, in fact, that $1$ query is sufficient to achieve polynomially small error in synthesizing arbitrary states and $2$-queries are sufficient for exponentially small error. Both the $1$-query and the $2$-query algorithms given here require exponential time and polynomial space.

To understand Aaronson's state synthesis algorithm, we first observe that we can write any quantum state in the form
\begin{align}
    \ket{\psi} = \sum_{x \in \{0,1\}^n} e^{i \theta_x} \, \sqrt{\Pr[X = x]} \, \ket{x}
\end{align}
where $\Pr[X = x]$ is the probability distribution of some $n$-bit random variable $X$ and $\{ \theta_x \}_{\bits^n}$ are a set of phases. The synthesis algorithm performs $2n$ queries to synthesize the ``QSample state''.
\begin{align}
     \sum_{x \in \{0,1\}^n} \sqrt{\Pr[X = x]} \, \ket{x}
\end{align}
and then performs two additional queries at the end to apply the phases $e^{i \theta_x}$ to each basis state $\ket{x}$. 

The $2n$-query procedure to build the QSample state works in $n$ stages. Inductively assume that after the $k$th stage, for $k < n$, the intermediate state of the algorithm is the $k$-qubit state
\begin{align}
    \sum_{y \in \{0,1\}^k} \sqrt{\Pr[X_{\leq k} = y]} \, \ket{y}
\end{align}
where $\Pr[X_{\leq k} = y]$ denotes the marginal probability of the first $k$ bits of $X$ are equal to $y$. Controlled on the prefix $\ket{y}$ the algorithm queries the oracle $f$ to obtain a (classical description of) the conditional probabilities $\Pr[X_{k+1} = 0 \mid X_{\leq k} = y]$ and $\Pr[X_{k+1} = 1 \mid X_{\leq k} = y]$, and prepares a $(k+1)$st qubit in the state
\begin{align}
\sqrt{\Pr[X_{k+1} = 0 \mid X_{\leq k} = y]} \, \ket{0} + \sqrt{\Pr[X_{k+1} = 1 \mid X_{\leq k} = y]} \, \ket{1}~.
\end{align}
The algorithm performs another query to $f$ to uncompute the descriptions of the conditional probabilities. The resulting $k+1$ qubit state is then equal to
\begin{align}
    &\sum_{y \in \{0,1\}^{k+1}} \sqrt{\Pr[X_{\leq k} = y_{\leq k}]} \cdot \sqrt{\Pr[X_{k+1} = y_{k+1} \mid X_{\leq k} = y_{\leq k}]} \, \ket{y} \\
    &=  \sum_{y \in \{0,1\}^{k+1}} \sqrt{\Pr[X_{\leq k+1} = y]} \, \ket{y}
\end{align}
which maintains the desired invariant. 
After the $n$th stage, a similar process applies the phases $\{\theta_x\}$ to generate the output state. The approximations come in when the conditional probabilities and phases are specified with $\poly(n)$ bits of precision, which result in the final state being at most $\exp(-n)$ far from the ideal target state $\ket{\psi}$.
With this $O(n)$-query state synthesis algorithm in mind, we now proceed to describe our results. 


\subsection{Our results}

\newcommand{\blu}[1]{\textcolor{blue}{\textbf{#1}}}

\newcommand{\magn}[1]{\textcolor{magenta}{\textbf{#1}}}
\newcommand{\grn}[1]{\textcolor{OliveGreen}{\textbf{#1}}}
\newcommand{\org}[1]{\textcolor{orange}{\textbf{#1}}}

\begin{table}[ht]
\begin{center}
\small
    \begin{tabularx}{\columnwidth}{p{2.5cm}|X|X|X}
    \toprule 
    Complexity class & 1 query & 2 queries & $O(n)$ queries \\
    \midrule
    $\NP$ & \blu{$\NP$ oracle, $\Omega(\inv{n})$ success probability, \newline \Cref{thm:qcma-query}} & $\leftarrow$ & $\NP$ oracle, \newline classical queries (folklore) \\ \midrule
    $\QCMA$ & \blu{$\QCMA$ oracle, $\Omega(\inv{n})$ success probability, \newline \Cref{thm:qcma-query}}& $\leftarrow$ & $\QCMA$ oracle, \newline classical queries (folklore) \\ \midrule
    $\QMA$ & \magn{$\PP$ oracle, $1/\poly(n)$ precision,  \Cref{thm:1-query-qma-informal}} & 
    $\leftarrow$ (\Cref{thm:two-query-informal} applies but is time-inefficient)
    & {$\PP$ oracle, $1/\exp(n)$ precision~\cite{aaronson2016complexity}} \\ \midrule
    $\QMA_{\exp}$\newline$(=\PSPACE)$ & \magn{$\PSPACE$ oracle \newline $\Omega(1)$ overlap, \newline \Cref{thm:1-query-qma-informal}} & 
    $\leftarrow$ (\Cref{thm:two-query-informal} applies but is time-inefficient)
    & $\PSPACE$ oracle, $1/\exp(n)$ precision~\cite{aaronson2016complexity}\\ \midrule
    Arbitrary states & \org{Arbitrary oracle, \newline $1/\poly(n)$ precision, \newline \Cref{thm:one-query-informal}}  & \grn{Arbitrary oracle, $1/\exp(n)$ precision, 2 queries, \newline \Cref{thm:two-query-informal}} & Arbitrary oracle, $1/\exp(n)$ precision~\cite{aaronson2016complexity}
    \end{tabularx}
\end{center}
\caption{Summary of past work and our results on upper bounds for search-to-decision reductions and state synthesis. The ``complexity class'' column refers to the complexity of the search problem (e.g.\ computing $\NP$ witnesses, or $\QMA$ witnesses). The other columns refer to the algorithmic results known for the specified number of queries; furthermore these are \emph{quantum} queries performed by quantum algorithms in superposition.}
\label{tab:summary}
\end{table}

\paragraph{A one-query search-to-decision algorithm for $\QMA$ with a $\PP$ oracle.} We show Section \ref{sec:QMA} that in the case of generating physically relevant states, i.e. solutions to $\QMA$ problems, such as the low-energy states of local Hamiltonians, that there exists a one-query search-to-decision algorithm using a $\PP$ oracle. While one would hope to find a search-to-decision reduction in which the oracle complexity is only $\QMA$, $\PP$ is the smallest complexity class containing $\QMA$ for which we can construct an oracular algorithm for search problems. Furthermore, given our no-go result for $\QMA$ search-to-decision (see below), this may be the optimal search-to-decision algorithm.

\begin{theorem}[$\QMA$-search to $\PP$-decision reduction]
There exists a probabilistic polynomial time quantum algorithm making a single query to a $\PP$ phase oracle such that, given as input a $\QMA$ problem, either aborts or outputs a witness $\ket{\phi}$. The algorithm will succeed in outputting a witness (i.e. not abort) with all but inverse exponential (in the system size) probability.
\label{thm:1-query-qma-informal}
\end{theorem}


To start sketching the proof, it is fruitful to notice that a single oracle query $\ket{x} \overset{\Oo_f}{\mapsto} (-1)^{f(x)} \ket{x}$ for $x \in \bits^n$ potentially contains $2^n$ bits of information and a quantum state requires $2^n$ complex numbers to describe. Furthermore, the collection of $2^{2^n}$ states 
\begin{align}
\ket*{p_f} \defeq \Oo_f H^{\otimes n} \ket{0^n} = \sum_{x \in \bits^n} (-1)^{f(x)} \ket{x}
\end{align}
defined for any function $\fn{f}{\bits^n}{\bits}$ are a diverse set of states in the Hilbert space. These states, referred to as \emph{phase states} henceforth, despite not forming an $\eps$-net for $\qubits{n}$, turn out to provide a good approximation for $\qubits{n}$ when considering the Haar-random distribution\footnote{Recall, the Haar-measure is the unique left- and right- invariant distribution over unitary matrices over $\qubits{n}$ and the Haar-random distribution is the distribution over quantum states $U \ket{0^n}$ where $U$ is sampled according to the Haar-measure.}. It follows that if we wanted to synthesize the witness to a $\QMA$-complete problem, such as a low-energy state $\ket{\tau}$ for a local Hamiltonian problem, it suffices to build phase state $\ket*{p_f}$ with constant overlap with the low-energy subspace. Finding a state with constant overlap with the target state is sufficient because $\QMA$ is efficiently verifiable, and given a state with constant overlap with the low-energy subspace, it is possible to distill a low-energy state with constant probability (by performing an energy measurement). However, it is not necessarily the case that a low-energy state of $\QMA$ problem will have a good approximation by a phase state. To solve this issue, we prove that for any state $\ket{\tau}$, with high probability $C \ket{\tau}$ will have a good approximation by a phase state where $C$ is a random Clifford unitary. Therefore, we can instead attempt to synthesize $C \ket{\tau}$ which is the result of Theorem \ref{thm:1-query-qma-informal}. In particular, if we can synthesize a phase state $\ket{p}$ that has constant overlap with $C \ket{\tau}$, then $C^{\dagger}\ket{p}$ will have constant overlap with the target $\ket{\tau}$.


Furthermore, we show that, using a slight modification of the same algorithm, we can perform a somewhat weaker one-query search-to-decision reduction for $\QMA_\exp$ (Theorem \ref{th:qma_exp}), the class of non - deterministic quantum computations with only an inverse exponential gap between completeness and soundness. $\QMA_\exp$ is known to equal $ \PSPACE$ \cite{qma_exp,spectral-gap-for-precise-qma}, and our algorithm prepares a witness state with constant overlap with a low-energy state with one query to a $\PSPACE$ oracle (note that here, we cannot efficiently amplify the overlap with an energy measurement due to the inverse-exponential energy gap). As a further observation, we also show that quantum query access to a classical oracle gives one-query search-to-decision reductions when the witness is classical: in particular, for $\QCMA$ (Theorem \ref{thm:qcma-query}), and also for $\NP$. The one-query algorithm preparing the witness first reduces $\QCMA$ to \emph{unique} $\QCMA$ ($\UQCMA$) using the Valiant-Vazirani reduction~\cite{quantum-valiant-vazirani}, and then uses the Bernstein-Vazirani algorithm to extract the unique polynomial length witness with a single query.

\paragraph{A no-go result for search-to-decision for $\QMA$.} The previous result shows that search-to-decision reductions for $\QMA$ are possible with a $\PP$ decision oracle. However, the optimal search-to-decision reduction for $\QMA$ is with a $\QMA$ decision oracle (rather than a stronger $\PP$ oracle). We provide evidence that this is unlikely to exist: we prove that there is a quantum oracle relative to which $\QMA$ search-to-decision is impossible. This stands in contrast to classes such as $\NP$, $\MA$, and $\QCMA$, which all have efficient search to decision reductions, relative to any oracle. 

More precisely, we show that there exists a \emph{quantum} oracle $\oracle$ relative to which \emph{all} efficient query algorithms fail to be a good search-to-decision reduction for $\QMA^\oracle$, the relativization of $\QMA$. The oracle $\oracle$ is a reflection $\II - 2\ketbra{\psi}{\psi}$ about a Haar-random state $\ket{\psi}$; we rely on the concentration of measure phenomenon of the Haar measure to prove this oracle no-go result. We formalize and prove this result in \Cref{sec:no-go}. 

\begin{theorem}[Oracle impossibility for $\QMA$ search-to-decision]
\label{thm:no-go-informal}
There exists a quantum oracle $\oracle$ relative to which \emph{all} $\poly(n)$-time query algorithms fail to be a good search-to-decision reduction for $\QMA^\oracle$.
\end{theorem}

\paragraph{A one-query state synthesis algorithm with inverse polynomial error.}
We also investigate the query complexity of synthesizing an arbitrary state, in the same spirit as Aaronson's adaptive state synthesis algorithm outlined in Section \ref{sec:start}. In particular, we show that that every state $\ket{\tau}$ can be encoded into a classical oracle $f_\tau$ such that by making one query to $\ket{\tau}$, a quantum algorithm can prepare $\ket{\tau}$ with inverse polynomial error.
The space complexity of the synthesis algorithm is polynomial in $n$, the number of qubits in the target state $\ket{\tau}$, but the time complexity is exponential. The starting point for the $1$-query algorithm is the same observation used in the protocol for synthesizing $\QMA$ witnesses, which is that a random state has an expected constant overlap with some phase state. We can think of the oracle function $f_{\tau}$ as hard-coding the target $\ket{\tau}$, but parameterized by unitary $U$ and standard basis state $x$. The oracle $f_{\tau}(U,x) = \text{sgn}(\Re(\bra{x}\ket{U \tau}))$ can be used to create a phase state $\ket{p_U}$ which has constant overlap with $U \ket{\tau}$ with high probability for random $U$. The state $U^{\dagger} \ket{p_U}$ is already then a decent approximation for $\ket{\tau}$. 

There are two remaining techniques to improve upon this basic synthesis protocol. First, we use a novel distillation procedure based on the swap test (explained below) to take a polynomial number of states generated in this manner, using unitaries
$U_{1}, \ldots, ,U_{m}$,
to create a single aggregated output state 
with greater overlap with the target state. Note that since the target state $\ket{\tau}$ is arbitrary, we do not have a means of measuring the overlap of an output state with $\ket{\tau}$ to boost the overlap as we did when the target state is a $\QMA$ witness.
Secondly,  we address the fact that the algorithm described above suffers from needing \emph{exponential space complexity}; this is because specifying a Haar-random unitary on $n$ qubits requires $\exp(\Omega(n))$ space, and thus the oracle $f_{\tau}(U,x)$ needs to act on exponentially many input bits. We derandomize this construction, and show via the probabilistic method that there exists a \emph{single} choice of unitaries $U_{\star,1}, \ldots, ,U_{\star,m}$ that works for \emph{all} $n$-qubit states. This will reduce the space complexity of the algorithm to polynomial, although implementing the unitaries  will still require exponential time.

\begin{theorem}
{\bf (One Query State Synthesis  - Informal)}
\label{thm:one-query-informal}
There is a $1$-query algorithm that uses polynomial space
and exponential time that synthesizes a state $\rho$ such that $\tr{\rho \ketbra{\tau}{\tau}} \ge 1 - 1/q(n)$ for some polynomial $q$ and an arbitrary target state $\ket{\tau}$. 
\end{theorem}

\paragraph{The Swap Test Distillation Algorithm.}
This procedure takes in a polynomial number of  states each of which has at least
a constant overlap with the target state and outputs a state whose overlap with
the target is at least $1 - 1/\text{poly}$.
In some sense, the Swap Test Distillation algorithm provides a way to take the ``mean'' of a collection of quantum samples where each state can be decomposed into a ``signal'' component and a ``noise'' component such that (1) the signal is some constant fraction of the mass and (2) the noise is roughly random.
This may be useful in other contexts in quantum algorithms.

For formally, the algorithm requires that the sequence of input states $\ket{\psi_1}, \ldots, \ket{\psi_m}$ satisfies two properties. The first is that there is a constant $a$ such that $|\braket{\psi_j}{\tau}|^2 \ge a$ for all $j$. (We also show that this condition can be relaxed so that the expected overlap of each input state with the target state is at least $a$, as long as the input states are independently generated.) The second condition is that for every pair of input states, their components orthogonal to $\ket{\tau}$ are close to orthogonal to each other:
\begin{align}
    |\bra{\psi_j}(\II - \ketbra{\tau}{\tau}) \ket{\psi_i}|^2 \le \delta,
\end{align}
for $\delta$ exponentially small in $n$. Intuitively, one can imagine that if the $\ket{\psi_j}$ are generated independently, then the error vectors (the components perpendicular to $\ket{\tau}$) would be random and uncorrelated. We prove that under these two conditions, if the number of states  is a sufficiently large polynomial, then the overlap of the resulting aggregated state with $\ket{\tau}$ is at least $1 - 1/\text{poly}$. The algorithm is based on the observation that if the swap test is applied to a pair of states which each have overlap at least $a$ with the target state, then conditioning on the swap test succeeding (measuring a $0$ in the output bit), the state in each register has an overlap with the target state that is strictly larger than $a$. 
In each round of the algorithm, the surviving states are paired up and the swap test is applied to each pair. One state from every pair that succeeds the swap test advances to the next round.


\paragraph{A two-query state synthesis algorithm with inverse exponential error.} While we do not know how to improve the error of the previous one-query algorithm beyond inverse polynomial, we show that there is a \emph{two-query} state synthesis algorithm that achieves inverse exponential error. 

\begin{theorem}
{\bf (Two Query State Synthesis  - Informal)}
\label{thm:two-query-informal}
There is a $2$-query algorithm that uses polynomial space
and exponential time that with high probability synthesizes a state $\rho$ such that $\tr{\rho \ketbra{\tau}{\tau}} \ge 1 - 1/r(n)$ for some function $r = \exp(n)$ and an arbitrary target state $\ket{\tau}$. 
\end{theorem}

Like with the one-query synthesis algorithm, we take advantage of the properties of Haar-random unitaries. Let $\ket{\tau}$ denote the target state to be synthesized. Whereas the basic building block of the one-query algorithm described is to synthesize the phase state corresponding to $U \ket{\tau}$ where $U$ is a Haar-random unitary, the two-query algorithm attempts to directly synthesize the state $U \ket{\tau}$, and then apply the inverse unitary $U^\dagger$ to recover $\ket{\tau}$. Since $U$ is Haar-random, the distribution of $U \ket{\tau}$ is that of a Haar-random state. 

We then argue that with overwhelmingly high probability, a Haar-random state can be synthesized via two queries to a classical oracle. This relies on the observation that the \emph{amplitude profile} of a Haar-random state concentrates extremely tightly around a fixed profile. By profile, we mean the list of absolute values of amplitudes of the state in sorted order. In other words, there exists a fixed, universal state $\ket{\theta} = \sum_x \beta_x \ket{x}$ such that, with very high probability, a Haar-random state $\ket{\psi} = \sum_x \alpha_x \ket{x}$ satisfies the following: there exists a permutation $\sigma$ on the set of basis states $\ket{x}$ such that the distance
\begin{align}
    \norm{ \ket{\theta} - \sum_x |\alpha_x| \ket{\sigma(x)} }
\end{align}
is exponentially small. To prove this, we utilize bounds from the theory of optimal transport that control the convergence of the \emph{Wasserstein distance} (also known as the \emph{Earth Mover Distance}) between a log-concave distribution and the empirical distribution resulting from sampling from the distribution. 

Given this, the two-query algorithm to synthesize $\ket{\tau}$ to exponential precision is clear: the algorithm first prepares the universal state $\ket{\theta}$. It then queries the classical oracle to determine how to permute the basis states $\ket{x}$ and what phase to apply to all the basis states. The algorithm applies the permutation and the phases in superposition. Finally, the algorithm queries the oracle again to uncompute the permutation/phase information.

Just as with the one-query algorithm, we also perform a derandomization step in order to make the query algorithm space-efficient (but not necessarily time-efficient). By expanding the dimension of the random unitary $U$, we show that there exists (via the probabilistic method) a \emph{single} unitary $U_\star$ that maps \emph{every} target state $\ket{\tau}$ to one whose amplitude profile is exponentially close to the universal one. 

\paragraph{Open Questions. }
We conclude with some open questions which are elaborated in greater detail in Section \ref{sec:openq}.
Can the $1$- and $2$-query algorithms for general state synthesis be improved to polynomial time by using random Cliffords instead of Haar-random unitaries? Is there a $1$-query algorithm for state synthesis that also achieves inverse exponential error?
What is the power of a $\QMA$ decision oracle? In particular, what states can be synthesized with queries to a $\QMA$ oracle in superposition? Is there a weaker oracle class than $\PP$ that can achieve search-to-decision for $\QMA$ witnesses?

%% file: oracle-no-go.tex
\section{Impossibility of search-to-decision for $\QMA$ in oracle model}
\label{sec:no-go}

In this section we show that efficient search-to-decision reductions for $\QMA$ do not exist in general in the oracle setting, perhaps providing some evidence that $\QMA$ does not have efficient search-to-decision reductions ``in the real world.'' More precisely, we show that there exists a \emph{quantum} oracle $\oracle$ relative to which \emph{all} polynomial-time quantum query algorithms fail to be a good search-to-decision reduction for $\QMA^\oracle$, the relativization of $\QMA$. Equivalently, $\QMA^\oracle$-search problems are not reducible to $\QMA^\oracle$-decision problems. 
We contrast this impossibility result with the fact that complexity classes like $\NP$, $\MA$ and $\QCMA$ all have efficient search-to-decision reductions, relative to any oracle (i.e. the reductions relativize)! For example, it is not hard to verify that the search-to-decision procedure for $\QCMA$ described in \Cref{sec:qcma-search-to-decision} relativizes. Thus, \Cref{thm:no-go} illustrates that, at least in the relativized setting, changing the proof model from classical to quantum nullifies the possibility of search-to-decision reductions.

We first define $\QMA^\oracle$ by way of a complete problem. Fix a small constant $\delta < \frac{1}{100}$. Define an \emph{$\oracle$-verifier} circuit $C$ to be a quantum circuit that can make queries to $\oracle$ (which can be viewed as applying a unitary gate for $\oracle$), and also takes as input a quantum proof state $\ket{\phi}$, as well as some ancilla qubits set to $\ket{0}$. Define the promise problem $\oracle$-$\qverifier$ whose YES instances consist of $\oracle$-verifier circuits $C$ for which there is a quantum proof state $\ket{\phi}$ such that $C(\ket{\phi})$ accepts with probability at least $1 - \delta$, and the NO instances are those circuits such that on all quantum witness states, $C$ accepts with probability at most $\delta$. Without access to $\oracle$, this is simply the canonical $\QMA$-complete problem $\qverifier$. The class $\QMA^\oracle$ is then the set of all promise decision problems that are polynomial-time reducible to $\oracle$-$\qverifier$.

Now we formalize the notion of search-to-decision reductions for $\QMA^\oracle$. Consider quantum circuits that can make queries in superposition to both the quantum oracle $\oracle$ and a \emph{classical} oracle $A^\oracle$ that decides the promise problem $\oracle$-$\qverifier$ as well as the controlled-versions of these oracles. Alternatively, we can consider a standard quantum circuit with special oracle ``gates'' implementing $\oracle$ and $A^\oracle$ unitary transformations. Specifically, the oracle $A^\oracle$ implements the unitary transformation
\begin{align}
	\ket{C}\ket{b} \mapsto \ket{C}\ket*{b \oplus A^\oracle(C)}
\end{align}
where $C$ is supposed to be a description of an $\oracle$-oracle circuit, $b$ $\in \bits$, $A^\oracle(C) \in \bits$ with $A^\oracle(C) = 1$ if $C$ is a YES instance of $\oracle$-$\qverifier$, $A^\oracle(C) = 0$ if $C$ is a NO instance, and otherwise $A^\oracle(C)$ is defined arbitrarily. This is sufficiently general as we previously remarked that all $\QMA^\oracle$ problems can be expressed as $\oracle$-oracle circuits $C$. 

{To recap, let $S$ be a quantum circuit describing such a search-to-decision reduction. This means that $S$ consists of standard 2-qubit unitary gates, gates for the oracle $\oracle$ as well as gates for implementing $A^\oracle$ as described previously. The input to $S$ is the description of a $\oracle$-oracle circuit $C$ which is, again, the description of a collection of 2-qubit unitary gates and gates for the oracle $\oracle$. The output of $S$ is a quantum state. In the following analysis, it will be useful to separate the description of $S$ from the oracles $\oracle$ and $A^\oracle$ being used in them. 
In our analysis, we will consider the circuit $S$ with access to different oracles $\oracle'$ and $A^{\oracle'}$. To visualize this, it may be helpful to think of $S$ as a circuit with ``holes'' for $\oracle$ and $A^\oracle$ gates that could be later filled in and swapped out with different oracle gates.}

We then say that such a quantum circuit $S$ is an \emph{$\epsilon$-good search-to-decision reduction} for the problem $\oracle$-$\qverifier$ -- or, alternatively, \emph{$\epsilon$-solves the search version} of $\oracle$-$\qverifier$ -- if when given a YES instance $C$ of $\oracle$-$\qverifier$, it outputs a state that is accepted by $C$ with probability at least $1 - \delta - \epsilon$.

We now state the main result of this section (the technical version of \Cref{thm:no-go-informal}).

\begin{theorem}
\label{thm:no-go}
There exists a constant $\epsilon > 0$ and a quantum oracle $\oracle$ relative to which there is no $\poly(n)$-sized $\epsilon$-good search-to-decision reduction for $\oracle$-$\qverifier$.\footnote{Technically, we should be considering an infinite family of oracles $\oracle$ and $A^\oracle$ where each oracle is parameterized by some input length $n$. However for simplicity we shall just deal with one input length; we will forgo the trouble of spelling out the details of stating our results for asymptotic $n$. To that end, let $\oracle$ be a unitary that acts on $n$ qubits, and we only consider $\oracle$-verifier circuits who accept $n$-qubit quantum proof states; the verifier circuits themselves will be of size $\poly(n)$. }
\end{theorem}

The quantum oracle we use to prove \Cref{thm:no-go} is a reflection unitary $\II- 2\ketbra{\psi}{\psi}$ where $\ket{\psi}$ is some $n$-qubit state\footnote{We conjecture that it may, in fact, be possible to relax this condition to reflection unitaries about only \emph{phase states} (See Definition \ref{def:phase-state}). The only ingredient one would need is a ``discrete'' version of Levy's lemma.} . The existence of a separating oracle $\oracle$ is established via the probabilistic method; we show that by picking $\ket{\psi}$ from the Haar measure, with high probability \emph{all} polynomial-size search-to-decision reduction algorithms fail to solve $\oracle$-$\qverifier$. The challenging part is to deal with the fact that $\oracle$-$\qverifier$ is a promise problem, and that a candidate search-to-decision algorithm may attempt to query the classical oracle $A^\oracle$ on an input $C$ that does not satisfy the promise. However, note that we have some freedom in defining $A^\oracle$ outside of the promise of $\oracle$-$\qverifier$ since any search-to-decision procedure should behave correctly regardless of the responses on invalid queries. We will make careful use of the freedom in defining the oracle $A^\oracle$ on inputs outside of the promise in order to prove the result.

This choice of oracle is inspired by the quantum oracle separation between $\QMA$ and $\QCMA$ obtained by Aaronson and Kuperberg~\cite{aaronson2007quantum}. In fact, one can view \Cref{thm:no-go}, in some sense, as a strengthening of their oracle separation, since the impossibility of search-to-decision for $\QMA$ (essentially) implies that $\QMA$ is different from $\QCMA$. \\

\begin{proof_of}{\Cref{thm:no-go}}
We use a combination of the probabilistic method and a diagonalization argument. Let $S_1,\ldots,S_M$ be an enumeration of all search-to-decision circuits of size at most $T$ for some $T$ to be determined later. Note that $M \leq \exp(\poly(n))$. We will assume each search-to-decision circuit $S_i$ is to take as input the description of a circuit with oracle gates. Pick $M+1$ Haar-random $n$-qubit states $\ket{\psi_1},\ldots,\ket{\psi_{M+1}}$, and define $\oracle_j \defeq \II - 2\ketbra*{\psi_j}$. We aim to show that there exists a choice of quantum oracle $\oracle_j$ such that each $S_i$ fails to be a good search-to-decision reduction for $\oracle_j$-$\qverifier$. To do this, we have to define the behavior of the oracles $A_j^{\oracle_j}$ on inputs $C$ that lie \emph{outside} of the promise of the problem $\oracle_j$-$\qverifier$. 

Let $\acc_\oracle(C)$ denote the maximum acceptance probability of an $\oracle$-verifier circuit $C$, over the choice of quantum proof state. If we treat $C$ as a unitary operator that maps $\hilb_{proof} \otimes \hilb_{anc}$ (proof and ancilla registers, respectively) to $\hilb_{out} \otimes \hilb_{junk}$ (output qubit and junk registers, respectively), then $\acc_\oracle(C)$ is equivalently defined as
\begin{equation}
\label{eq:ckt}
	\acc_\oracle(C) = \left \| ( \bra{1}^{out} \otimes I) C (I \otimes \ket{0 \cdots 0}^{anc}) \right \|^2
\end{equation}
where the $\bra{1}^{out}$ vector acts on the output qubit of the circuit $C$, and $\ket{0 \cdots 0}^{anc}$ acts on the ancillary qubits of $C$. The norm $\| \cdot \|$ denotes the operator norm (i.e.\ maximum singular value).



\textbf{Defining the first oracle $A^{\oracle_1}$.} We now define the \emph{first} classical oracle $A^{\oracle_1}$: for every valid circuit description $C$, we set $A^{\oracle_1}(C) = 1$ if and only if $\acc_{\oracle_1}(C) > 1/2$, otherwise we set it to $0$.
For strings $C$ that don't correspond to valid circuits, we define $A^{\oracle_1}(C) = 0$.

\textbf{Defining the rest of the oracles $A^{\oracle_j}$ for $j > 1$.} The other oracles $A^{\oracle_j}$ are defined exactly in the same fashion, except for circuits $C$ such that $\delta < \acc_{\oracle_j}(C) < 1 - \delta$ (i.e. the $C$ lie outside of the promise of $\oracle_j$-$\qverifier$), we define $A^{\oracle_j}(C) = A^{\oracle_1}(C)$.

\vspace{10pt}

\newcommand{\leTable}{\mathcal{T}}

Draw a table $\leTable$ whose columns are indexed by the circuits $S_1,\ldots,S_M$, and the rows are indexed by $\oracle_1,\ldots,\oracle_{M+1}$. Place a \cmark in the entry $(S_i,\oracle_j)$ if $S_i$ $\epsilon$-solves the search version of $\oracle_j$-$\qverifier$ when given access to $\oracle_j$ and $A_j^{\oracle_j}$, and put an \xmark otherwise.

\begin{lemma}[Main Lemma]
\label{lem:main-no-go}
	With high probability over the choice of oracles $\oracle_1,\ldots,\oracle_{M+1}$, there is at most one \cmark in every column of $\leTable$.
\end{lemma}

\Cref{thm:no-go} follows from the Main Lemma, because it implies that with high probability there exists an oracle $\oracle_{j^*}$ for which there is no \cmark in its row (because there is one more row than there are columns of $\leTable$) -- implying that there is \emph{no} search-to-decision reduction for $\oracle_{j^*}$-$\qverifier$. 

We now prove the Main Lemma:
	Fix a candidate search-to-reduction circuit $S_i$ (i.e., the $i$'th column of table $\leTable$). Fix two distinct row indices $j < k \in [M+1]$. We calculate the probability that $(S_i,\oracle_j)$ and $(S_i,\oracle_k)$ both have \cmark's in them. 
	
	Consider the following verifier circuit $C^*$: given a supposed proof state $\ket{\phi}$, it applies the unitary oracle $\oracle_j$ conditioned on a control qubit $\ket{-}$. Then it projects the control qubit on the state $\ket{+}$, and accepts if the projection succeeds. It is easy to see that it accepts with probability $\left | \ip{\phi}{\psi_j} \right |^2$. Thus for all oracles $\oracle_j$, the circuit $C^*$ is a YES instance of $\oracle_{j}$-$\qverifier$ because there is always a proof that $C^*$, equipped with oracle $\oracle_j$, accepts with probability $1$ (namely, the proof is $\ket{\psi_j}$). 
	
	We first show that with very high probability, the circuit $S_i$ cannot tell the difference when given access to the oracle pair $\oracle_j,A^{\oracle_j}$ versus the oracle pair $\oracle_k,A^{\oracle_k}$. Let $\ket{\phi_j}$ (resp. $\ket{\phi_k}$) denote the output of $S_i$ on input $C^*$ given query access to $\oracle_j$ and $A^{\oracle_j}$ (resp. $\oracle_k$ and $A^{\oracle_k}$). 
	
	\begin{lemma}
	\label{lem:no-go-closeness}
	\begin{align}
	  \Pr \left [ \Big \| \ket{\phi_j} - \ket{\phi_k} \Big \|^2 \geq O(\exp(-\Omega(2^{n/4})))  \right] \leq O(\exp(-\Omega(2^{n/4})))
	\end{align}
	where the probability is over the choice of states $\ket{\psi_j} $ and $\ket{\psi_k}$.
	\end{lemma}

	
	
	Given the lemma, we then have that, with high probability, it cannot be that $S_i$ simultaneously $\epsilon$-solves the search versions of $\oracle_j$-$\qverifier$ and $\oracle_k$-$\qverifier$:
	\begin{align}
	    &\Pr \left [ \text{$S_i$ $\epsilon$-solves $\oracle_j$-$\qverifier$ and $\oracle_k$-$\qverifier$} \right]\\ &\leq \Pr \left [ \left | \braket{\phi_j}{\psi_j} \right|^2 \geq 1 - \delta - \epsilon \text{ and } \left | \braket{\phi_k}{\psi_k} \right|^2 \geq 1 - \delta - \epsilon \right] \label{eq:no-go-double-solve} \\
	    &\leq \Pr
	        \begin{bmatrix}
	              \left | \braket{\phi_j}{\psi_j} \right|^2 \geq 1 - \delta - \epsilon \text{ and } \\  
	              \left | \braket{\phi_k}{\psi_k} \right|^2 \geq 1 - \delta - \epsilon \text{ and } \\
	              \Big \| \ket{\phi_j} - \ket{\phi_k} \Big \|^2 \leq O(\exp(-\Omega(2^{n/4})))
	        \end{bmatrix} \notag \\
	    &\qquad \qquad \qquad +   \Pr \left [ \Big \| \ket{\phi_j} - \ket{\phi_k} \Big \|^2 \geq O(\exp(-\Omega(2^{n/4}))) \right] \label{eq:no-go-union-bound} \\
	    &\leq \Pr \left [ \left | \braket{\psi_j}{\psi_k} \right|^2 \geq 1 - O(\delta + \epsilon) \right] + \Pr \left [ \Big \| \ket{\phi_j} - \ket{\phi_k} \Big \|^2 \geq O(\exp(-\Omega(2^{n/4}))) \right] \label{eq:no-go-1}
	\end{align}
	Eq.~(\ref{eq:no-go-double-solve}) follows because $\eps$-solving the search versions of both $\oracle_j$- and $\oracle_k$-$\qverifier$ implies that $S_i$ can solve those problems on input $C^*$, which means that the output states $\ket{\phi_j},\ket{\phi_k}$ are close to the states $\ket{\psi_j},\ket{\psi_k}$, respectively. Eq.~(\ref{eq:no-go-union-bound}) follows from the union bound, and \eqref{eq:no-go-1} follows from the triangle inequality. 
	
	On the other hand, with high probability over the choice of $\oracle_j,\oracle_k$, we have that $\ket{\psi_j}$ is \emph{far} from $\ket{\psi_k}$. This follows from Levy's Lemma (\Cref{lem:levy}): 
	\begin{align}
	    \Pr \left [ \left | \braket{\psi_j}{\psi_k} \right|^2 \geq \frac{1}{2} \right] \leq \exp \left (-\Omega(2^{n}) \right)~.
	\end{align}
	This, combined with \Cref{lem:no-go-closeness}, implies that ~\eqref{eq:no-go-1} is at most $O(\exp(-\Omega(2^{n/4})))$. Union bounding over all pairs $j < k$ of rows and all columns $i$, we have that the probability there is a column with more than a single \cmark is at most 
	$O(M^3) \cdot O(\exp(-\Omega(2^{n/4}))) \leq O(\exp(-\Omega(2^{n/4})))$, which proves the Main Lemma.
	
	
	Thus it just remains to show \Cref{lem:no-go-closeness}. 
	The circuit $S_i$ can be expressed as a product of $R \leq T$ unitary operations for
    \begin{align}
		S_i = U_R U_{R-1} \cdots U_1
	\end{align}
	where $U_t$ is either a unitary that's independent of the choice of $\oracle$, a (controlled) call to $\oracle$, or a (controlled) call to $A^\oracle$. The oracles may act on different subsets of qubits each time.
	
	Fix a $U = U_t$ for some $t \in [R]$. We argue that, when $U$ is either a controlled call to $\oracle$ or $A^\oracle$, for a fixed input state $\ket{\alpha}$, the state $U \ket{\alpha}$ is very close to a \emph{fixed} vector that is independent of the oracle $\oracle$, with high probability over $\oracle = \II- 2\ketbra{\psi}{\psi}$. (This is also trivially true when $U$ is independent of $\oracle$.)
	
	Suppose $U$ is a query to $\oracle = \II- 2\ketbra{\psi}{\psi}$. Suppose for now that $\oracle$ is applied to the first $n$ qubits of $\ket{\alpha}$. Then
	\begin{align}
		\bra{\alpha} \oracle \ket{\alpha} = 1 -2  \Tr \Big ( \ketbra{\alpha}{\alpha} \ketbra{\psi}{\psi} \Big ).
	\end{align}
	Note that $\ket{\psi}$ is an $n$-qubit state, while $\ket{\alpha}$ is a state on potentially more qubits, because it is supposed to represent the intermediate state the circuit $S_i$. 
	Define the function $f : (\C^2)^{\otimes n} \to \RR$ be defined as
	\begin{align}
		f(\ket{\psi}) \defeq \Tr\left ( \ketbra{\alpha}{\alpha} \ketbra{\psi}{\psi} \right ).
	\end{align}
	The expectation of $f$ over a Haar-random $\ket{\psi}$ is simply $2^{-n}$. By Levy's Lemma, this implies that 
	\begin{align}
		\Pr \left [ \left | f(\ket{\psi}) - 2^{-n} \right| > \eta \right ] \leq 2\exp (-\Omega(2^n \eta^2)).
	\end{align}
	Set $\eta = 2^{-n/3}$. Translating from inner products to squared Euclidean distance and taking a union bound we have
	\begin{align}
	    &\Pr \left [ \max \left \{ \left \| \ket{\alpha} - \oracle_j \ket{\alpha} \right\|^2 , \left \| \ket{\alpha} - \oracle_k \ket{\alpha} \right\|^2 \right \}  \geq O(\eta) \right] \\
	    & \qquad \leq 2 \Pr \left [ \left \| \ket{\alpha} - \oracle \ket{\alpha} \right\|^2 \geq O\eta) \right ]\\
	    & \qquad = 2 \Pr \left [ 2 - 2 \bra{\alpha} \oracle \ket{\alpha}  \geq O(\eta) \right ] \\
	    & \qquad = 2 \Pr \left [ f(\ket{\psi})  \geq O(\eta) \right ] \\
	    &\qquad \leq O(\exp(-\Omega(2^{n/3})))
	\end{align}
	where the probability is over the choice of $\oracle_j,\oracle_k$. 
	

	
	Now suppose $U$ is a query to $A^\oracle$. Consider an input state 
	\begin{align}
	\ket{\alpha} = \sum_{C,b} \alpha_{C,b} \ket{C,b} \ket{\varphi_{C,b}}
	\end{align}
	where the registers $\ket{C,b}$ correspond to the query registers for $A^\oracle$, and the $\{ \ket{\varphi_{C,b}} \} $ are arbitrary states that depend on $C$ and $b$, but are not acted upon by $A^\oracle$. The sum runs over all verifier circuit instances $C$ of size at most $T$ (many of which are not proper encodings of circuits), and $\ket{b}$ is a qubit register.
	
	Let $\ket{\beta_j} = A^{\oracle_j} \ket{\alpha}$ and $\ket{\beta_k} = A^{\oracle_k} \ket{\alpha}$ be the result of running $A^{\oracle_j}$ or $A^{\oracle_k}$ on the fixed input state $\ket{\alpha}$. We have that
	\begin{align}
		\ket{\beta_j} &= \sum_{C,b} \alpha_{C,b} \ket{C,b \oplus A^{\oracle_j}(C)} \ket{\varphi_{C,b}} \qquad \text{ and } \\ \qquad \ket{\beta_k} &= \sum_{C,b} \alpha_{C,b} \ket{C,b \oplus A^{\oracle_k}(C)} \ket{\varphi_{C,b}}~.
	\end{align}
	Therefore
	\begin{align}
		\Exp \left \| \ket{\beta_j} - \ket{\beta_k} \right \|^2  
		 &= 2 \Exp \sum_{C}  \left | \alpha_{C,0} - \alpha_{C,1} \right |^2   \cdot \mathbf{1} \{ A^{\oracle_j} (C) \neq A^{\oracle_k}(C) \} \\
		 &\leq 4 \Exp \sum_{C}  \left ( |\alpha_{C,0}|^2 + |\alpha_{C,1} |^2 \right)   \cdot \mathbf{1}{\{ A^{\oracle_j} (C) \neq A^{\oracle_k}(C) \}} \\
		 &= 4 \sum_{C}  \left ( |\alpha_{C,0}|^2 + |\alpha_{C,1} |^2 \right)   \cdot \Pr \left [ A^{\oracle_j} (C) \neq A^{\oracle_k}(C) \right ]
	\end{align}
	Here, the expectation is over the choice $\oracle_j,\oracle_k$. Now we argue that for any fixed circuit $C$ of size at most $T$, the probability $\Pr \left [ A^{\oracle_j} (C) \neq A^{\oracle_k}(C) \right ]$
	is small. If $C$ is not a valid verifier circuit, then this probability is $0$ (since both $A^{\oracle_j}$ and $A^{\oracle_k}$ by construction give answer $0$). 
	
	Now suppose that $C$ is a properly encoded verifier circuit. Let us define $\acc_{\psi}(C)$ as $\acc_{\oracle}(C)$ for $\oracle = \II - 2 \ketbra{\psi}$ and $\acc_{\oracle}(C)$ defined in equation~\eqref{eq:ckt}; we abbreviate this as $	\acc_{\psi}(C) = \| \bra{1} C_\psi \ket{0} \|^2$ where $C_\psi$ denotes the unitary operator corresponding to circuit $C$ that makes calls to oracles $\oracle$ and $A^\oracle$ that depend on $\ket{\psi}$.
	Define $avg(C) \defeq \Exp \acc_\psi (C)$, where the expectation is over a Haar-random $\ket{\psi}$. Define the function $f(\ket{\psi}) = \acc_\psi(C)$. We calculate its Lipschitz constant. We have for all $\ket{\psi},\ket{\theta}$,
	\begin{align}
		\left | f(\ket{\psi}) - f(\ket{\theta}) \right |  &= \left | \| \bra{1} C_\psi \ket{0} \|^2 - \| \bra{1} C_\theta \ket{0} \|^2 \right | \\
		&\leq 2 \left \|\bra{1} C_\psi \ket{0} - \bra{1} C_\theta \ket{0} \right \|^2 \\
		&\leq 2 \left \| C_\psi -  C_\theta \right \|^2. 
	\end{align} 
	The second line follows from the triangle inequality for the operator norm. The third line follows from the fact that $\| A B \| \leq \|A \| \cdot \|B\|$ for operators $A,B$. Next, suppose we write $C$ as $V_T V_{T-1} \cdots V_1$ where $V_t$ are unitary operators that are either independent of the oracle $\oracle$, or $V_t = \oracle$. We can then write
	\begin{align}
	    C_\psi = V^\psi_R V^\psi_{R-1} \cdots V^\psi_1 \qquad \text{ and } \qquad C_\theta = V^\theta_R V^\theta_{R-1} \cdots V^\theta_1~.
	\end{align}
	By a hybrid argument we have
	\begin{align}
	    \left \| C_\psi -  C_\theta \right \| \leq \sum_{t = 1}^R \| V_t^\psi - V_t^\theta \|~.
	\end{align}
	If $V_t^\psi$ and $V_t^\theta$ are independent of $\psi,\theta$, respectively, then the $t$'th term is $0$. If $V_t^\psi = \II- 2\ketbra{\psi}{\psi}$ and $V_t^\theta = \II- 2\ketbra{\theta}{\theta}$, then we have 
	\begin{align}
	\| V_t^\psi - V_t^\theta \| \leq 2 \| \ketbra{\psi}{\psi} - \ketbra{\theta}{\theta} \|~.
	\end{align}
	Thus, we have
	\begin{align}
	\left | f(\ket{\psi}) - f(\ket{\theta}) \right | \leq O \Big ( T^2 \| \ketbra{\psi}{\psi} - \ketbra{\theta}{\theta} \|^2 \Big) \leq O \Big (T^2 \| \ket{\psi} - \ket{\theta} \| \Big)
	\end{align}
	where we used that $\| \ketbra{\psi}{\psi} - \ketbra{\theta}{\theta} \|^2 \leq 4\| \ket{\psi} - \ket{\theta} \|^2 \leq 16 \| \ket{\psi} - \ket{\theta} \|$. 
	This implies that the Lipschitz constant for $f$ is at most $O(T^2)$. Hence by Levy's Lemma,
	\begin{align}
	\Pr \left [ \left | f(\ket{\psi}) - avg(C) \right | \geq \eta \right ] \leq 2 \exp \left ( -\Omega (2^n  \eta^2/16T^4) \right).
	\end{align}
	Define $\gamma \defeq 2 \exp \left ( -O(2^n \eta^2/16T^4) \right)$. 
	
	We can now evaluate the probability that $A^{\oracle_j}(C) \neq A^{\oracle_k}(C)$. We consider two cases. The first is that $avg(C)$ is in the ``grey zone'', meaning that $\delta + \eta < avg(C) < 1 - \delta - \eta$. This means that, on average, the acceptance probability of $C$ is outside of the promise of $\oracle$-$\qverifier$. Then with probability at least $1 - 2\gamma$, the acceptance probabilities $\acc_{\psi_j}(C)$ and $\acc_{\psi_k}(C)$ are going to remain strictly between $\delta$ and $1 - \delta$. By construction, this means that $A^{\oracle_j}(C) = A^{\oracle_k}(C) = A^{\oracle_1}(C)$, so in this case $\Pr \left [A^{\oracle_j}(C) \neq A^{\oracle_k}(C) \right ] \leq 2\gamma$.
	
	The second case is that $avg(C)$ is not in the grey zone. Consider the case that $avg(C) \leq \delta + \eta$. With probability at least $1 - 3\gamma$, all of $\acc_{\psi_1}(C),\acc_{\psi_j}(C),\acc_{\psi_k}(C)$ are at most $\delta + 2\eta \ll \frac{1}{2}$. This means that $A^{\oracle_1}(C) = 0$, and thus regardless of whether $C$ lies outside the promise of $\oracle_j$-$\qverifier$ or $\oracle_k$-$\qverifier$, we have that $A^{\oracle_j}(C) = A^{\oracle_k}(C) = A^{\oracle_1}(C) = 0$ by construction. A similar argument goes through for when $avg(C) > 1 - \delta - \eta$.
	
	Therefore in all cases $\Pr \left [ A^{\oracle_k} (C) \neq A^{\oracle_j}(C) \right ] \leq 3\gamma$. This implies that $\Exp \left \| \ket{\beta_j} - \ket{\beta_k} \right \|^2 \leq 12\gamma$. By Markov's inequality with probability at least $1 - O(\sqrt{\gamma})$, we have $\left \| \ket{\beta_j} - \ket{\beta_k} \right \|^2 \leq O(\sqrt{\gamma})$. 
	
	To summarize, we have showed that for any $t \in [R]$, for any fixed state $\ket{\alpha}$, with probability at least $1 - \exp(-\Omega(2^{n/3}/T^4))$, we have that 
	\begin{align}
	\| U_t^{\psi_j} \ket{\alpha} - U_t^{\psi_k} \ket{\alpha} \| \leq O(\exp(-\Omega(2^{n/3}/T^4)))
	\end{align}
	where $U_t^{\psi_j}$ and $U_t^{\psi_k}$ denote the $t$'th unitary operation of the candidate search-to-decision circuit $S_i$ when querying oracles $\oracle_j,A^{\oracle_j}$ and $\oracle_k,A^{\oracle_k}$, respectively. Summing over all $R$ time steps of the circuit $R$, we get that with probability at least 
	\begin{align}
	    1 - O(R \, \exp(-\Omega(2^{n/3}/T^4)))
	\end{align}
	the outputs of $S_i$ with either oracle $\oracle_j$ or $\oracle_k$ are going to be at most $O(R \, \exp(-\Omega(2^{n/3}/T^4)))$-far from each other. Plugging in $R \leq T = \poly(n)$, we obtain \Cref{lem:no-go-closeness}. 
	
	

\end{proof_of}

%% file: one-query.tex
\section{$1$-query state synthesis algorithm with polynomially small error}
\label{sec:one-query}

We describe here a $1$-query, polynomial-space algorithm that achieves polynomially small error.
The state synthesis algorithm will not be efficient. We will start with a first attempt, which has exponential space complexity and then fix it so that it has polynomial space complexity.
The algorithm makes use of the Swap Test Distillation algorithm described in Section \ref{sec:swaptest}
that takes as input a polynomial number of states, each with at least constant overlap with the target state, and uses successive applications of the Swap Test to produce a final state whose overlap with the target state is at least $1 - 1/\poly(n)$.

\subsection{A space-inefficient algorithm}

Let $d = 2^n$, $n' = n^2$, and $d' = 2^{n'}$.
The $m$ applications of the $1$-query algorithm along with the Swap Test Distillation algorithm will be applied to $n'$-qubit registers with target state
$\ket{\tau'} = \ket{\tau} \otimes \ket{0}^{\otimes (n'-n)}$.
The expansion of the space is important for derandomizing the algorithm later on. In particular, we will show that there is a fixed sequence of unitaries that works for all $\ket{\tau'}$
of the form $\ket{\tau} \otimes \ket{0}^{\otimes (n'-n)}$. This will allow us to hard-code the unitaries into the oracle function. The resulting algorithm
will still require exponential time to implement the unitaries, but the derandomized algorithm will require only polynomial space.

We will define a function $f_{\tau'}: \text{U}(d') \times \{0,1\}^{n'} \rightarrow \{0,1\}$,  where $\text{U}(d')$ is the space of all unitaries on a $d'$-dimensional Hilbert space, and
\begin{align}
    f_{\tau'} (U, x) &\defeq \sgn \left( \Re \left( \bra{x} U \ket{\tau'} \right) \right) \label{eq:1q-fn}
\end{align}
The corresponding phase state is
\begin{align}
    \ket{p_U} = \sum_{x \in \{0,1\}^{n'}} (-1)^{f_{\tau'}(U,x)} \ket{x}
\end{align}

\begin{figure}[ht]
\noindent
\begin{center}
\fbox{\begin{minipage}{\textwidth}
\begin{tabbing}
{\sc OneQueryStateSynthesis} (\emph{space inefficient version})\\
(1) \FOR $j = 1, \ldots, m$ in parallel:\\
(2) ~~~~~~~\= Sample Haar-random $n'$-qubit unitary $U_j$. \\
(3) \> In the $j$th $n$'-qubit register, prepare the equal superposition $\sum_{x \in \{0,1\}^{n'}} \ket{x}$. \\
(3) \> Controlled on basis state $\ket{x}$, query the oracle on input $(U_j,x)$ to apply $f_{\tau'}(U_j,x)$ \\
\> ~~~~~~ and produce phase state $\ket{p_{U_j}}$ on $n'$ qubits. \\
(4) \> Apply $U_j^{\dagger}$ to the phase state. \\
(5) Apply the {\sc SwapTestDistillation} Algorithm (Figure \ref{fig:IterativeSwap}) to the $m$ resulting states $\ket{\psi_1}, \ldots, \ket{\psi_m}$\\
(6) Output the first $n$ qubits of any surviving register.
\end{tabbing}
\end{minipage}}

\end{center}
\caption{Pseudo-code for the {\sc OneQueryStateSynthesis} query algorithm that uses exponential space complexity.}
\label{fig:onequery}
\end{figure}

The algorithm will output $m$ expanded registers on $n'$ qubits.
We will apply the Swap Test Distillation algorithm to $m$ states on
$n'$ qubits generated by $m$ parallel (and independent) applications of
the $1$-query algorithm and analyze the probability
that the resulting state has at least $1 - 1/\poly(n)$ overlap with
$\ket{\tau'} $.
The mixed state $\rho$ in the first $n$-qubits will also have
$\tr{\rho \ketbra{\tau}{\tau}} \ge 1 - 1/\poly(n)$.

The following lemma establishes that with high probability after step $(4)$
of the algorithm, the conditions for the Swap Test Distillation Algorithm are met.

\begin{lemma}
\label{lem:swaptestconditions}
{\bf (Probability Conditions Satisfied for Swap Test Distillation)}
Let $\ket{\psi_1} \otimes \cdots \otimes \ket{\psi_m}$ be the states in the $m$ registers after Step $(4)$. There is a  constant  $C$  such that
\begin{enumerate}
\item $\Pr_{U_1, \cdots, U_m} \left[ \min_j \{ | \braket{\psi_j}{\tau'} |^2 \} \le 1/8 \right] \le m \cdot \exp(-C d')$ 
\item $\Pr_{U_1, \cdots, U_m} \left[ \max_{i \neq j} \{ | \bra{\psi_i} (I - \ketbra{\tau'}{\tau'}) \ket{\psi_j} |^2 \} \ge  (d')^{-1/4} \right] \le  m^2 \cdot \exp(-C (d')^{1/2})$  
\end{enumerate}
\end{lemma}

\begin{proof}
For the first part, consider a single application of the $1$-query protocol with Haar random unitary $U$.
Let $\ket{p_U}$ be the phase state corresponding to $U \ket{\tau'}$.
The output state is $U^{\dagger} \ket{p_U}$ and we would like to upper bound
the probability
that 
$|\bra{p_U} U \ket{\tau'}|^2$ is less than $1/8$:
\begin{align}\Pr\left[ |\bra{p_U} U \ket{\tau'}|^2 \le 1/8 \right] &=
\Pr\left[ |\bra{p_U} U \ket{\tau'}| \le 1/2\sqrt{2} \right]
\\
&\le \Pr\left[ |\bra{p_U}  \ket{\Re(U\tau')}| \le 1/2\sqrt{2} \right].\end{align}
Using Fact \ref{fact:general-state-overlap}, 
\begin{align}\Pr\left[ |\bra{p_U}  \ket{\Re(U\tau')}| \le 1/2\sqrt{2} \right]
= \Pr\left[ \norm{\Re(U\tau')}_1 \le \sqrt{d'}/2\sqrt{2} \right]\end{align}
Since $U \ket{\tau'}$ is a Haar-random state, the expected $L_1$ norm of
$\ket{\Re(U\tau')}$ is $\sqrt{d'/2}$. We can now use Levy's Lemma (\ref{lem:levy})
with the dimension $N = d'$ and $\delta = \sqrt{d'/2}$. If the function $f$ in Levy's Lemma is the $L_1$ norm, then the value of $K$ is $\sqrt{d'}$.
Thus, 
\begin{align}\Pr\left[ \norm{\Re(U\tau')}_1 \le \sqrt{d'}/2\sqrt{2} \right] &\le
\Pr\left[ \left| \norm{\Re(U\tau')}_1 - \sqrt{d'/2} \right| \le \sqrt{d'}/2\sqrt{2} \right] \\
&\le \exp(-C_1 d'),\end{align}
for some constant $C_1$. The probability that the protocol fails to produce a state with overlap at least $1/8$ with the target state
in a single run is at most $\exp(-C_1 d')$. Since the $1$-protocol is run $m$ times in parallel, the probability
that any of the resulting states fails to have overlap at least $1/8$ is
at most $m \cdot \exp(-C_1 d')$.

For the second part, define $\ket{\phi_j}$ to be the normalized component of $\ket{\psi_j}$ that is orthogonal to $\ket{\tau'}$ and $\ket{r_j}$ the component of $\ket{p_U}$ that is orthogonal to $U \ket{\tau'}$. Equivalently,
\begin{align}
\ket{\phi_j} = \frac{ (\II - \ketbra{\tau'}{\tau'}) \ket{\psi_j}}{ \norm{(\II - \ketbra{\tau'}{\tau'}) \ket{\psi_j}}}~~~~~
\text{and}~~~~~
\ket{r_j} = \frac{  \ket{p_U} - \bra{\tau'}  U^{\dagger} \ket{p_U} U \ket{\tau'}}{\norm{\ket{p_U} - \bra{\tau'} U^{\dagger} \ket{p_U} U \ket{\tau'}}}.
\end{align}
We will bound
$|\braket{\phi_i}{\phi_j}|^2$ which is in turn an upper bound for $| \bra{\psi_i} (I - \ketbra{\tau'}{\tau'}) \ket{\psi_j} |^2$. Note that $U^{\dagger} \ket{r_j} = \ket{\phi_j}$.

Now fix $\ket{\phi_i}$ from the $i^{th}$ run of the $1$-query algorithm and consider the selection of $U_j$ on the $j^{th}$ run. We can think of selecting $U_j$ in two stages:
\begin{enumerate}
    \item First select the target of state $\ket{\tau}$ under $U_j$, which we will call $\ket{s_j} \defeq U_j \ket{\tau}$.
    \item Then select a Haar-random isometry $V_j$ that maps $\ket{s_j}^{\perp}$ to $\ket{\tau}^{\perp}$, where $\ket{s_j}^{\perp}$ is the subspace perpendicular to $\ket{s_j}$ and
    $\ket{\tau}^{\perp}$ is the subspace perpendicular to $\ket{\tau}$.
\end{enumerate}
Then $U_j = \ketbra{s_j}{\tau} + V_j^{\dagger}$. The state $\ket{\phi_i}$ is a fixed state (independent of the choice of $U_j$) that lies in $\ket{\tau}^{\perp}$. The selection of $\ket{s_j}$ in Step $1$ determines $\ket{r_j}$. Then, the unitary $V$ maps $\ket{s_j}$ to a random state in $\ket{\tau}^{\perp}$. 
We can apply Levy's Lemma again with 
$f(V \ket{s_j}) = |\bra{\phi_i} V \ket{s_j}|^2$.
The value of $K$ is $O(1)$, and 
the expectation of $f$ over the choice of $V_j$ is then $1/(d'-1)$. 
Setting $\delta = (d')^{-1/4}$, we have that
\begin{align}\Pr\left[ |\bra{\phi_i} V \ket{s_j}|^2 \ge (d')^{-1/4} \right] \le \exp(-C_2 (d')^{1/2}).\end{align}
There are at most $m^2$ pairs of states. The probability that any of them have $|\braket{\phi_i}{\phi_j}|^2 \ge (d')^{-1/4}$
is at most $m^2 \exp(-C_2 (d')^{1/2}) $.
The final constant $C$ in the statement of the theorem can be taken to be $\min\{C_1, C_2\}$.
\end{proof}

\subsection{A space-efficient algorithm}

The algorithm described above suffers from needing \emph{exponential space complexity}; this is because specifying a Haar-random unitary on $n'$ qubits requires $\exp(\Omega(n'))$ space, and thus the oracle $f(U,x)$ needs to act on exponentially many input bits. We derandomize this construction, and show via the probabilistic method that there exists a \emph{single} choice of unitaries $U_{\star,1}, \ldots, ,U_{\star,m}$ that works for \emph{all} $n$-qubit states --- this is why we expanded the space to dimension $d'$.

Let $\ket{v_1},\ldots,\ket{v_D}$ denote an $\epsilon$-net for the space of $n$-qubit quantum states where $\epsilon = d^{-1}$. Then there at most $D \leq \epsilon^{-d} = d^{d}$ states in this enumeration. Fix an index $1 \leq i \leq D$. Imagine running the $1$-query protocol in Figure \ref{fig:onequery} in parallel $m$ times with target state $\ket{v_i} \otimes \ket{0}$. The probability that the protocol fails to satisfy the conditions for the Swap Test Distillation algorithm from Lemma \ref{lem:swaptestconditions} is at most $2 m^2 \exp(-\Omega( (d')^{3/4}))$ over the choice of $U_1, \ldots, U_m$. By a union bound, the probability that a random choice of $U_1, \ldots, U_m$ fails to satisfy the conditions from Lemma \ref{lem:swaptestconditions}
for \emph{a single one} of the $\ket{v_1},\ldots,\ket{v_D}$ is at most
\begin{align}
	d^d \cdot 2m^2 \exp(-\Omega( (d')^{1/2})) \leq 2^{n2^n} \cdot 2m^2 \exp(-\Omega( 2^{n^2/2})).
\end{align}
Since $m$ is polynomial in $n$, for sufficiently large $n$, this probability is less than $1$. Thus there exists a choice of unitaries $U_{\star,1}, \ldots, ,U_{\star,m}$ that results in a set of $m$ states that satisfy the conditions for the Swap Test Distillation algorithm for  \emph{all} the $\ket{v_1},\ldots,\ket{v_D}$. Hardcode these unitaries into the algorithm and oracles:
$f_{\tau,i}(x) = f_{\tau}(U_{\star,i},x )$.
Now the oracles  only take $n'$ bits as input each, and the resulting query algorithm  now only requires $\poly(n)$ space. Note that the implementation of the unitaries $U_{\star,1}, \ldots, ,U_{\star,m}$ will not be time-efficient in general, but they are still fixed unitary operators that act on $n'$ qubits. 

Thus for an arbitrary target state $\ket{\tau}$, use the oracles $f_{v_i}(U_{\star,1},x), \ldots, f_{v_i}(U_{\star,m},x)$ corresponding to the nearest state $\ket{v_i}$ in the $\eps$-net, which is within $d^{-1}$ of $\ket{\tau}$. Therefore, the one-query algorithm using
unitaries $U_{\star,1}, \ldots, ,U_{\star,m}$, followed by the Swap Distillation Algorithm will incur an additional
$O(d^{-1})$ error.

\begin{theorem}
{\bf (One Query State Synthesis Performance)}
For every polynomial $q$, there is a polynomial $p$ and constant $C'$ such that if the {\sc OneQueryStateSynthesis} is run with $m \ge p(n)$ registers, then with probability at least $1 - \exp(-C' n)$, the algorithm produces a state $\rho$ such that
$\tr{\rho \ketbra{\tau}{\tau}} \ge 1 - 1/q(n)$. The oracle queried by the algorithm will depend on the closest state to $\ket{\tau}$ in the $\epsilon$-net.
\label{thm:onequerystatesynthesis}
\end{theorem}

\begin{proof}
Those choice of $U_{\star,1}, \ldots, ,U_{\star,m}$ guarantees that the conditions from Lemma \ref{lem:swaptestconditions} hold for the nearest state to the target state $\ket{\tau}$ from the $\epsilon$-net. The overlap between the $\ket{\phi_j}$'s will
be $\delta = C (d')^{-1/4} = C 2^{-n^2/4}$. The overlap $a$ between each $\ket{\phi_j}$ and the target state will be at least $1/8$.

From Theorem \ref{th:one-query-small-overlap}, we can select $\ell = c \log_{5/4}(2n) + 1 /2a^2$ and $m \ge n 6^\ell$ to obtain overlap to the nearest state in the $\epsilon$-net
that is at least $1 - 1/n^c - 8m \sqrt{\delta}$ with probability at least $1 - 2 \exp(-n/12) - 4m \sqrt{\delta}$. The overlap with the desired target state is then at least
$1 - 1/n^c - 8m \sqrt{\delta} - 2^{-n}$.
Select a constant $c$ so that the overlap is at least $1 - 1/q(n)$ which then determines $\ell$ and the polynomial $p(n) \ge n 6^l$. 
\end{proof}

%% file: wasserstein.tex
\renewcommand{\cal}[1]{\mathcal{#1}}
\newcommand{\normal}{\cal{N}}
\newcommand{\sort}{\mathrm{sort}}
\newcommand{\rv}[1]{#1}
\newcommand{\reg}[1]{\mathsf{#1}}

\section{$2$-query state synthesis algorithm with exponentially small error}

We now describe a $2$-query state synthesis algorithm that achieves \emph{exponentially small} error. Like with the $1$-query algorithm from \Cref{sec:one-query},  it will be space-efficient, but not time-efficient. And also like in \Cref{sec:one-query}, we first describe a version of the algorithm with exponential space complexity, and then describe how to reduce the space complexity to polynomial.

\subsection{A space-inefficient algorithm} 
Let $d = 2^n$, $n' = n^2$ and $d' = 2^{n'}$. 
Let $\{ \sigma_{U,V} \}_{U,V}$ denote a set of permutations on $\{0,1\}^{n'}$, indexed by unitaries $U,V$ on $n'$ qubits. For all unitaries $U,V$ and $n'$-bit strings $x$, let $\phi(U,V,x)$ denote a number in $[0,2\pi)$, representable using $(n')^2$ bits. We will specify $\{ \sigma_{U,V} \}$ and $\phi$ later. Define the oracles
\begin{align}
	f(U,V,x) \defeq (\phi(U,V,x),\sigma_{U,V}(x))	\quad \text{and} \quad g(U,V,y) = (\sigma_{U,V}^{-1}(y),\phi(U,V,\sigma_{U,V}^{-1}(y)))~.
\end{align}
These oracles have the property that if $f(U,V,x) = (\phi,y)$, then $g(U,V,y) = (x,\phi)$.

\begin{figure}[ht]
\noindent
\begin{center}
\fbox{\begin{minipage}{\textwidth}
\begin{tabbing}
{\sc TwoQueryStateSynthesis} (\emph{space inefficient version})\\
(1) Sample Haar-random $n'$-qubit unitaries $U,V$. \\
(2) Prepare the state $\ket{\theta} = U \ket{0}^{\otimes n'}$ in an $n'$-qubit register $\reg{A}$. \\
(3) Controlled on basis state $\ket{x}$ in register $\reg{A}$, call the oracle $f$ on input $(U,V,x)$ to obtain \\ 
~~~~~~ $\phi \in [0,2\pi)$ in register $\reg{B}$ and $y \in \{0,1\}^{n'}$ in register $\reg{C}$.  \\
(4) Controlled on basis state $\ket{\phi}$ in register $\reg{B}$, apply the phase $e^{i \phi}$.\\
(5) Controlled on basis state $\ket{y}$ in register $\reg{C}$, call the oracle $g$ on input $(U,V,y)$ to \\ 
~~~~~~  uncompute $\ket{x} \otimes \ket{\phi}$ in registers $\reg{A}$ and $\reg{B}$. \\
(6) Apply the inverse unitary $V^\dagger$ on register $\reg{C}$.\\
(7) Output the first $n$ qubits of register $\reg{C}$.
\end{tabbing}
\end{minipage}}

\end{center}
\caption{Pseudo-code for the {\sc TwoQueryStateSynthesis} query algorithm that uses exponential space complexity.}
\label{fig:twoquery}
\end{figure}

The algorithm has the following behavior. Fix unitaries $U,V$. Let
\begin{align}
	\ket{\theta} = U \ket{0}^{\otimes n'} = \sum_{x \in \{0,1\}^{n'}} \rv{u}_x \ket{x}
\end{align}
where $\rv{u}_x \in \C$ are amplitudes that depend on the unitary $U$. 
After Step 3 of the algorithm, the state of registers $\reg{A} \reg{B} \reg{C}$ is
\begin{align}
	\sum_x \rv{u}_x \ket{x}_{\reg{A}} \otimes \ket{\phi(U,V,x)}_{\reg{B}} \otimes \ket{\sigma_{U,V}(x)}_{\reg{C}}~.
\end{align}
After Step 4, the state becomes
\begin{align}
	\sum_x e^{i \phi(U,V,x)} \, \rv{u}_x\, \ket{x}_{\reg{A}} \otimes \ket{\phi(U,V,x)}_{\reg{B}} \otimes \ket{\sigma_{U,V}(x)}_{\reg{C}}~.
\end{align}
After Step 5, the state becomes
\begin{align}
	\ket{0,0}_{\reg{A} \reg{B}} \otimes \sum_x e^{i \phi(U,V,x)} \, \rv{u}_x \, \ket{\sigma_{U,V}(x)}_{\reg{C}}~.
\end{align}
The state of register $\reg{C}$ at the end of the algorithm is thus
\begin{equation}
\label{eq:final-state}
V^\dagger \sum_x e^{i \phi(U,V,x)} \, \rv{u}_x \, \ket{\sigma_{U,V}(x)}_{\reg{C}}~.
\end{equation}
We now argue that for every choice of $n$-qubit state $\ket{\psi}$, there exist permutations $\{ \sigma_{U,V} \}_{U,V}$ and a function $\phi(U,V,x)$ such that the state in~\eqref{eq:final-state}, with high probability over $U,V$, is exponentially close to $\ket{\psi} \otimes \ket{0}^{\otimes n' - n}$. 
Write
\begin{align}
	V (\ket{\psi} \otimes \ket{0}^{\otimes n' - n}) = \sum_{y \in \{0,1\}^{n'}} \rv{v}_y \ket{y}~.
\end{align}
Let $\rv{u} = (\rv{u}_x)_{x \in \{0,1\}^{n'}}$ and $\rv{v} = (\rv{v}_y)_{y \in \{0,1\}^{n'}}$. Let $|\rv{u}|$ and $|\rv{v}|$ denote the entry-wise absolute values of $\rv{u}$ and $\rv{v}$, respectively. For a vector $v \in \RR^m$, let $\sort(v) \in \RR^m$ denote the entries of $v$ sorted in non-increasing order.

Define the permutation $\sigma_{U,V}$ on $\{0,1\}^{n'}$ to be such that $\sigma_{U,V}(x) = y$ if and only if $|\rv{u}_x|$ and $|\rv{v}_y|$ have the same \emph{rank} in the sorted vectors $\sort(|\rv{u}|)$ and $\sort(|\rv{v}|)$, respectively. In other words, if $|\rv{u}_x|$ is the $k$'th largest entry in $\sort(|\rv{u}|)$, then $|\rv{v}_y|$ must also be the $k$'th largest entry in $\sort(|\rv{v}|)$.

Define $\phi(U,V,x)$ to be the best $(n')^2$-bit approximation of the phase $\theta(U,V,x) \in [0,2\pi)$ such that $e^{i \theta(U,V,x)} \, \rv{u}_x$ has the same phase as $\rv{v}_y$ where $y = \sigma_{U,V}(x)$. In other words, for all $U,V,x$, we have
\begin{align}
    | \phi(U,V,x) - \theta(U,V,x) | \leq \exp(-\Omega (n')^2)~.
\end{align}
Since $|e^{ib} - e^{ia}| \leq |a - b|$ for all $a,b \in \RR$, we have
\begin{equation}
\label{eq:wasserstein-0}
    | e^{i \phi(U,V,x)} - e^{i \theta(U,V,x)} | \leq \exp(-\Omega (n')^2)~.
\end{equation}
We now analyze the distance between~\eqref{eq:final-state} and $\ket{\psi} \otimes \ket{0}^{\otimes n' - n}$, or equivalently, the distance

\begin{align}
	&\left \| \sum_x e^{i \phi(U,V,x)} \, \rv{u}_x \, \ket{\sigma_{U,V}(x)} - V (\ket{\psi} \otimes \ket{0}^{\otimes n' - n}) \right \|^2 \\
	&= \left \| \sum_x \Big( e^{i \phi(U,V,x)} \, \rv{u}_x - \rv{v}_{\sigma_{U,V}(x)} \Big) \, \ket{\sigma_{U,V}(x)} \right \|^2 \\
	&= \sum_x \left | e^{i \phi(U,V,x)} \, \rv{u}_x - \rv{v}_{\sigma_{U,V}(x)} \right |^2 \\
	&\leq \sum_x \Big( \left | e^{i \theta(U,V,x)} \, \rv{u}_x - \rv{v}_{\sigma_{U,V}(x)} \right | + | e^{i \theta(U,V,x)} -  e^{i \phi(U,V,x)} | \Big)^2 \label{eq:wasserstein-1} \\
	&\leq \sum_x 2  \left | |\rv{u}_x| - |\rv{v}_{\sigma_{U,V}(x)}| \right |^2 + 2 | e^{i \theta(U,V,x)} -  e^{i \phi(U,V,x)} |^2 \\
	&\leq 2 \| \sort(|\rv{u}|) - \sort(|\rv{v}|) \|^2 + 2^{n'+1} \exp(-\Omega(n')^2) 
\end{align}
where \Cref{eq:wasserstein-1} follows from the definition of $\phi(U,V,x)$, and the last line follows from the definition of the permutation $\sigma_{U,V}$ as well as \Cref{eq:wasserstein-0}.

We now argue that, with overwhelming probability, the distance $\| \sort(|\rv{u}|) - \sort(|\rv{v}|) \|^2$ is exponentially small. Notice that the vectors $\rv{u}, \rv{v}$ correspond to the amplitudes of two independent Haar-random states on $n'$ qubits; this allows us to use the following Proposition.

\begin{proposition} \label{prop:sort} 
Let $\rv{u}, \rv{v}$ denote two independent Haar-random unit vectors in $\C^d$. There exist universal constants $C,C' > 0$ such that 
\begin{align}
\Pr \left [ \| \sort(|\rv{u}|) - \sort(|\rv{v}|) \| \geq C d^{-1/4} \right ] \leq \exp(-C' d^{1/4})
\end{align}
where the probability is over the choice of $\rv{u},\rv{v}$. 
\end{proposition}

Applying \Cref{prop:sort} shows that with probability at least $1 - \exp(-\Omega((d')^{1/4}))$ over the choice of $U,V$, the output of the algorithm is $O((d')^{-1/4})$-close to $\ket{\psi}$. 

%

\subsection{A space-efficient algorithm}

The algorithm described above suffers from needing \emph{exponential space complexity}; this is because specifying a Haar-random unitary on $n'$ qubits requires $\exp(\Omega(n'))$ space, and thus the oracles $f,g$ need to act on exponentially many input bits. We derandomize this construction, and show via the probabilistic method that there exists a \emph{single} choice of unitary $U_\star,V_\star$ that works for \emph{all} $n$-qubit states --- this is why we expanded the space to dimension $d'$.

Let $\ket{\psi_1},\ldots,\ket{\psi_D}$ denote an $\epsilon$-net for the space of $n$-qubit quantum states. Set $\epsilon = d^{-1}$. Then there at most $D \leq \epsilon^{-d} = d^{d}$ states in this enumeration. Fix an index $1 \leq i \leq D$. Imagine running the aforementioned protocol on state $\ket{\psi_i}$. As discussed, the probability that the protocol fails to synthesize a $(d')^{-1/4}$-approximation to $\ket{\psi_i}$ is at most $\exp(-\Omega( (d')^{1/4}))$ over the choice of $U,V$. By a union bound, the probability that a random choice of $U,V$ fails to synthesize a $(d')^{-1/4}$-approximation to \emph{a single one} of the $\ket{\psi_1},\ldots,\ket{\psi_D}$ is at most
\begin{align}
	d^d \cdot \exp(-\Omega( (d')^{1/4})) \leq 2^{n2^n} \cdot \exp(-\Omega( 2^{n^2/4}))
\end{align}
which, for sufficiently large $n$, is less than $1$. Thus there exists a choice of unitaries $U_\star,V_\star$ that enables successful synthesis of \emph{all} the $\ket{\psi_1},\ldots,\ket{\psi_D}$. Hardcode these unitaries into the algorithm and oracles in \Cref{fig:twoquery}; i.e., the oracles $f$ and $g$ only take $n'$ bits as input. The resulting query algorithm now only requires $\poly(n)$ space. Note that the implementation of the unitaries $U_\star,V_\star$ will not be time-efficient in general, but they are still fixed $n'$-qubit unitary operators that are independent of the state being synthesized.

%
%

Thus for an arbitrary state $\ket{\psi}$, by letting the oracles $A,B$ correspond to the nearest state $\ket{\psi_i}$ in the $\eps$-net that is within $d^{-1}$ of $\ket{\psi}$, the two-query algorithm synthesizes $\ket{\psi}$ with $O(d^{-1})$ error.

\begin{theorem}
[Two Query State Synthesis Performance]
For all $n$-qubit states $\ket{\tau}$, the algorithm {\sc TwoQueryStateSynthesis} uses $\poly(n)$ space, makes two queries to a classical oracle depending on $\ket{\tau}$, and outputs a mixed state that is $\exp(-\Omega(n))$-close in trace distance to $\ketbra{\tau}{\tau}$. 
\label{thm:twoquerystatesynthesis}
\end{theorem}

%
%
%
%

\subsection{Proof of \Cref{prop:sort}}

We now prove \Cref{prop:sort}. 

Sample a Haar-random unit vector $\rv{u} \in \C^d$ by sampling i.i.d. complex Gaussians $\rv{g} = (\rv{g}_1,\ldots,\rv{g}_d)$ where $\rv{g}_i \sim \C\normal(0,1)$ \footnote{A complex Gaussian $\rv{z} \sim \C\normal(0,1)$ can be sampled by sampling two independent real Gaussians $\rv{x},\rv{y} \sim \normal(0,\frac{1}{2})$ and then setting $\rv{z} = \rv{x} + i\rv{y}$.} for all $i$, and then defining
\begin{align}
	\rv{u} \defeq \frac{\rv{g}}{\| \rv{g} \|}~.
\end{align}

Sample another independent Haar-random unit vector $\rv{v} = \rv{h}/\| \rv{h} \|$ in the same way. Define $|\rv{u}|$ and $|\rv{v}|$ to be the entry-wise absolute value of the vectors $\rv{u}$ and $\rv{v}$, respectively. 


We prove \Cref{prop:sort} by arguing that $\| \sort(|\rv{g}|) - \sort(|\rv{h}|) \|$ is small with overwhelming probability, and then argue that dividing by the norms of $\rv{g}$ and $\rv{h}$ respectively does not change the distance by much. 



Let $\hat{\rv{g}} = |\rv{g}|$ and $\hat{\rv{h}} = |\rv{h}|$. We prove this Proposition by first observing that the entries of $\hat{\rv{g}}$, $\hat{\rv{h}}$ are distributed according to a \emph{Rayleigh distribution}~\cite{wiki:rayleigh}. In other words, since each entry is distributed as the absolute value of a standard complex Gaussian $\rv{z} \sim \C\normal(0,1)$, we have
\begin{align}
	\sqrt{\rv{z} \rv{z}^*} = \sqrt{(\rv{x} + i \rv{y})(\rv{x} - i \rv{y})} = \sqrt{ \rv{x}^2 + \rv{y}^2}
\end{align}
where $\rv{x},\rv{y} \sim \normal(0,\frac{1}{2})$. We now relate $\| \sort(\hat{\rv{g}}) - \sort(\hat{\rv{h}}) \|^2$ to the \emph{$2$-Wasserstein distance} between the empirical distributions on $\RR$ induced by $\hat{\rv{g}}$ and $\hat{\rv{h}}$.

%

We explain the $2$-Wasserstein distance. Let $\mu, \nu$ denote two probability measures on the real line $\RR$. Then we define the $2$-Wasserstein distance between them as 
\begin{align}
	W_2^2(\mu,\nu) \defeq  \inf_{\omega} \int \left | \mu(x) - \nu(y) \right|^2 \mathrm{d}\omega(x,y)
\end{align}
where the infimum is over all probability measures $\omega$ on $\RR \times \RR$ such that the marginal on the first coordinate is $\mu$, and the marginal on the second coordinate is $\nu$. This is also known as the ``$\ell_2$-Earth Mover Distance'', because it captures the minimum amount of ``earth'' that has to be moved if we want to morph the distribution $\mu$ to $\nu$, viewing the distributions as mounds of dirt. We now list some useful properties of the Wasserstein distances.

Let $\mu,\nu,\tau$ denote probability distributions on $\RR$. Then
\begin{enumerate}
	\item (Symmetry) $W_2(\mu,\nu) = W_2(\nu,\mu)$.
	\item (Triangle inequality) $W_2(\mu,\tau) \leq W_2(\mu,\nu) + W_2(\nu,\tau)$.
	\item (Discrete distributions) Suppose that $\mu$ (resp. $\nu$) is a discrete distribution that assigns $\frac{1}{d}$ mass on distinct points $x_1,\ldots,x_d \in \RR$ (resp. $y_1,\ldots,y_d \in \RR$). Then 
	\begin{align}
		W_2^2(\mu,\nu) = \frac{1}{d} \sum_{i=1}^d |\tilde{x}_i - \tilde{y}_i|^2
	\end{align}
	where $(\tilde{x}_1,\ldots,\tilde{x}_d) = \sort(x_1,\ldots,x_d)$ and similarly $(\tilde{y}_1,\ldots,\tilde{y}_d) = \sort(y_1,\ldots,y_d)$. A reference for this can be found in Lemma 4.2 of~\cite{bobkov2019one}.
\end{enumerate}

Finally, we will use the fact that the empirical distribution of $d$ i.i.d. Rayleigh samples converges to the Rayleigh distribution $\cal{R}$ with respect to the Wasserstein distance. The first lemma states that the \emph{expected} value of the Wasserstein distance between the two distributions is small:

\begin{lemma}
\label{lem:convergence}
Let $\cal{R}$ denote the Rayleigh distribution. Let $\mu_{\hat{\rv{g}}}$ denote the discrete distribution that assigns $1/d$ mass to each point $\hat{\rv{g}}_1,\ldots,\hat{\rv{g}}_d$. Then 
\begin{align}
	\Big [ \Exp W_2(\mu_{\hat{\rv{g}}},\cal{R}) \Big]^2 \leq \Exp W_2^2(\mu_{\hat{\rv{g}}},\cal{R}) \leq O \Big ( \frac{\log d}{d} \Big)
\end{align}
where the expectation is over the choice of $\hat{\rv{g}}$.
\end{lemma}
\begin{proof}
This is a consequence of Corollary 6.12 of~\cite{bobkov2019one}, which states that for a log-concave density $\nu$ with standard deviation $\sigma$, we have
\begin{align}
\Big [ \Exp W_2(\nu_d,\nu) \Big]^2 \leq \Exp W_2^2(\nu_d,\nu) \leq \frac{C \sigma^2 \log d}{d}
\end{align}
for some universal constant $C > 0$. Here $\nu_d$ denotes the empirical distribution arising from $d$ independent samples from $\nu$. 

The lemma is established by using that the Rayleigh distribution $\cal{R}$ is log-concave with standard deviation $\sqrt{ \Big(2 - \frac{\pi}{2} \Big) \cdot \frac{1}{\sqrt{2}} }$~\cite{wiki:rayleigh}.
\end{proof}

The next lemma  shows that $W_2(\mu_{\hat{\rv{g}}},\cal{R})$ \emph{concentrates} around $\Exp W_2(\mu_{\hat{\rv{g}}},\cal{R})$:
\begin{lemma}[{\cite[Theorem 7.1]{wiki:rayleigh}}]
\label{lem:convergence2}
For all log-concave densities $\nu$ with standard deviation $\sigma$, for all $\delta > 0$,
\begin{align}
	\Pr \left [ \left | W_2(\nu_d,\nu) - \Exp W_2(\nu_d,\nu) \right| \geq \delta \right ] \leq O\Big (\exp(-\frac{d^{1/2} \delta}{6 \sigma^2}) \Big)~.
\end{align}
\end{lemma}

Now we put everything together to prove \Cref{prop:sort}. Let $\mu_{\hat{\rv{g}}},\mu_{\hat{\rv{h}}}$ denote the empirical distributions corresponding to the random vectors $\hat{\rv{g}},\hat{\rv{h}}$. Then 
\begin{align}
	\| \sort(\hat{\rv{g}}) - \sort(\hat{\rv{h}}) \| = \sqrt{d} \,\, W_2(\mu_{\hat{\rv{g}}},\mu_{\hat{\rv{h}}}) \leq \sqrt{d} \Big( W_2(\mu_{\hat{\rv{g}}},\cal{R}) + W_2(\mu_{\hat{\rv{h}}},\cal{R}) \Big)
\end{align}
where we used the triangle inequality for the Wasserstein distance. Using \Cref{lem:convergence} and \Cref{lem:convergence2} with $\delta = d^{-1/4}$ we get that 
\begin{align}
	&\Pr \left [ \| \sort(\hat{\rv{g}}) - \sort(\hat{\rv{h}}) \| \geq 2 (d^{1/4} + O(\log d))  \right ] \\
	&\leq \Pr \left [ W_2(\mu_{\hat{g}},\cal{R}) + W_2(\mu_{\hat{h}},\cal{R})) \geq 2 (d^{-1/4} + O(d^{-1/2} \log d))  \right ] \\
	&\leq 2\Pr \left [ W_2(\mu_{\hat{g}},\cal{R}) \geq d^{-1/4} + O(d^{-1/2} \log d))  \right] \\
	&\leq O \Big( \exp( -\Omega(d^{1/4})) \Big)~.	\label{eq:Gaussian-concentration}
\end{align}


We now relate this to the distance $\| \sort(|\rv{u}|) - \sort(|\rv{v}|) \|^2$:
\begin{align}
\| \sort(|\rv{u}|) - \sort(|\rv{v}|) \|^2 &= \| \rv{u} \|^2 + \| \rv{v} \|^2 - 2 \langle \sort(|\rv{u}|), \sort(|\rv{v}|) \rangle \\
&= 2 \Big ( 1 - \frac{1}{ \| \rv{g} \| \cdot \| \rv{h} \|} \, \langle \sort(\hat{\rv{g}}), \sort(\hat{\rv{h}}) \rangle \Big) \\
&\leq \frac{ \| \sort(\hat{\rv{g}}) - \sort(\hat{\rv{h}}) \|^2}{ \| \rv{g} \| \cdot \| \rv{h} \|}
\end{align}
where we used that $\rv{u},\rv{v}$ are unit vectors, and the last inequality follows from
\begin{align}
2 \langle \sort(\hat{\rv{g}}), \sort(\hat{\rv{h}}) \rangle = \| \rv{g} \|^2 + \| \rv{h} \|^2 - \| \sort(\hat{\rv{g}}) - \sort(\hat{\rv{h}}) \|^2  \geq 2 \| \rv{g} \| \cdot \|\rv{h} \| - \| \sort(\hat{\rv{g}}) - \sort(\hat{\rv{h}}) \|^2~.
\end{align}

We now argue that $\|\hat{\rv{g}} \|, \| \hat{\rv{h}} \|$ concentrate around $\sqrt{d}$. This follows from standard (sub-)Gaussian concentration bounds. We use the following bound:
\begin{lemma}[Concentration of sum of squared Gaussians {\cite[Section 1.5.2]{gupta-mohanthy}}]
\label{lem:Gaussian}
Let $\rv{x}_1,\ldots,\rv{x}_m \sim \cal{N}(0,1)$ denote i.i.d. standard Gaussians. Let $\rv{z} = \rv{x}_1^2 + \cdots + \rv{x}_m^2$. Then 
\begin{align}
	\Pr \left [ \left | \rv{z} - m \right | \geq \epsilon m \right ] \leq 2 \exp \left ( - \frac{m \epsilon^2 }{8} \right)~.
\end{align}
\end{lemma}
The norms $\| \rv{g} \|^2 ,\| \rv{h} \|^2$ are sums of squared Gaussians; this is because 
\begin{align}
	\| \rv{g} \|^2 = \sum_j | \rv{g}_j|^2 = \sum_j \rv{g}_j \rv{g}_j^* = \sum_j (\rv{x}_j + i \rv{y}_j)(\rv{x}_j - i \rv{y}_j) = \sum_j \rv{x}_j^2 + \rv{y}_j^2
\end{align}
where $\rv{x}_j,\rv{y}_j$ are independent real Gaussians drawn from $\normal(0,\frac{1}{2})$. Thus $2 \| \rv{g} \|^2$ is a sum of $2d$ independent real Gaussians drawn from $\normal(0,1)$, and thus \Cref{lem:Gaussian} applies. We get
\begin{align}
	\Pr \left [ \| \rv{g} \| \leq \sqrt{d/2} \right ] = \Pr \left [ 2 \| \rv{g} \|^2 \leq d \right ] \leq 2 \exp \left ( - \frac{d}{16} \right)~.
\end{align}
Thus, by union bound,
\begin{align}
\Pr \left [ \| \rv{g} \| \cdot \|\rv{h}\| \leq d/2 \right ] \leq 4\exp \left ( - \frac{d }{16} \right)~.
\end{align}
Combining this with~\eqref{eq:Gaussian-concentration} we get
\begin{align}
\| \sort(|\rv{u}|) - \sort(|\rv{v}|) \| \leq O \Big ( d^{-1/4} + d^{-1/2} \log d \Big) = O(d^{-1/4})
\end{align}
with probability at least $1 - \exp(-\Omega(d^{1/4}))$.

%% file: prelims.tex
\section{Preliminaries}
\label{sec:prelim}

We assume the reader has a standard knowledge of quantum computation lexicon. The following are definitions important for the results of this paper.

\paragraph{Circuits} Quantum circuits on $n$ qubits are defined as an ordered lists of unitaries, each acting on 2-qubits, and a specification $(i,j) \in [n]^2$ of the two qubits being acted on. $\oracle$-oracle quantum circuits are quantum circuits augmented with list terms corresponding to the mulit-qubit unitary $\oracle$ and a specification of the qubits $\subseteq [n]$ being acted on. This can be generalized to multiple oracles.

\subsection{Proximity of phase states}
\label{subsec:prelim-phase-state-prelims}

In this subsection, we define the notion of a phase state and prove results on how they are good approximations for random states. These definitions are pertinent for Section \ref{sec:QMA} (where we discuss search-to-decision for $\QMA$) as well as \Cref{sec:one-query} (where we discuss the $1$-query algorithm for general state synthesis). 

\begin{definition}
\label{def:phase-state}
  A \emph{phase state} on $n$ qubits is a state of the form
  \begin{align} \ket*{p_f} = 2^{-n/2} \sum_{x\in\bits^n} (-1)^{f(x)} \ket{x}, \end{align}
  for some boolean function $f: \{0,1\}^n \to \{0,1\}$.
\end{definition}

Phase states are an interesting collection of states as they form good approximations for states of large $\ell_1$ norm, i.e. $\norm{\cdot}_1 \approx \Omega(\sqrt{2^n})$. While basis states do not have this property (their $\ell_1$ norm is 1), it is a property of Haar-random states and even outputs of 2-designs. We show this in the following statements.


\begin{fact}[Good overlap with phase states]
Let $\ket{a} \in \RR^d$ be a vector. There exists a phase state $\ket{p_f}$ defined by $f(x) =\sgn(a_x)$ and 
\begin{align}
    \abs{ \braket{a}{p_f} } = \sum_x \frac{a_x \cdot \sgn(a_x)}{\sqrt{d}} = \frac{\norm{\ket{a}}_1}{\sqrt{d}}.
\end{align}
\label{fact:general-state-overlap}
\end{fact}
\noindent Furthermore, the previous phase state is the optimal phase state in terms of overlap (up to a global phase).

\begin{fact}
Let $\Cc$ be a unitary 2-design on $n$ qubits and $\ket{\tau}$ be an
arbitary $n$-qubit state. Then,
\begin{align}
    \Exp_{C \in \Cc} \abs{\bra x C \ket \tau}^2 = \frac{1}{2^n}, \textrm{ and}\qquad \Exp_{C \in \Cc} \abs{\bra x C \ket \tau}^4 = \frac{2}{2^n(2^n + 1)}.
\end{align}
Furthermore, for any $x \in [2^n]$,
\begin{align}
    \Pr_{C \in \Cc} \left( \abs{\bra x C \ket{\tau}}^2 \geq \frac{\theta}{2^n} \right) \geq \frac{(1-\theta)^2}{2}.
\end{align}
\label{fact:ell_4_norm_of_haar}
\end{fact}


\begin{proof}
First suppose that $C$ is a Haar random unitary. For the first two statements, calculations of second and fourth moments for Haar random states in $\CC^d$ are well-known and can be found in \cite[Chapter 4.2]{semicircle-law}, which in general gives
\begin{align}
    \Exp_{\ket{\psi} \sim \Hh} \abs{ \braket{x}{\psi}}^2 = \frac{1}{d}, \textrm{ and}\qquad \Exp_{\ket \psi \sim \Hh} \abs{ \braket{x}{\psi}}^4 = \frac{2}{d(d+1)}.
\end{align}
For the third statement, by the Payley-Zygmund inequality, we have
\begin{align}
    \Pr_{\ket{\psi} \sim \Hh} \left( \abs{\braket{x}{\psi}}^2 \geq \frac{\theta}{d} \right) \geq (1-\theta)^2 \cdot \frac{\left(\Exp \abs{ \braket{x}{\psi}}^2\right)^2}{\Exp \abs{ \braket{x}{\psi}}^4} = (1-\theta)^2 \cdot \frac{d^2 + d}{2d^2} \geq \frac{(1-\theta)^2}{2}.
\end{align}
The expectations only involve polynomials of degree 2 in the entries of $C$ and $C^{\dagger}$, and so have the same value when $C$ is chosen uniformly from $\mathcal{C}$.
We can furthermore translate these results to unitary 2-designs. For example,
\begin{align}
    \Exp_{C \in \Cc} \abs{\bra{x} C \ket{\tau}}^4 &= \Exp_{C \in \Cc} \abs{\bra{x} C \ketbra{\tau} C^\dagger \ketbra{x} C \ketbra{\tau} C^\dagger \ket{x} }\\
    &= \int_{U \sim \Hh} dU  \abs{ \bra{x} U \ketbra{\tau} U^\dagger \ketbra{x} U \ketbra{\tau} U^\dagger \ket{x} }\\
    &= \Exp_{\psi \in \Hh} \abs{ \braket{x}{\psi}\braket{\psi}{x}\braket{x}{\psi}\braket{\psi}{x}} \\
    &= \Exp_{\psi \in \Hh} \abs{ \braket{x}{\psi}}^4 \\
    &= \frac{2}{2^{n}(2^n + 1)}.
\end{align}
Here we defined $\ket{\psi} = U \ket{\tau}$ which is Haar-randomly distributed and used the result for Haar states. Likewise, we can use similar arguments for the second moment and the probability bound.
\end{proof}

Then, this can be easily extended to showing that
\begin{align}
    \Exp_{\substack{C \in \Cc \\ \ket{\psi} = C \ket{\tau}}} \sum_x \abs{\braket{x}{\psi}}^4 = \frac{2}{2^n+1}.
\end{align}
By a standard Markov's inequality argument and the following fact, Fact \ref{fact:holder-usecase},
\begin{align}
    \Pr_{\substack{C \in \Cc \\ \ket{\psi} = C \ket{\tau}}} \left( \norm{\ket{\psi}}_1 \geq \frac{\sqrt{2^n}}{2\sqrt{\alpha}} \right) &\geq \Pr_{\substack{C \in \Cc \\ \ket{\psi} = C \ket{\tau}}} \left( \frac{1}{\sqrt{\sum_x \abs{\psi_x}^4}} \geq \sqrt{\frac{2^n+1}{2\alpha}} \right) \\
    &= \Pr_{\substack{C \in \Cc \\ \ket{\psi} = C \ket{\tau}}} \left( \sum_x \abs{\psi_x}^4 \leq
    \frac{2\alpha}{2^n+1} \right) \\
    &\geq 1 - \frac{1}{\alpha}.
    \label{eq:l1-bound}
\end{align}

\begin{fact}
Let $\ket{a} \in \CC^d$ be a unit vector. Then,
\begin{align}
    \norm{\ket{a}}_1 \geq \frac{1}{\norm{\ket{a}}_4^2} = \frac{1}{\sqrt{\sum_x \abs{a_x}^4}}.
\end{align}
\label{fact:holder-usecase}
\end{fact}
\begin{proof}
Let $\ket{b} = \sum_x \abs{a_x}^{4/3} \ket{x}$ and $\ket{c} = \sum_x \abs{a_x}^{2/3} \ket{x}$. Then by H\"older's inequality,
\begin{align}
    1 = \sum_x \abs{a_x}^2 &= \sum_x \abs{b_x c_x} \\
    &\leq \left( \sum_x \abs{b_x}^3 \right)^{1/3} \cdot \left( \sum_x \abs{c_x}^{3/2} \right)^{2/3} \\
    &= \left( \sum_x \abs{a_x}^4 \right)^{1/3} \cdot \left( \sum_x \abs{a_x} \right)^{2/3} \\
    &= \norm{\ket{a}}_4^{4/3} \cdot \norm{\ket{a}}_1^{2/3}.
\end{align}
The statement follows after rearrangement.
\end{proof}

Now we can prove our main lemma about the approximability of states by 2-design unitaries. We recall that the collection of Clifford unitaries form a 2-design \cite{cliffords-are-2-designs}. We denote the Clifford group as $\Cliff$.

\overlap

%

\begin{proof}
For any state $\ket \psi \in \qubits{n}$, we can write $\ket \psi$ as $\ket a + i \ket b$ where $\ket a, \ket b \in \RR^{2^n}$. With probability $\half$, $\norm{\ket a}_1 \geq \norm{\ket b}_1$. Combining this with the previous statement \eqref{eq:l1-bound},
\begin{align}
\Pr_{\substack{C \in \Cc \\ \ket{a} = \Re(C \ket{\tau})}} \left( \norm{\ket{a}}_1 \geq \sqrt{\gamma d} \right) \geq \frac{1}{2} - 2\gamma.
\end{align}
Therefore, since $\abs{\braket*{\psi}{p_f}}^2 \geq \abs{\braket*{a}{p_f}}^2$, we can use Fact \ref{fact:general-state-overlap} to complete the proof:
\begin{align}
    \Pr_{\substack{C \in \Cc \\ \ket{\psi} = C \ket{\tau}}} \left(  \abs{\braket*{\psi}{p_f}}^2 \geq \gamma \right) &\geq \Pr_{\substack{C \in \Cc \\ \ket{a} = \Re(C \ket{\tau})}} \left(  \abs{\braket*{a}{p_f}}^2 \geq \gamma \right) \\
    &= \Pr_{\substack{C \in \Cc \\ \ket{a} = \Re(C \ket{\tau})}} \left( \norm{\ket{a}}_1 \geq \sqrt{\gamma d} \right).
\end{align}
\end{proof}






\subsection{Phase estimation algorithm results}
\label{subsec:prelim-phase-estimation}

Our algorithm in Section \ref{sec:QMA} will rely on a finer analysis of the phase estimation algorithm. We use the following lemmas.

\begin{lemma}[{\cite[Corollary 16]{rall-phase-estimation}}]
  Let $H$ be a local Hamiltonian on $n$ qubits and $\delta > 0$. Then
  the Improved Energy Estimator of~\cite{rall-phase-estimation} has
  query complexity $O(\log(\delta^{-1}) \cdot 2^m)$ that returns an
  estimate to the energy that is accurate up to $1/2^m$. That is, on
  input that is an eigenstate of $H$ with eigenvalue
  $E_j$, the output state of the algorithm is within trace
  distance $\delta$ of 
  \begin{align} \rho_j = (p_j \proj{\lfloor 2^m E_j \rfloor} + (1
    - p_j) \proj{\lfloor 2^m E_j \rfloor - 1}) \otimes \proj{E_j},
    \end{align}
  where the outcome register is interpreted mod $2^m$, and $0 \leq p_j
  \leq 1$. 
  \label{lem:rall-energy}
\end{lemma}
\begin{proof}
  This follows from the cited lemma of \cite{rall-phase-estimation},
  applying the case without a rounding promise, and converting the
  stronger diamond norm bound on the map to a trace distance bound on
  the output state.
\end{proof}

The following fact describes the use of phase estimation to improve a
low-quality $\QMA$ witness. This technique was introduced by Abrams
and Lloyd~\cite{abrams-lloyd}.

\begin{lemma}
 Let $H$, $m$, and $\delta$ be as in the previous lemma, let
 $\ket{\psi}$ be an arbitrary $n$-qubit state. Let the eigenvalues of
 $U$ be $\lambda_i = e^{2\pi i \phi_i}$ where $\phi_i \in
 [0,1)$. Suppose that $\ket{\psi}$ has overlap at least $C > 0$ on the
 subspace spanned by eigenvectors corresponding to $\phi_i \in
 [a,b]$. Then, the Improved Energy Estimator of~\cite{rall-phase-estimation} returns a value $\theta \in [a - \frac{1}{2^m}, b + \frac{1}{2^m}]$ with probability at least $C - \delta$. Moreover, conditioned on phase estimation returning a value $\theta$ in this interval, the residual state has overlap
 at least $1 - 2\delta/C$ with the subspace spanned by eigenvectors of $U$ corresponding to
 $\phi_i \in [a - \frac{2}{2^m}, b + \frac{2}{2^m}]$.
 \label{lem:rall-energy-post}
\end{lemma}
\begin{proof}
  Write the input state as
  \begin{align} \ket{\psi} = \sum_{j} c_j \ket{E_j}. \end{align}
  Then, by \Cref{lem:rall-energy}, the state
  after the energy estimation algorithm is run is $\delta$-close in
  trace distance to the following ideal state:
  \begin{align} \rho_{ideal} = \sum_{j} |c_j|^2 (p_j \proj{\lfloor 2^m E_j \rfloor} + (1
    - p_j) \proj{\lfloor 2^m E_j \rfloor - 1}) \otimes \proj{E_j},\end{align}
  where the outcome register is interpreted mod $2^m$. The claim about the probability of returning $\theta$ in the specified range then follows by inspection, and the given bound that the overlap of $\ket{\psi}$ with the $[a,b]$-eigenspace is at least $C$.
  
  Now, for the claim about the residual state conditioned on obtaining $\theta$ in the desired range, consider jointly measuring both registers of the \emph{true}
  output state in the computational
  basis. This yields a probability distribution over outcomes
  \begin{align} \Pr[ \theta, E = E_j] \approx_{\delta} |c_j|^2 ( p_j \cdot \mathbf{1}[\theta =
    \lfloor 2^m E_j \rfloor] + (1 - p_j) \cdot \mathbf{1}[\theta =
    \lfloor 2^m E_j \rfloor - 1]), \end{align}
  where the $\approx_\delta$ notation means that the distributions on
  the two sides are $\delta$-close in total variational distance.
  
Define $E_{good} = [a - \frac{2}{2^m}, b + \frac{2}{2^m}]$ and
$\Theta_{good} = [a - \frac{1}{2^m}, b + \frac{1}{2^m}]$. The quantity
we now wish to bound from below is $\Pr[ E_j  \in E_{good} | \theta
\in \Theta_{good}]$. We may do this using Bayes' rule:

\begin{align}
  \Pr[ E \in E_{good}| \theta \in \Theta_{good}]
  &= \frac{\Pr[E \in E_{good} , \theta \in \Theta_{good}]}{\Pr[\theta
    \in \Theta_{good}]}. \\
  &= \frac{\Pr[ \theta \in \Theta_{good} | E \in E_{good} ] \Pr[
    E \in E_{good}]}{\Pr[\theta \in \Theta_{good}]} \\
  &\geq \frac{C - \delta}{C + \delta} \geq 1 - \frac{2 \delta}{C}.
\end{align}
\end{proof}

\subsection{Concentration Inequalities}
\begin{lemma}[Levy's Lemma]
\label{lem:levy}
	Let $\mathcal{S}(\C^d)$ denote the set of unit vectors in $\C^d$. Let $K$, $f: \mathcal{S}(\C^d) \to \RR$ be such that for all $\ket{\psi},\ket{\phi} \in \mathcal{S}(\C^d)$ we have
	\begin{align}
		| f(\ket{\psi}) - f(\ket{\phi}) | \leq K \cdot \| \ket{\psi} - \ket{\phi} \|.
	\end{align}
	Then there exists a universal constant $C >0$ where
	\begin{align}
		\Pr \left [ \left | f(\ket{\psi}) - \Exp f \right | \geq \delta \right ] \leq 2 \exp \left ( - C d \delta^2/K^2 \right)
	\end{align}
	where the probability is over the choice of $\ket{\psi}$ drawn from the Haar measure over $\mathcal{S}(\C^d)$, and $\Exp f$ is the average of $f$ over the Haar measure.
\end{lemma}

\subsection{Azuma's Inequality}

\begin{lemma}[Azuma's Inequality]
\label{lem:azuma}
Suppose that the sequence of random variables $S_0, S_1, S_2, \ldots$ is a
Martinagale ($E[S_{j+1}|S_1, \ldots, S_j] = S_j$ for all $j \ge 0$).
Suppose also that $|S_{j} - S_{j-1}| \le c_j$. Then
$$\Pr[S_L - S_0 < - \epsilon] \le \exp \left( \frac{- \epsilon^2}{2 \sum_{j=1}^{L} c_j} \right)$$
\end{lemma}